%% file: manuscript_games_product_form_demand_response_modelling.tex
\documentclass[a4paper,12pt]{book}
\usepackage[T1]{fontenc}
\usepackage[utf8]{inputenc}
\usepackage{float}
\usepackage{amsmath}
\usepackage{amsfonts}
\usepackage{indentfirst}
\usepackage{tabularray}
\usepackage{multirow}
\usepackage{euscript}
\usepackage[margin=2cm]{geometry}
\usepackage{graphicx}
\usepackage[export]{adjustbox}
\usepackage{wrapfig}
\usepackage{amssymb}
\usepackage{enumitem}
\usepackage{stmaryrd}
\usepackage{xcolor}
\usepackage{cancel}
\usepackage{optsys}
\usepackage{macros_demand_response_modelling}
\usepackage{tabularx} 
\usepackage[round]{natbib}
\usepackage{fancyhdr}
\usepackage{subcaption}
\usepackage{fontawesome5}
\usepackage{authblk} 
\usepackage{tikz} 
\usetikzlibrary{positioning}
\usetikzlibrary{calc}
\usetikzlibrary{decorations.pathmorphing}
\usepackage{hyperref}

\title{\bfseries Games in Product Form for\\Demand Response Modelling}

\author[1]{Thomas Buchholtzer}
\author[2]{Michel De Lara}

\affil[1]{CERMICS, École nationale des ponts et chaussées, IP Paris, thomas.buchholtzer@enpc.fr}
\affil[2]{CERMICS, École nationale des ponts et chaussées, IP Paris; michel.delara@enpc.fr}

\date{}

\begin{document}

\begin{titlepage}
    \vspace*{1cm}
    
    {\huge \bfseries \noindent Games in Product Form for\\  Demand Response Modelling \par}
    \vspace{0.4cm}

    {\Large \noindent Thomas Buchholtzer\textsuperscript{1}, Michel De Lara\textsuperscript{2}\par}
    \vspace{0.4cm}
    { \itshape \noindent \textsuperscript{1}CERMICS, École nationale des ponts et chaussées, IP Paris, thomas.buchholtzer@enpc.fr\\ 
    \textsuperscript{2}CERMICS, École nationale des ponts et chaussées, IP Paris, michel.delara@enpc.fr
    \par}

    \vspace{2cm}
    \noindent\rule{\textwidth}{1pt} 
    \vspace{0.01cm}

    {\bfseries \centering Abstract\par}
    \vspace{0.5cm}
   \noindent Energy systems are changing rapidly. More and more, energy production is becoming decentralized, highly variable and intermittent (solar, wind), while demand is diversifying (electric vehicles). As a result, balancing supply and demand is becoming more complex, making the adjustment of demand an interesting tool. Demand response is a typical leader-follower problem: a consumer (follower) adjusts his energy consumption based on the prices (or any other incentive) set by the supplier (leader). We propose a versatile and modular framework to address any leader-follower problem, focusing on the handling of often overlooked informational issues. First, we introduce a model that defines the rules of the game (W-model): agents are decision-makers, and Nature encapsulates everything beyond their control, such as private knowledge and exogenous factors. Following the so-called Witsenhausen intrinsic model, we present an efficient way to represent — on a product set, equipped with a product $\sigma$-algebra — the information available to agents when making decisions. Next, we introduce Games in Product Form (W-games) by equipping each player (a group of agents) with preferences (objective function and belief) over different outcomes. Thereby, we incorporate an additional layer of information, the characteristics of the preferences linked to players, which affects the possible definitions of an equilibrium. We make this explicit in Nash and Stackelberg equilibria. Equipped with this framework, we reformulate several papers on demand response, highlighting overlooked informational issues. We also provide an application based on the Thailand demand response program.

    \vspace{0.5cm}

    \vspace{0.35cm}
    \noindent\rule{\textwidth}{1pt} 
    
\end{titlepage}

\newpage
\thispagestyle{plain}

\tableofcontents


\pagestyle{fancy}
\fancyhf{} 
\fancyhead[L]{\leftmark} 
\fancyfoot[C]{\thepage}
\setlength{\headheight}{14.6pt}
\newenvironment{illustration}
{\par\vspace{1em}\noindent$\star$\hspace{0.5em}\sffamily}
  {\par\vspace{1em}}

\newcommand{\nb}[3]{
{\colorbox{#2}{\bfseries\sffamily\tiny\textcolor{white}{#1}}}
{\textcolor{#2}{$\blacktriangleright${#3}$\blacktriangleleft$}}}

\newcommand{\mdl}[1]{\nb{Michel}{red}{#1}}

\chapter{Introduction}

\section{An overview of demand response}

The increasingly visible consequences of climate change are pushing many countries, including those in the European Union, to aim for carbon neutrality by 2050.
This involves a greater reliance on renewable energy sources to decarbonize electricity production.
However, this transition complicates the management of the electricity grid, as renewable energies, such as solar and wind, are intermittent and difficult to predict.
In the past, energy systems could be easily controlled (e.g., with nuclear or gas), making it possible to match supply to demand.
Today, with these variable energy sources, it is necessary to rethink the approach and modulate demand rather than simply adjusting supply.

Moreover, the emergence of new actors, such as demand aggregators or prosumers (individuals who, for example, produce energy through solar panels installed on their roofs), is changing the role of consumers.
These individuals are no longer just passive users but are becoming producers, storers, and exchangers of energy within the grid. 
This shift requires rethinking the interactions between all these actors to ensure the smooth functioning of the system.

Faced with an increasingly complex and unpredictable energy market, technological advances now enable real-time adjustments to energy demand, a critical aspect of \emph{Demand Side Management} (DSM). 
DSM encompasses strategies designed to optimize energy consumption by influencing when and how energy is used. 
These strategies aim to enhance grid stability and efficiency, reduce the need for costly infrastructure upgrades, and minimize the environmental impact of energy consumption.

A central component of DSM is \emph{Demand Response} (DR), which adjusts consumer energy consumption in response to price signals or grid conditions, as defined in \cite{Albadi-El-Saadany:2008}.
DR can be achieved through price-based methods, where dynamic pricing incentivizes consumers to shift energy usage (e.g. reducing consumption during peak demand periods to prevent grid overloads).
Alternatively, incentive-based methods offer financial rewards to consumers who modify their energy use during high-demand periods, contributing to grid stability.

Such demand response problems can be modeled as \emph{leader-follower problems}, where the energy supplier (leader) sets prices, and consumers (followers) adapt their energy use accordingly \cite{Laffont-Martimort:2002}.
We propose analyzing these interactions using the \emph{Witsenhausen Intrinsic Model}, or \emph{W-model}, initially introduced in \citet{Witsenhausen:1971a} and extended in \citet{Witsenhausen:1975}. 
The W-model provides a structured framework to address decision-making under uncertainty, incorporating the private knowledge of both players (e.g. the producer's production costs and the consumer's unwillingness to shift consumption) as well as exogenous factors (e.g. weather conditions).

Building on this, we explore \emph{games in product form}, or \emph{W-games}, a concept developed by \citet{Heymann-DeLara-Chancelier:2022}, to formalize interactions between leaders and followers. 
This game-theoretic approach links decision-making processes to bilevel optimization, as highlighted in \citet{Wogrin-Pineda-Tejada-Arango:2020}, and examines equilibria such as Nash and Stackelberg, explored in \cite{Basar-Olsder:1998}.
Extending this framework, we address more complex scenarios, including multi-leader-multi-follower dynamics, and propose systematic reformulations to tackle demand response problems in smart grids, building on the general frameworks provided in \citet{Besancon:2020} and \citet{Antunes-Alves-Ecer:2020}.

\section{Structure of the document}

This document is organized into three parts.

\subsection{Part I: theoretic framework}

In Part~\ref{W-model_W-games_leader-follower_problems}, we introduce the mathematical tools. 
In Chap.~\ref{Background_on_W-model_and_W-games}, we start by presenting the Witsenhausen model.
Building on this foundation, we introduce a game concept known as the W-game, or game in product form, where we define the notion of game data to address issues of information asymmetry between players.
Finally, we express these games in normal form to facilitate discussions on Nash equilibrium, with particular attention to the role of game data.
After examining this general framework, we explain in Chap.~\ref{W-games_leader_follower_problems} how this formalism is particularly useful for addressing leader-follower problems.
To do this, we introduce the necessary notation and explain that the modularity of W-games allows for a unified approach to problems, whether static or dynamic, involving only two actors or a larger group of participants. 
The leader-follower structure also refines the Nash equilibrium concept by introducing the notion of a Nash-Stackelberg equilibrium.

\subsection{Part II: reformulating existing literature}

Following this theoretical foundation, in Part~\ref{reformulation_energy_management_problems_W-games} to the literature on leader-follower problems in the energy sector, transforming their typical bilevel optimization formulation into a W-game.
This shift enables us to focus on the information structure, which is often overlooked.
We split this part in two: in Chap.~\ref{Single-leader-single-follower_problems}, we address single-leader-single-follower problems, focusing on cases with only two players and in Chap.~\ref{Multi-leader-multi-follower_problems}, we discuss multi-leader-multi-follower problems, that is cases with multiple leaders and one follower, one leader and multiples followers or other variations like a trilevel case.
In both chapters, we begin with a detailed examination of selected articles that highlight different aspects of these problems. 
We then offer reading suggestions for further exploration using our approach and conclude by identifying similarities across the literature.

\subsection{Part III: modelling new problems}

Finally, in Part~\ref{Design_W-games_energy_management}, we propose a direct application of the W-game model on demand response on the Thai power grid.


\part[W-model and W-games for leader-follower problems]{W-model and W-games \\ for leader-follower problems}
\label{W-model_W-games_leader-follower_problems}


\chapter{W-model and W-games}
\label{Background_on_W-model_and_W-games}

Game theory is a mathematical field focused on the study of strategic interactions among rational decision-makers. 
The outcome for each player depends not only on their own actions but also on the actions of other players and random factors collectively referred to as Nature.
Each player is characterized by their payoff (the outcome they receive from the game), their information (what they know about other players' decisions), and their beliefs about the state of Nature. 
The representation of who knows what and when can be depicted using extensive games, which are often represented as trees (possibly infinite). 
Here, we utilize another method for efficiently representing information asymmetry, known as the \emph{Witsenhausen Intrinsic Model}, or \emph{W-model}, as introduced in \citet{Witsenhausen:1975} and further applied in \citet{Carpentier-Chancelier-Cohen-DeLara:2015} for modeling multi-level optimization problems.

Chap.~\ref{Background_on_W-model_and_W-games} details the different layers of our model to study leader-follower optimization problems in energy.
Sect.~\ref{W-model} introduces the Witsenhausen model, laying out the key elements required for its definition, along with the concepts of strategy and playability.
It leads us to a \emph{game form}.
Then, Sect.~\ref{W-games} expands on the model by introducing the idea of players, endowed with objective functions and beliefs, so that we can speak of W-games. 
Finally, Sect.~\ref{Nash_equilibrium} explores the concept of solutions within this framework, focusing on how players evaluate their strategies and how the concept of Nash equilibrium can be formulated.

\begin{illustration}
    Text in this font corresponds to concrete illustrations in energy management of the concepts being introduced.
\end{illustration}


\section{W-model}
\label{W-model}

Sect.~\ref{W-model} introduces the W-model by presenting the necessary elements for its definition 
in \S~\ref{Agents_Nature_actions_configuration_space_and_information_fields}, followed by the concepts of strategy and playability in \S~\ref{Strategies_playability_and_solution_map}.


\subsection{Agents, Nature, actions, configuration space \\ and information fields}
\label{Agents_Nature_actions_configuration_space_and_information_fields}

The W-model consists of a set of agents, of Nature, of a collection of action sets and information fields.

\subsubsection*{Agents}
Let $\AgentSet$ be a set (which can be infinite or finite). 
Its elements, denoted as $\agent$, are called \emph{agents}.
An agent is a decision-maker taking only one decision.
Later on, these agents should not be confused with players, who are individuals or corporations taking several decisions and endowed with an objective function and a risk measure, or belief.

\begin{illustration}
    We consider the problem of an electricity producer and a consumer over the course of a year.
    The electricity producer decides on peak and off-peak prices at the beginning of the year, and each day the consumer decides on his energy consumption.
    We model the problem using 366 agents: 1 for the electricity producer and 365 for the consumer.
\end{illustration}

\subsubsection*{Nature}
Let $(\Nature, \trib)$ be a measurable space. 
Recall that a measurable space is a pair $(\mathcal{X}, \mathfrak{T})$ where $\mathcal{X}$ is a set and $\mathfrak{T}$ is a $\sigma$-field on $\mathcal{X}$, that is a nonempty collection of subsets of $\mathcal{X}$ closed under complement and countable unions.
This last property is particularly useful for being able to work with probabilities.
We will always denote sets with uppercase calligraphic letters and $\sigma$-fields with uppercase Gothic letters.

The set $\Nature$ is referred to as \emph{Nature} and it encompasses all uncertainties, meaning everything except decisions.
Each element $\nature \in \Nature$ is called a \emph{state of Nature}.

\begin{illustration}
    Nature can be big and contain many factors.
    For instance, it can include spot market prices, temperature, production technologies (like solar panels and batteries), the consumer's reluctance to change their consumption, and many other elements that influence energy production and consumption.
\end{illustration}

\subsubsection*{Actions}
Each agent $\agent \in \AgentSet$ comes with a measurable space $(\CONTROL_{\agent}, \tribu{\Control}_{\agent})$.
Each element $\control_{\agent} \in \CONTROL_{\agent}$ is called an \emph{action}, or \emph{decision}, of the agent $\agent$.
Thus, $\CONTROL_{\agent}$ is called the \emph{action set}, or \emph{decision set}, of the agent $\agent$.


\begin{illustration}
    If an agent $\agent$ decides to accept or refuse an electricity contract, his action set writes $\CONTROL_{\agent} = \{ 0, 1 \}$, associated with the $\sigma$-field $\tribu{\Control}_{\agent} = \mathfrak{P}(\{ 0, 1 \})$.
    Similarly, if an agent $\bgent$ decides on an electricity price, his action set is $\CONTROL_{\bgent} = \RR_+$ equipped with the Borel $\sigma$-field $\tribu{\Control}_{\bgent} = \borel{ \RR_+ }$.
\end{illustration}

\subsubsection*{Configuration space}
The \emph{configuration space} is defined as the product space 
\begin{subequations}
\begin{align}
    \HISTORY 
    = \Nature \times \prod_{\agent \in \AgentSet}{\CONTROL_{\agent}} \eqsepv
\end{align}
equipped with the corresponding product $\sigma$-field
\begin{align}
    \tribu{\History} 
    = \trib \otimes \bigotimes_{\agent \in \AgentSet} \tribu{\Control}_{\agent} \eqfinp 
\end{align}
\end{subequations}

Elements of the configuration space $\HISTORY$ are called \emph{configurations}, or \emph{outcomes}.

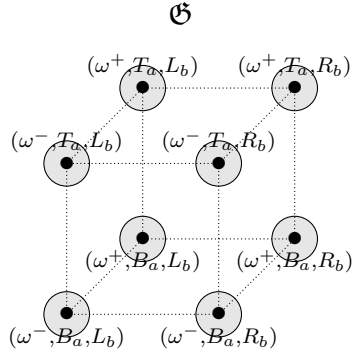
\begin{figure}[ht]
    \begin{center}
    \begin{tikzpicture}

          \foreach \x/\y in {0/0, 2/0, 0/2, 2/2, 1/1, 3/1, 1/3, 3/3} {
              \draw[fill=gray!20] (\x,\y) circle (0.3cm);
          }
          
          \draw[densely dotted] 
          (0,0) node[anchor=north]{$\begin{smallmatrix} (\nature^-, \Bot_\agent, \Lef_\bgent) \end{smallmatrix}$} -- 
          (2,0) node[anchor=north]{$\begin{smallmatrix} (\nature^-,\Bot_\agent, \Rig_\bgent) \end{smallmatrix}$}  -- 
          (2,2) node[anchor=south]{$\begin{smallmatrix} (\nature^-,\Top_\agent, \Rig_\bgent) \end{smallmatrix}$} --
          (0,2) node[anchor=south]{$\begin{smallmatrix} (\nature^-, \Top_\agent, \Lef_\bgent) \end{smallmatrix}$} -- 
          (0,0)
          (0,0) -- (1,1) node[anchor=north]{$\begin{smallmatrix} (\nature^+,\Bot_\agent, \Lef_\bgent) \end{smallmatrix}$}
          (2,0) -- (3,1) node[anchor=north]{$\begin{smallmatrix} (\nature^+,\Bot_\agent, \Rig_\bgent) \end{smallmatrix}$}
          (0,2) -- (1,3) node[anchor=south]{$\begin{smallmatrix} (\nature^+,\Top_\agent, \Lef_\bgent) \end{smallmatrix}$}
          (2,2) -- (3,3) node[anchor=south]{$\begin{smallmatrix} (\nature^+,\Top_\agent, \Rig_\bgent) \end{smallmatrix}$}
          (1,1) -- (1,3)  -- (3,3) -- (3,1) -- (1,1);
          \draw (1.5,4) node{$\trib$};
          
          \draw (2,0) node{$\bullet$};
          \draw (2,2) node{$\bullet$};
          \draw (0,2) node{$\bullet$};
          \draw (0,0) node{$\bullet$};
          \draw (1,1) node{$\bullet$};
          \draw (3,1) node{$\bullet$};
          \draw (1,3) node{$\bullet$};
          \draw (3,3) node{$\bullet$};
          
    \end{tikzpicture}
    \end{center}
    \caption{Illustration of the configuration space}
    \label{fig:illustration_configuration_space}
\end{figure}

\begin{illustration}
    We illustrate these definitions in the simple case of two agents, namely $\AgentSet = \{ \agent, \bgent \}$.
    Agent $\agent$ can play Top ($\Top_\agent$) or Bottom ($\Bot_\agent$) while agent $\bgent$ chooses between Left ($\Lef_\bgent$) and Right ($\Rig_\bgent$).
    It means we consider $\CONTROL_{\agent} = \{ \Top_\agent, \Bot_\agent \}$ and $\CONTROL_{\bgent} = \{ \Lef_\bgent, \Rig_\bgent \}$, equipped with the complete $\sigma$-fields.
    Similarly, we consider a Nature only made of a positive and a negative state, that is $\Nature = \{ \nature^+, \nature^- \}$, also equipped with the complete $\sigma$-field.
    Figure \ref{fig:illustration_configuration_space} depicts the corresponding configuration space $\HISTORY = \{ \nature^+, \nature^- \} \times \{ \Top_\agent, \Bot_\agent \} \times \{ \Lef_\bgent, \Rig_\bgent \}$.
\end{illustration}

\subsubsection*{Information fields}
Each agent $\agent \in \AgentSet$ is equipped with a subfield
\begin{equation}
    \tribu{\Information}_{\agent} \subset \tribu{\History} = \trib \otimes \bigotimes_{\agent \in \AgentSet} \tribu{\Control}_{\agent} \eqsepv
\end{equation}
called the \emph{information field} of the agent, which describes efficiently which other agents' decisions and which parts of Nature are observed when decision $\control_{\agent}$ is made.

\begin{figure}[ht]
    \begin{center}
    \begin{tikzpicture}[scale=0.8]

          \draw[rotate around={45:(1.5,1.5)},fill=gray!20] (1.5,1.5) ellipse (2.25cm and 1.6cm);
          
          \draw[densely dotted] 
          (0,0) node[anchor=north]{$\begin{smallmatrix} (\nature^-, \Bot_\agent, \Lef_\bgent) \end{smallmatrix}$} -- 
          (2,0) node[anchor=north]{$\begin{smallmatrix} (\nature^-,\Bot_\agent, \Rig_\bgent) \end{smallmatrix}$}  -- 
          (2,2) node[anchor=south]{$\begin{smallmatrix} (\nature^-,\Top_\agent, \Rig_\bgent) \end{smallmatrix}$} --
          (0,2) node[anchor=south]{$\begin{smallmatrix} (\nature^-, \Top_\agent, \Lef_\bgent) \end{smallmatrix}$} -- 
          (0,0)
          (0,0) -- (1,1) node[anchor=north]{$\begin{smallmatrix} (\nature^+,\Bot_\agent, \Lef_\bgent) \end{smallmatrix}$}
          (2,0) -- (3,1) node[anchor=north]{$\begin{smallmatrix} (\nature^+,\Bot_\agent, \Rig_\bgent) \end{smallmatrix}$}
          (0,2) -- (1,3) node[anchor=south]{$\begin{smallmatrix} (\nature^+,\Top_\agent, \Lef_\bgent) \end{smallmatrix}$}
          (2,2) -- (3,3) node[anchor=south]{$\begin{smallmatrix} (\nature^+,\Top_\agent, \Rig_\bgent) \end{smallmatrix}$}
          (1,1) -- (1,3)  -- (3,3) -- (3,1) -- (1,1);
          
          \draw (2,0) node{$\bullet$};
          \draw (2,2) node{$\bullet$};
          \draw (0,2) node{$\bullet$};
          \draw (0,0) node{$\bullet$};
          \draw (1,1) node{$\bullet$};
          \draw (3,1) node{$\bullet$};
          \draw (1,3) node{$\bullet$};
          \draw (3,3) node{$\bullet$};
          \draw (1.5,4.5) node{\small $\{\emptyset , \Nature\} \otimes  \{\emptyset , \CONTROL_{\agent}\} \otimes  \{ \emptyset , \CONTROL_{\bgent} \}$};

          \draw[fill=gray!20] (6.5,0.5) ellipse (2cm and 0.95cm);
          \draw[fill=gray!20] (6.5,2.5) ellipse (2cm and 0.95cm);
          
          \draw[densely dotted] 
          (5,0) node[anchor=north]{$\begin{smallmatrix} (\nature^-, \Bot_\agent, \Lef_\bgent) \end{smallmatrix}$} -- 
          (7,0) node[anchor=north]{$\begin{smallmatrix} (\nature^-,\Bot_\agent, \Rig_\bgent) \end{smallmatrix}$}  -- 
          (7,2) node[anchor=south]{$\begin{smallmatrix} (\nature^-,\Top_\agent, \Rig_\bgent) \end{smallmatrix}$} --
          (5,2) node[anchor=south]{$\begin{smallmatrix} (\nature^-, \Top_\agent, \Lef_\bgent) \end{smallmatrix}$} -- 
          (5,0)
          (5,0) -- (6,1) node[anchor=north]{$\begin{smallmatrix} (\nature^+,\Bot_\agent, \Lef_\bgent) \end{smallmatrix}$}
          (7,0) -- (8,1) node[anchor=north]{$\begin{smallmatrix} (\nature^+,\Bot_\agent, \Rig_\bgent) \end{smallmatrix}$}
          (5,2) -- (6,3) node[anchor=south]{$\begin{smallmatrix} (\nature^+,\Top_\agent, \Lef_\bgent) \end{smallmatrix}$}
          (7,2) -- (8,3) node[anchor=south]{$\begin{smallmatrix} (\nature^+,\Top_\agent, \Rig_\bgent) \end{smallmatrix}$}
          (6,1) -- (6,3)  -- (8,3) -- (8,1) -- (6,1);

          \draw[rotate around={45:(10.5,0.5)},fill=gray!20] (10.5,0.5) ellipse (0.9cm and 0.3cm);
          \draw[rotate around={45:(12.5,0.5)},fill=gray!20] (12.5,0.5) ellipse (0.9cm and 0.3cm);
          \draw[rotate around={45:(10.5,2.5)},fill=gray!20] (10.5,2.5) ellipse (0.9cm and 0.3cm);
          \draw[rotate around={45:(12.5,2.5)},fill=gray!20] (12.5,2.5) ellipse (0.9cm and 0.3cm);
      
          \draw (7,0) node{$\bullet$};
          \draw (7,2) node{$\bullet$};
          \draw (5,2) node{$\bullet$};
          \draw (5,0) node{$\bullet$};
          \draw (6,1) node{$\bullet$};
          \draw (8,1) node{$\bullet$};
          \draw (6,3) node{$\bullet$};
          \draw (8,3) node{$\bullet$};
          \draw (6.5,4.5) node{\small $\{\emptyset , \Nature\} \otimes \tribu{\Control}_{\agent} \otimes  \{ \emptyset , \CONTROL_{\bgent} \}$};

          \draw[densely dotted] 
          (10,0) node[anchor=north]{$\begin{smallmatrix} (\nature^-, \Bot_\agent, \Lef_\bgent) \end{smallmatrix}$} -- 
          (12,0) node[anchor=north]{$\begin{smallmatrix} (\nature^-,\Bot_\agent, \Rig_\bgent) \end{smallmatrix}$}  -- 
          (12,2) node[anchor=south]{$\begin{smallmatrix} (\nature^-,\Top_\agent, \Rig_\bgent) \end{smallmatrix}$} --
          (10,2) node[anchor=south]{$\begin{smallmatrix} (\nature^-, \Top_\agent, \Lef_\bgent) \end{smallmatrix}$} -- 
          (10,0)
          (10,0) -- (11,1) node[anchor=north]{$\begin{smallmatrix} (\nature^+,\Bot_\agent, \Lef_\bgent) \end{smallmatrix}$}
          (12,0) -- (13,1) node[anchor=north]{$\begin{smallmatrix} (\nature^+,\Bot_\agent, \Rig_\bgent) \end{smallmatrix}$}
          (10,2) -- (11,3) node[anchor=south]{$\begin{smallmatrix} (\nature^+,\Top_\agent, \Lef_\bgent) \end{smallmatrix}$}
          (12,2) -- (13,3) node[anchor=south]{$\begin{smallmatrix} (\nature^+,\Top_\agent, \Rig_\bgent) \end{smallmatrix}$}
          (11,1) -- (11,3)  -- (13,3) -- (13,1) -- (11,1);
          
          \draw (12,0) node{$\bullet$};
          \draw (12,2) node{$\bullet$};
          \draw (10,2) node{$\bullet$};
          \draw (10,0) node{$\bullet$};
          \draw (11,1) node{$\bullet$};
          \draw (13,1) node{$\bullet$};
          \draw (11,3) node{$\bullet$};
          \draw (13,3) node{$\bullet$};
          \draw (11.5,4.5) node{\small $\trib \otimes \tribu{\Control}_{\agent} \otimes  \{ \emptyset , \CONTROL_{\bgent} \}$};

    \end{tikzpicture}
    \end{center}
    \caption{Illustration of three different information fields}
    \label{fig:illustration_information_fields}
\end{figure}
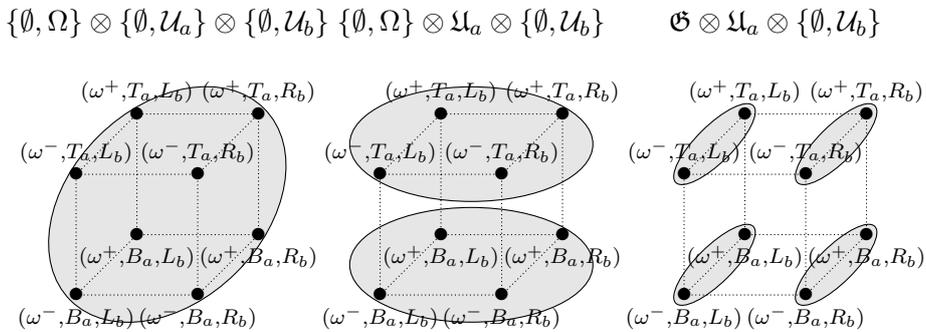

\begin{illustration}
    We still talk about the example depicted in Fig.~\ref{fig:illustration_configuration_space}. 
    We consider agent $\agent$ takes a decision without observing anything: $\tribu{\Information}_{\agent} = \{\emptyset , \Nature\} \otimes  \{\emptyset , \CONTROL_{\agent}\} \otimes  \{ \emptyset , \CONTROL_{\bgent} \}$.
    Then, we consider two different information fields for agent $\bgent$.
    He can take his decision observing simply what agent $\agent$ does : $\tribu{\Information}_{\bgent} = \{\emptyset , \Nature\} \otimes  \tribu{\Control}_{\agent} \otimes  \{ \emptyset , \CONTROL_{\bgent} \}$.
    Or, he can observe Nature, resulting in $\tribu{\Information}_{\bgent} = \trib \otimes  \tribu{\Control}_{\agent} \otimes  \{ \emptyset , \CONTROL_{\bgent} \}$.
    These information fields are represented in Fig.~\ref{fig:illustration_information_fields}.
\end{illustration}

\subsubsection*{Absence of self-information}
A natural property is the so-called \emph{absence of self-information}, meaning that each agent's information does not depend on his own decision.
This is expressed as:
\begin{equation}
    \forall \agent \in \AgentSet \eqsepv \tribu{\Information}_{\agent} \subset \trib \otimes \{\emptyset, \CONTROL_{\agent}\} \otimes \bigotimes_{\bgent \in \AgentSet \setminus \{ \agent \} }\tribu{\Control}_{\bgent} \eqfinp
\label{eq:absence_of_self_information}
\end{equation}

\subsubsection*{W-model}
A \emph{W-model}, or \emph{Witsenhausen Intrinsic Model} as introduced in \cite{Witsenhausen:1971a}, is a collection $(\AgentSet, (\Nature, \trib), (\CONTROL_{\agent}, \tribu{\Control}_{\agent})_{\agent \in \AgentSet}, (\tribu{\Information}_{\agent})_{\agent \in \AgentSet}) $ where
\begin{itemize}
    \item $\AgentSet$ is the set of agents,
    \item $(\Nature, \trib)$ represents Nature, equipped with a $\sigma$-field,
    \item $(\CONTROL_{\agent}, \tribu{\Control}_{\agent})_{\agent \in \AgentSet}$ is the collection of agents' action sets equipped with $\sigma$-fields,
    \item $(\tribu{\Information}_{\agent})_{\agent \in \AgentSet}$ is the collection of agents' information fields, subfields of $\tribu{\History} = \trib \otimes \bigotimes_{\agent \in \AgentSet} \tribu{\Control}_{\agent}$,
    \item the absence of self information property \eqref{eq:absence_of_self_information} holds.
\end{itemize}


\subsection{Strategies, playability and solution map}
\label{Strategies_playability_and_solution_map}

After establishing a W-model, we look at how agents choose to act by discussing the strategies of an agent and explaining the notion of playability, leading to the existence of the solution map.

\subsubsection*{Strategies}
A mapping 
\begin{equation}
   \policy_{\agent}: (\HISTORY, \tribu{\History}) \to (\CONTROL_{\agent}, \tribu{\Control}_{\agent}) \eqsepv
\end{equation}
is called a \emph{W-strategy}, or simply \emph{strategy}, of agent $\agent \in \AgentSet$ if it is measurable with respect to his information field $\tribu{\Information}_{\agent}$, that is,
\begin{equation}
    \policy_{\agent}^{-1} (\tribu{\Control}_{\agent}) \subset \tribu{\Information}_{\agent} \eqfinp
\end{equation}
It expresses the fact that a strategy for an agent may only depend upon his available information.

The \emph{strategy set} of an agent $\agent \in \AgentSet$ is denoted 
\begin{equation}
    \Policy_{\agent} = 
    \Bset{\policy_{\agent} : (\HISTORY, \tribu{\History}) \to (\CONTROL_{\agent}, \tribu{\Control}_{\agent})}{\policy_{\agent}^{-1} (\tribu{\Control}_{\agent}) \subset \tribu{\Information}_{\agent}}  \eqfinp 
\end{equation}


\begin{illustration}
    An interesting example is the case of an agent $\agent$ who observes nothing, meaning his information field is trivial. 
    Since the only measurable functions with respect to the trivial $\sigma$-field are constants, his strategy set $\Policy_{\agent}$ consists of the set of constant functions taking values in $\CONTROL_{\agent}$.
\end{illustration}

\subsubsection*{Playability}
Briefly speaking, \emph{playability} is the property that, for each state of Nature, the agents' strategies uniquely determine their decisions. 
It can be expressed as:
\begin{equation}
\begin{split}
    &\forall \nature \in \Nature \eqsepv 
    \forall \, \policy = (\policy_{\agent})_{\agent \in \AgentSet} \in \prod_{\agent \in \AgentSet} \Policy_{\agent} \eqsepv    \exists! \, \control = (\control_{\agent})_{\agent \in \AgentSet} \in \prod_{\agent \in \AgentSet} \CONTROL_{\agent} \eqsepv \\
    &\underbrace{\forall \agent \in \AgentSet \eqsepv
    \control_{\agent} = \policy_{\agent} \bigl(\nature, (\control_{\bgent})_{\bgent \in \AgentSet}\bigr)}_{\control = \policy(\nature, \control)} \eqfinp
\end{split}
\label{eq:playability}
\end{equation}

Braces offer a more concise version and reveal a closed-loop equation.
Remark that this notion depends solely on the information structure, since the W-model does not introduce any notion of timing.

\subsubsection*{Solution map}
Let fix a strategy profile $\policy \in \prod_{\agent \in \AgentSet} \Policy_{\agent}$.
Suppose that playability property \eqref{eq:playability} holds true.
We define the \emph{solution map} as the mapping 
\begin{equation}
   \SolutionMap_{\policy}: \Nature \to \HISTORY = \Nature \times \prod_{\agent \in \AgentSet} \CONTROL_{\agent} \eqsepv
\end{equation}
which gives for every state of Nature the unique outcome of the game, that is,
\begin{equation}
    \SolutionMap_{\policy}(\nature) = (\nature, \control) \iff
    \control = \policy(\nature, \control) \eqsepv
    \forall (\nature, \control) \in \Nature \times \prod_{\agent \in \AgentSet} \CONTROL_{\agent} \eqfinp
\end{equation}

\subsubsection*{Game form}
The mapping
\begin{equation}
    \policy \in \prod_{\agent \in \AgentSet} \Policy_\agent \mapsto \SolutionMap_{\policy} \in \HISTORY^\Nature \eqsepv
\end{equation}
defines a \emph{game form}.


\section{W-games}
\label{W-games}

Sect.~\ref{W-games} presents W-games, which are W-models with players endowed with preferences, namely objectives functions and risk measures or beliefs. 
We present these new objects in \S~\ref{Players_objective_functions_beliefs_and_risk_measures} and we introduce the notion of \emph{game data} to represent information asymmetries among the players.
Then, for clarity purposes, we introduce new notations in \S~\ref{Focus_on_players}.


\subsection{Players, objective functions and risk measures/beliefs}
\label{Players_objective_functions_beliefs_and_risk_measures}

A player is a rational decision-maker equipped with an objective function and a belief on Nature, or more broadly, a risk measure.

\subsubsection*{Players}
Let $\PLAYER$ be a set (finite or infinite).
Each element $\player \in \PLAYER$ is called a player and is associated with a set of agents $\AgentSet^{\player} \subset \AgentSet$.
Thus, all players form a partition of the set of agents $\AgentSet$:
\begin{equation}
    \AgentSet = \bigsqcup_{\player \in \PLAYER} \AgentSet^{\player} \eqfinp
\end{equation}

\begin{illustration}
    In the case of the electricity supplier who sets prices at the beginning of the year, and each day, the consumer, adjusts his consumption, we can write the set of agents as $\AgentSet = \{ \agent_{0}, \agent_{1}, \ldots, \agent_{365} \}$. However, it is clear that there are only two actors in this problem: the supplier, denoted player $\player$, and the consumer, denoted player $\qlayer$. 
    Thus, we have $\PLAYER = \{ \player, \qlayer \}$.
    The link between agents and players is made by writing $\AgentSet^{\player} = \{ \agent_{0} \}$ and $\AgentSet^{\qlayer} = \{ \agent_{1}, \ldots, \agent_{365} \}$.
\end{illustration}

\subsubsection*{Objective functions}
Each player $\player \in \PLAYER$ comes with a $\tribu{\History}$-measurable function
\begin{equation}
    \criterion^{\player} : \HISTORY \to \barRR \eqsepv
\end{equation}
called her\footnote{We use masculine pronouns for agents and feminine pronouns for players.}
\emph{objective function}, or \emph{criterion}, which represents her preferences over the different outcomes.
When the objective function is to be maximized, it is referred to as a \emph{payoff}, and when it is to be minimized, it is referred to as a \emph{cost}.

The objective function can take infinite values. 
Later on, if we want to represent an impossible configuration, we can then assign it $-\infty$ in the case of a payoff and $+\infty$ in the case of a cost.

\begin{illustration}
    The objective function of an electricity supplier is made of the profit she earns minus her production cost.
    The objective function of a consumer if made of her bills and her unwillingness to shift her consumption.
\end{illustration}

\subsubsection*{Risk measures/beliefs}

Each player perceives risk in a different way.
The concept of risk measure reflects this in the general case.
However, in most cases we will encounter, the more specific notion of beliefs is sufficient.

\subsubsubsection{Risk measures}
Each player $\player \in \PLAYER$ is equipped with a pair $(\RiskSpace^{\player}, \RiskMeasure^{\player})$ made of a set of random variables
\begin{equation}
   \RiskSpace^{\player} \subset \barRR^{\Nature}  \eqsepv
\end{equation}
and of a \emph{risk measure} on that space 
\begin{equation}
    \RiskMeasure^{\player} : \RiskSpace^{\player} \to \barRR \eqfinp
\end{equation}

A risk measure is a function\footnote{verifying certain properties} that assigns a numerical value to a random variable.
In our case, the random variables considered are random variables on Nature.

\begin{illustration}
    A popular risk measure in risk management is the Conditional Value at Risk (CVaR) which is an average focusing on the worst-case scenarios. 
    It can be useful for an electricity producer to evaluate her investments. 
    Nevertheless, the simplest risk measure remains the expected value,  for which we rather introduce the notion of beliefs.
\end{illustration}

\subsubsubsection{Beliefs}
To equip a player $\player \in \PLAYER$ with a \emph{belief}, that is a probability distribution on Nature
\begin{equation}
    \Belief^{\player} : \; \trib \to \ClosedIntervalClosed{0}{1} \eqsepv
\end{equation}
means that we implicitly consider the following risk measure over the set of integrable random variables:
\begin{equation}
    \left\{
    \begin{aligned}
        \RiskSpace^{\player} &= L^1(\Nature, \trib, \Belief^{\player}) \eqsepv \\
        \RiskMeasure^{\player} &= \expect{}[\; \cdot \;] \eqfinp
    \end{aligned}
    \right.
\end{equation}

In other words, it means that when a player does not know a random variable, she is going to consider its expected value instead, which is a very frequent case.
It is easy to interpret a player's belief as a measure of what she knows about the uncertainties she faces.

\begin{illustration}
    If Nature corresponds to the electricity demand for the day, an electricity supplier is equipped with an estimate of that demand, which will be her belief.
    And she focuses on her average payoff over the different demand scenarios.
\end{illustration}

\subsubsection*{W-game}
A \emph{W-game}, or \emph{game in product form} as introduced in \cite{Heymann-DeLara-Chancelier:2022}, is a collection $(\AgentSet, (\Nature, \trib), (\CONTROL_{\agent}, \tribu{\Control}_{\agent})_{\agent \in \AgentSet}, (\tribu{\Information}_{\agent})_{\agent \in \AgentSet}, \PLAYER, (\criterion^{\player})_{\player \in \PLAYER}, (\RiskSpace^{\player}, \RiskMeasure^{\player})_{\player \in \PLAYER}) $ where
\begin{itemize}
    \item $(\AgentSet, (\Nature, \trib), (\CONTROL_{\agent}, \tribu{\Control}_{\agent})_{\agent \in \AgentSet}, (\tribu{\Information}_{\agent})_{\agent \in \AgentSet})$ is a W-model,
    \item $\PLAYER$ is a partition of the agents' set whose atoms are the players,
    \item $(\criterion^{\player})_{\player \in \PLAYER}$ is the collection of players' objective functions,
    \item $(\RiskSpace^{\player}, \RiskMeasure^{\player})_{\player \in \PLAYER}$ is the collection of players' risk measures with the corresponding spaces.
\end{itemize}

\subsubsection*{W-game data}
For every player $\player \in \PLAYER$, we call \emph{player $\player$'s data} the pair
\begin{equation}
    \data^{\player} = (\criterion^{\player}, (\RiskSpace^{\player}, \riskmeasure^{\player})) \eqsepv
\end{equation}
consisting of her objective function and her risk measure, along with the corresponding set.

By extension, \emph{W-game data} refer to the collection of the players' data, denoted by
\begin{equation}
    \data^{\PLAYER} = (\data^{\player})_{\player \in \PLAYER} \eqfinp
\end{equation}

While the W-model is something shared by everyone, we introduce W-game data to raise the question of their knowledge among the players, since they encapsulate purely personal data.
Symbolically, we could write that
\begin{equation}
    \text{W-game} = \text{W-model} + \text{W-game data} \eqfinp
\end{equation}

\begin{illustration}
    A consumer optimizes a certain function when managing his electricity consumption and has an estimate of the uncertainties he faces.
    They are respectively his cost function and his belief about Nature and together they form his data.
    It is worth considering whether the electricity provider has access to this information or not.
\end{illustration}


\subsection{Notations relative to players}
\label{Focus_on_players}

This paragraph emphasizes the differences between an agent and a player and introduces new notations, keeping in mind that a player is simply a group of agents.

\subsubsection*{Agent vs player}
A player should not be confused with an agent.
To facilitate the reader's understanding, we draw attention to the chosen convention:
\begin{itemize}
    \item Objects related to agents are denoted with subscripts: $\control_{\agent}, \CONTROL_{\agent}, \tribu{\Information}_{\agent}, \policy_{\agent}, \Policy_{\agent}$.
    \item Objects related to players are denoted with superscripts: $\AgentSet^{\player}, \criterion^{\player}, \Belief^{\player}, \data^{\player}$.
\end{itemize}

Moreover, we say that an agent \emph{observes}, or \emph{sees}, something to refer to his information field, whereas we say that a player \emph{knows} something when referring to her belief.

\subsubsection{Writing the strategy profile of a player}
When dealing with game theory, we want to speak of a player's strategy.
Therefore, we define the strategy set for a player $\player \in \PLAYER$ by the product space
\begin{equation}
    \Policy^{\player} = \prod_{\agent \in \AgentSet^{\player}} \Policy_\agent \eqfinp
\end{equation}
We insist that this is a formal notation since there is no corresponding information field of a player.
More generally, we can define the strategy set of any subset of players $\QLAYER \subset \PLAYER$ by
\begin{equation}
    \Policy^{\QLAYER} = \prod_{\qlayer \in \QLAYER} \Policy^{\qlayer} = \prod_{\qlayer \in \QLAYER} \prod_{\agent \in \AgentSet^\qlayer} \Policy_\agent \eqfinp
\end{equation}

Thanks to this, we can adopt a common notation in game theory, which consists of denoting the strategy set of all players except a player $\player \in \PLAYER$ by
\begin{equation}
    \Policy^{-\player} = \Policy^{\PLAYER \setminus \{ \player \}} \eqsepv
\end{equation}
or even the strategy set of all players except a group $\QLAYER \subset \PLAYER$ by
 \begin{equation}
    \Policy^{-\QLAYER} = \Policy^{\PLAYER \setminus \QLAYER } \eqfinp
 \end{equation}

Thus, when we consider a strategy profile, we can decompose it from the perspective of a player $\player \in \PLAYER$ like so:
\begin{equation}
    \policy = 
    (\underbrace{\policy^{\player}}_{\in \Policy^{\player}},
    \underbrace{\policy^{-\player}}_{\in \Policy^{-\player}}) 
    \in \Policy^{\PLAYER} \eqsepv
\end{equation}
or from the perspective of a group of players $\QLAYER \subset \PLAYER$
\begin{equation}
    \policy = 
    (\underbrace{\policy^{\QLAYER}}_{\in \Policy^{\QLAYER}},
    \underbrace{\policy^{-\QLAYER}}_{\in \Policy^{-\QLAYER}}) 
    \in \Policy^{\PLAYER} \eqfinp
\end{equation}


\section{Nash equilibrium in W-games}
\label{Nash_equilibrium}

From now on, we only deal with strategies, which contain the information structure.
Sect.~\ref{Nash_equilibrium} discusses the meaning that can be given to a solution in a W-game.
We first write a W-game in normal form in \S~\ref{W-games_normal_form}.
Then, \S~\ref{Best_response_and_Nash_equilibrium} introduces the notion of best response, enabling to define a Nash equilibrium.


\subsection{W-games in normal form}
\label{W-games_normal_form}

In this section, we fix a player $\player \in \PLAYER$.
To analyze her strategic decision-making process, the concept of the normal form objective function is introduced, providing a tool to assess strategies under uncertainty, based on the players' risk measures or beliefs.



\subsubsection*{From the solution map...} 
Let $\policy = (\policy^{\player}, \policy^{-\player}) \in \Policy^{\PLAYER}$ be a strategy profile.
Assuming playability (as described in equation \eqref{eq:playability}), we have the solution map $\SolutionMap_{\policy^{\player}, \policy^{-\player}}$ that connects the state of Nature to the game's unqiue outcome based on this strategy profile:
\begin{equation}
    \Nature \xrightarrow{\SolutionMap_{\policy^{\player}, \policy^{-\player}}} \HISTORY \eqfinp
\end{equation}

\subsubsection*{... and the objective function...}
Having mapped Nature to the game outcomes, player $\player$ takes into account her preferences over these outcomes represented by her objective function $\criterion^{\player}$:
\begin{equation}
    \Nature \xrightarrow{\SolutionMap_{\policy^{\player}, \policy^{-\player}}} \HISTORY
    \xrightarrow{\criterion^{\player}} \barRR \eqfinp
\end{equation}

By composition, this allows player $\player$ to assign a numerical value to each state of Nature:
\begin{equation}
    \Nature \xrightarrow{\criterion^{\player} \compo \SolutionMap_{\policy^{\player}, \policy^{-\player}}} \barRR \eqfinp
\end{equation}

\subsubsection*{... to the normal form objective function}
Since we have something that depends on Nature, we need to examine how player $\player$ perceives Nature, first in the general case of risk measures, and then in the simpler case of beliefs. 

\subsubsubsection{With risk measures}
Without delving too deeply into measurability issues, we can consider that $\criterion^{\player} \compo \SolutionMap_{\policy^{\player}, \policy^{-\player}}$ belongs to the set of random variables $\RiskSpace^{\player}$. 
Hence, player $\player$ can rely on her risk measure $\RiskMeasure^{\player}$ to map it to a numerical value:
\begin{equation}
    \left[ \Nature \xrightarrow{\criterion^{\player} \compo \SolutionMap_{\policy^{\player}, \policy^{-\player}}} \barRR 
    \right]
    \xrightarrow{\RiskMeasure^{\player}} \barRR \eqfinp
\label{strategy_assessment}
\end{equation}

In summary, player $\player$ can associate a numerical value with each strategy profile $\policy = (\policy^{\player}, \policy^{-\player}) \in \Policy^{\PLAYER}$ by the mapping
\begin{subequations}
\begin{align}
    \Criterion^{\player} : 
    & \; \Policy^{\player} \times \Policy^{-\player} \to \barRR \eqsepv \\
    & \; (\policy^{\player}, \policy^{-\player}) \mapsto \RiskMeasure^{\player}
    [\criterion^{\player} \compo \SolutionMap_{\policy^{\player}, \policy^{-\player}}] \eqfinp
\end{align}
\end{subequations}
which we call her \emph{normal form objective function}.

\subsubsubsection{With beliefs}
In the particular case of beliefs, the normal form objective function writes 
\begin{equation}
    \Criterion^{\player}( \policy^{\player}, \policy^{-\player}) = \expect{\Belief^{\player}}
    [\criterion^{\player} \compo \SolutionMap_{\policy^{\player}, \policy^{-\player}}]  \eqfinp
\end{equation}

If the objective function is actually a cost, the corresponding normal form objective function is referred to as a \emph{normal form cost}.
And if the objective function is actually a payoff, the corresponding normal objective function is referred to as a \emph{normal form payoff}.

\subsubsection{Focus on the W-game data}
It is important to note that player $\player$'s data is hidden within their normal form objective function, since the latter depends on the objective function and the risk measure or belief of the player.
Therefore, we explicitly write the dependence of the normal form objective function to the game data by writing them as arguments of the function, that is,
\begin{equation}
    \Criterion^{\player}( \policy^{\player}, \policy^{-\player}) =
    \Criterion^{\player}( \policy^{\player}, \policy^{-\player}; \underbrace{\data^{\player}}_{\text{data}}) \eqfinp
\end{equation}

\subsubsection{Game in normal form}

The triplet $(\PLAYER, (\Policy^{\player})_{\player \in \PLAYER}, (\Criterion^{\player})_{\player \in \PLAYER})$ where
\begin{itemize}
    \item $\PLAYER$ is the set of players,
    \item $(\Policy^{\player})_{\player \in \PLAYER}$ is the collection of players' strategies,
    \item $(\Criterion^{\player})_{\player \in \PLAYER}$ is the collection of players' normal form objective functions,
\end{itemize}
is an (infinite) \emph{normal-form game}.

\begin{illustration} 
    The prisoner's dilemma is the most well-known example of a normal-form game. 
    Two individuals are arrested and interrogated separately by the police. Each player can either betray the other by defecting (D) or remain silent to cooperate (C). Table~\ref{tab:prisoners_dilemma} shows the jail time, in years, associated with each possible outcome.
\end{illustration}

\begin{table}[ht]
    \begin{center}
    \begin{tabular}{|l|l|l|}
    \hline
                        & C (cooperate) & D (defect) \\ \hline
    C (cooperate)       & 1/2, 1/2       & 10, 0   \\ \hline
    D (defect)          & 0, 10        & 5, 5   \\ \hline
    \end{tabular}
    \end{center}
    \caption{Prisoner's dilemma}
    \label{tab:prisoners_dilemma}
\end{table}



\subsection{Best response and Nash equilibrium}
\label{Best_response_and_Nash_equilibrium}

We explain how to define a Nash equilibrium for W-games in normal form by utilizing the concept of best response.

\subsubsection*{Best response}
Contrary to common practice in game theory, we define the \emph{set of best responses} for player $\player \in \PLAYER$ to a strategy of the other players $\underline{\policy}^{-\player} \in \Policy^{-\player}$ by
\begin{equation}
    \Nash{\Policy}^{\player}(\underline{\policy}^{-\player}; \data^{\player}) = \argmin_{\policy^{\player} \in \Policy^{\player}} 
    \Criterion^{\player}(\policy^{\player}, \underline{\policy}^{-\player}; \data^{\player}) \subset \Policy^{\player} \eqsepv
\end{equation}
because later, the players who will choose their strategies from the set of best responses are players for whom the normal form objective function is a cost.
Any strategy in the set of best responses is called a \emph{best response}.

We define subsets of strategies, called best responses, depending on the normal objective function.
It is then necessary to also explicitly include player $\player$'s data in the notation of the best responses.

\begin{illustration} 
    To clarify the ideas, we underline the values that are the best responses for each player to the other's strategy in the prisoner's dilemma in Table~\ref{tab:Best_responses_prisoner_dilemma}.
\end{illustration}

\begin{table}[ht]
    \begin{center}
    \begin{tabular}{|l|l|l|}
    \hline
                        & C (cooperate)      & D (defect) \\ \hline
    C (cooperate)       & 1/2, 1/2             & 10, \underline{0}   \\ \hline
    D (defect)          & \underline{0}, 10  & \underline{5}, \underline{5}   \\ \hline
    \end{tabular}
    \end{center}
    \caption{Best responses in the prisoner's dilemma}
    \label{tab:Best_responses_prisoner_dilemma}
\end{table}

\subsubsection*{Nash equilibrium}
A strategy profile $\underline{\policy} = (\underline{\policy}^{\player})_{\player \in \PLAYER} \in \Policy^{\PLAYER}$ is said to be a \emph{Nash equilibrium} if and only if each player's strategy is a best response to the other players' strategies, that is
\begin{equation}
    \forall \player \in \PLAYER \eqsepv
    \underline{\policy}^{\player} \in \Nash{\Policy}^{\player}(\underline{\policy}^{-\player}; \data^{\player}) \eqfinp
\end{equation}

A Nash equilibrium represents a stable state in a game where no player has an incentive to deviate unilaterally from their chosen strategy.

\begin{illustration} 
    Underlining the best responses clearly reveals the Nash equilibrium in Table.~\ref{tab:Nash_equilibrium_prisoner_dilemma}. 
    It is interesting to note that the Nash equilibrium is the strategy profile (D, D) even though (C, C) is more favorable for both.
\end{illustration}

\begin{table}[ht]
    \begin{center}
    \begin{tabular}{|l|l|l|}
    \hline
                        & C (cooperate) & D (defect) \\ \hline
    C (cooperate)       & 1/2, 1/2                & 10, \underline{0}   \\ \hline
    D (defect)          & \underline{0}, 10   & $(\underline{5}, \underline{5})^{\mathcal{N}}$   \\ \hline
    \end{tabular}
    \end{center}
    \caption{Nash equilibrium in the prisoner's dilemma}
    \label{tab:Nash_equilibrium_prisoner_dilemma}
\end{table}



\section*{Conclusion}

We have developed the concept of W-games, based on the concept of W-model, and we have given sense to a notion of equilibrium in such games.
We will now see to what extent W-games is an efficient solution to model leader-follower problems.


\chapter{Leader-follower problems as W-games}
\label{W-games_leader_follower_problems}

Leader-follower problems arise in systems where one agent, the \emph{leader} acts first, and the other agent, the \emph{follower}, responds based on the leader’s actions.
These problems are common in control theory, economics, and game theory, often involving strategic interactions and information asymmetry.
The leader must anticipate the followers' responses, while followers adjust their strategies according to the leader’s initial move.
The W-model and its extension with W-games, provides a formal framework to address such problems when the information structure is complex.

\begin{figure}[ht]
    \begin{center}
    \begin{tikzpicture}
        
        \tikzset{
            block/.style = {rectangle, draw, text centered, minimum height=3em, minimum width=10em},
            arrow/.style = {thick,->,>=stealth},
            darrow/.style = {thick, dashed,->,>=stealth},
            zigzag/.style = {decorate, decoration=zigzag, thick,->,>=stealth}
        }

        \node[block] (Weather) {
            \begin{tabular}{c}
                exogenous factors \\
            \end{tabular}
        };
        \node[block, below=1.2cm of Weather, xshift = -2.5cm] (Electricity producer) {
            \begin{tabular}{c}
                Maximizes: payoff \\
                Decides: energy prices \\
                Has: private knowledge 
            \end{tabular}
        };
        \node[block, below=1.8cm of Electricity producer, xshift = 2cm] (Consumer) {
            \begin{tabular}{c}
                Minimizes: cost \\
                Decides: consumption \\
                Has: private knowledge
            \end{tabular}
        };

        \draw[arrow] ($(Electricity producer.south east) + (-1.6,0)$) -- ($(Consumer.north) +(-1.1,0)$)
        node[midway, left] {Reaction};
        
        \draw[darrow] (Consumer.south) |- ++(0,-0.8) -| ($(Electricity producer.south west) + (0.5,0)$)
        node[pos=0.25, below] {Anticipation
        };

        \draw[zigzag] ($(Weather.south) + (0.9,0)$) -- ($(Consumer.north) + (1.4, 0)$);
        \draw[zigzag] ($(Weather.south) + (-1.1,0)$) -- ($(Electricity producer.north) +(1.4,0)$);
        
        \node[right=1.8cm of Electricity producer] {
        \begin{tabular}{c}
                Leader \\ 
                producer, supplier, aggregator 
            \end{tabular}
        };
        \node[right=0.6cm of Consumer] {
        \begin{tabular}{c}
                Follower \\ 
                consumer, prosumer 
            \end{tabular}
        };
        \node[right=1cm of Weather] {
        \begin{tabular}{c}
                ``pure randomness'' 
            \end{tabular}
        };
        
    \end{tikzpicture}
    \end{center}
    \caption{Leader follower problems in energy management}
    \label{leader_follower_problems_energy_management}
\end{figure}
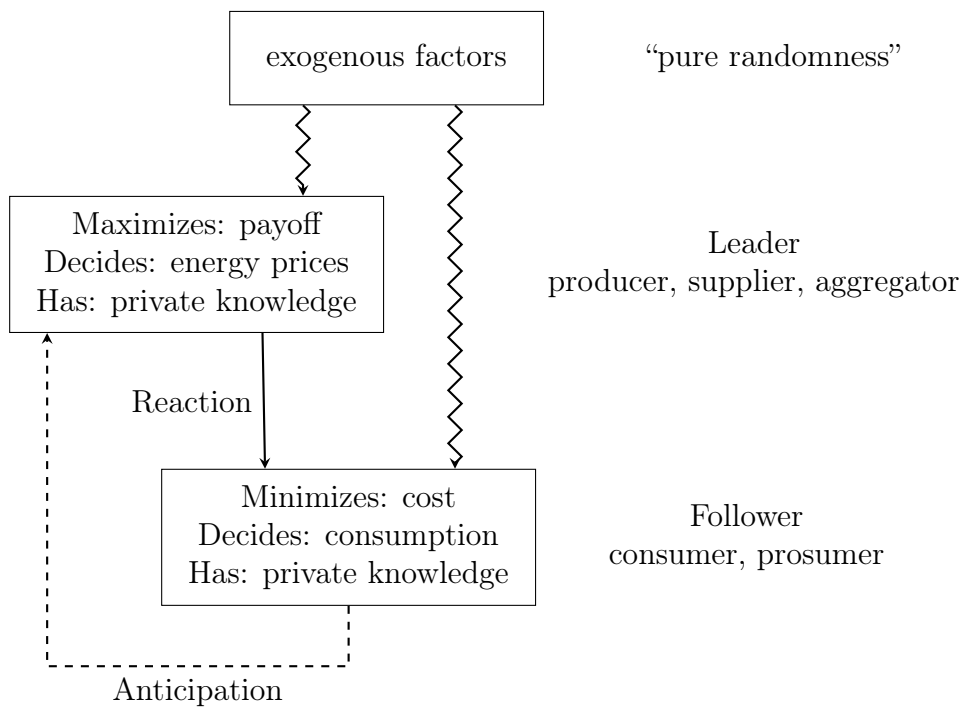

In energy management, encountering leader-follower problems is common. 
The leader is typically an aggregator, an electricity supplier, or a producer, who determines energy price signals.
The follower, which can be a consumer or a prosumer, reacts based on this signal and adjusts his consumption behavior accordingly.
In this situation, the leader aims to maximize his profit, while the follower seeks to minimize his cost.
We emphasize the private information of both the leader and the follower, as well as the role of exogenous parameters in the problem.
This can be summarized in a diagram, see Fig.~\ref{leader_follower_problems_energy_management}.

In Chap.~\ref{W-games_leader_follower_problems}, we explain how to construct W-games specifically for leader-follower problems by introducing the appropriate notation and highlighting the unique features of these problems.
First, in Sect.~\ref{SLSF_setup}, we focus on simpler cases involving only two players, each represented by a single agent. 
Then, in Sect.~\ref{MLMFS_setup}, we show how to adapt this framework to handle more complex situations, such as cases where each player consists of multiple agents, as well as cases with multiple players.
After explaining the construction of these games, we introduce in Sect.~\ref{Nash-Stackelberg_equilibria_W-games} Nash-Stackelberg equilibria, a concept based on Stackelberg strategies that is better suited to address the special information structure in leader-follower setups.

\begin{illustration}
     Text in this font corresponds to the same example of time-of-use pricing problem in the language of W-games.
\end{illustration}


\section{Static single-leader-single-follower setup}
\label{SLSF_setup}

We follow the structure of Chap.\ref{Background_on_W-model_and_W-games} while attempting to conduct a comprehensive study of the case where there are two players and each player has only one agent.
In \S.~\ref{SLSF_agents_Nature_actions_configuration_space_information_fields}, we first address the construction of the W-model, then we focus on strategies and solutions in \S.~\ref{SLSF_strategies_playability_solution_map}, before discussing the W-game in \S.~\ref{SLSF_players_objective_functions_beliefs}.


\subsection{Agents, Nature, actions, configuration space \\ and information fields}
\label{SLSF_agents_Nature_actions_configuration_space_information_fields}

We revisit the structure of Chap.~\ref{Background_on_W-model_and_W-games} by exploring the specifics of the single-leader-single-follower case.
We begin by introducing the formalism for leader-follower problems in terms of agents, then define a specific form for Nature, and finally address the particular information field.

\subsubsection*{Agents}
We are dealing here with a specific case where the set of agents writes
\begin{equation}
    \AgentSet = \{ \leader, \follower \} \eqsepv
\end{equation}
where $\leader$ is called the \emph{leader} agent and $\follower$ the \emph{follower} agent.


\begin{illustration}
    The leader ($\leader$) is an electricity producer and the follower ($\follower$) is a consumer.
\end{illustration}

\subsubsection*{Actions}
The leader agent comes with his (measurable) action set $(\CONTROL_{\leader}, \tribu{\Control}_{\leader})$, and the follower agent comes with his (measurable) action set $(\CONTROL_{\follower}, \tribu{\Control}_{\follower})$.

\begin{figure}[ht]
    \begin{center}
        \begin{tikzpicture}
        
        \tikzset{
            block/.style = {rectangle, draw, text centered, minimum height=3em, minimum width=10em},
            arrow/.style = {thick,->,>=stealth},
            darrow/.style = {thick, dashed,->,>=stealth}
        }
        
        \node[block] (Electricity producer) {
            \begin{tabular}{c}
                Decides: (peak, off-peak) prices \\
                \includegraphics[width=0.3\textwidth]{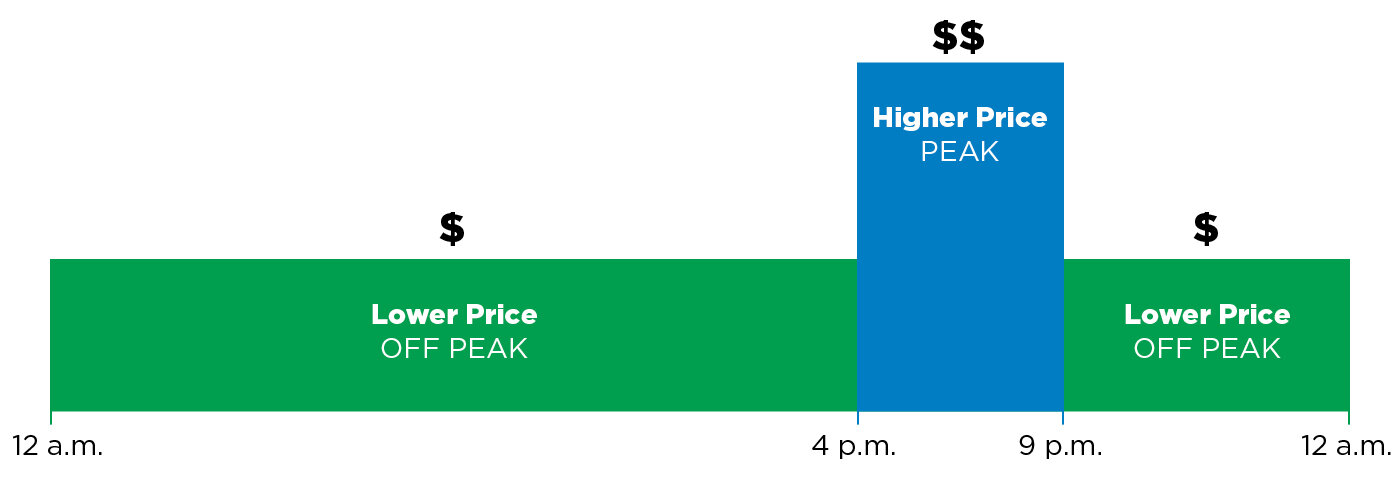} 
            \end{tabular}
        };
        \node[block, below=0.8cm of Electricity producer, xshift = 1.8cm] (Consumer) {
            \begin{tabular}{c}
                Decides: consumption shift \\
                \includegraphics[width=0.4\textwidth]{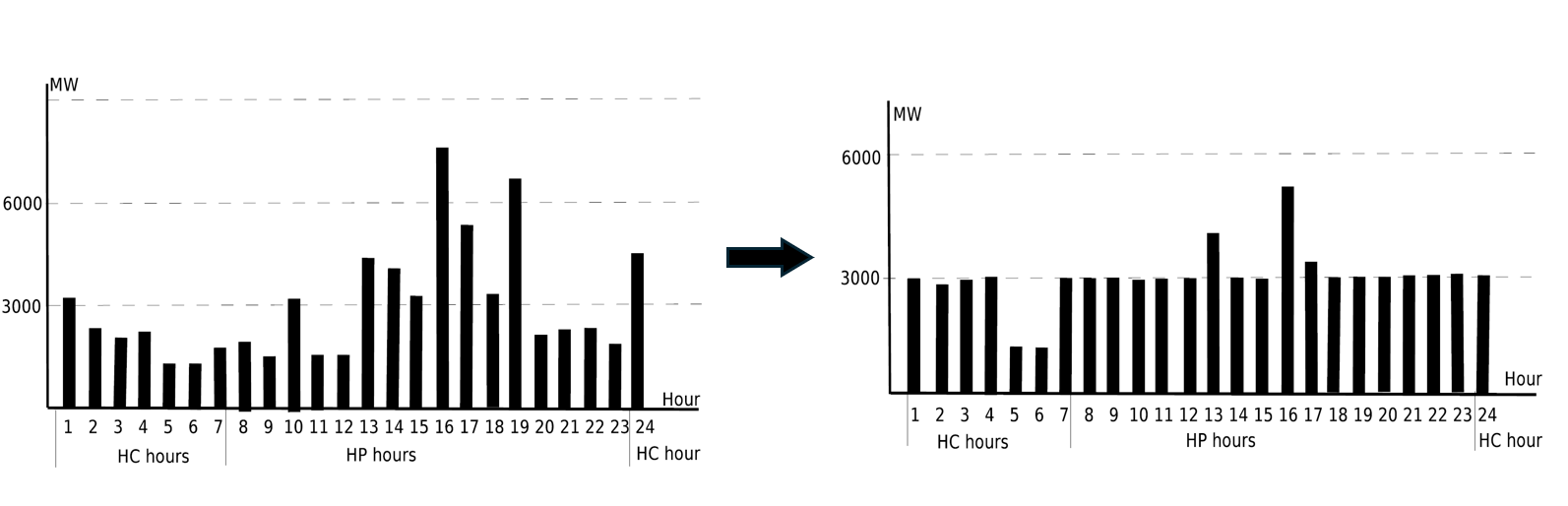} 
            \end{tabular}
        };
        
        \draw[arrow] ($(Electricity producer.south west) + (4.9,0)$) -- (Consumer)
        node[midway, right] {
        };
        
        \draw[darrow] (Consumer.south) |- ++(0,-0.5) -| ($(Electricity producer.south west) + (0.4,0)$)
        node[pos=0.25, below] {
        };
        
        \node[right=0.3cm of Electricity producer] {
        \begin{tabular}{c}
                Leader (agent) \\ 
                electricity producer  \faIndustry
        \end{tabular}
        };
        \node[right=0.3cm of Consumer] {
        \begin{tabular}{c} 
                Follower (agent) \\
                consumer  \faHome
        \end{tabular}
        };
        
    \end{tikzpicture}
    \end{center}
    \caption{Example schematic representation \\ Images from https://www.cleanpowersf.org/to (top) and \cite{Alekseeva-Brotcorne-Lepaul-Montmeat:2019} (bottom)}
    \label{fig:schematic_representation_time_of_use_pricing_problem}
\end{figure}

\begin{illustration}
    The leader (the electricity producer) sets peak and off-peak prices for electricity.
    We can model this action as 
    \begin{subequations}
        \begin{align}
        \control_{\leader} = (\overline{\control}_{\leader}, \underline{\control}_{\leader}) \in \CONTROL_{\leader} = \left\{ (x, y) \in \RR^2 \mid x \geq y \right\} \subset \RR^2 \eqsepv
        \end{align}
    where $\overline{\control}_{\leader}$ is the peak price (€) and $\underline{\control}_{\leader}$ is the off-peak price (€).  
    The follower (the consumer) does not decide on his consumption, as it is considered random and dependent on external factors, however he decides when to consume.
    Therefore his action is written as 
        \begin{align}
        \control_{\follower} = (\overline{\control}_{\follower}, \underline{\control}_{\follower}) \in \CONTROL_{\follower} = \left\{ (\alpha, \beta) \in \RR_+^2 \mid \alpha + \beta = 1 \right\} \subset \RR_+^2 \eqsepv    
        \end{align}
    \label{eq:ex_action_set}
    \end{subequations}
    where $\overline{\control}_{\follower}$ is the fraction of consumption (\%) during peak hours and $\underline{\control}_{\follower}$ is the fraction of consumption (\%) during off-peak hours.
    Fig.\ref{fig:schematic_representation_time_of_use_pricing_problem} gives a schematic representation of decisions of the leader and the follower.    
\end{illustration}

\subsubsection*{Nature}
In leader-follower problems, we identify three sources of uncertainty.
Thus, we express Nature as a product space of three sets, each endowed with its corresponding $\sigma$-field. 
Specifically, we write Nature as 
\begin{subequations}
\begin{align}
   &\Nature = \underbrace{\exo{\Nature}}_{\substack{\text{exogenous} \\ \text{Nature}}} \times \underbrace{\Nature^{\leader}}_{\substack{\text{leader} \\ \text{type}}} 
   \times \underbrace{\Nature^{\follower}}_{\substack{\text{follower} \\ \text{type}}} \eqsepv 
\end{align}
equipped with the corresponding $\sigma$-field
\begin{align}
   &\trib = \exo{\trib} \otimes \trib^{\leader} \otimes \trib^{\follower} \eqfinp
\end{align}
\end{subequations}

We decide to write the indices as superscript because we attach types to players and not agents.
The leader type $\Nature^{\leader}$ contains what is related to the leader player.
The follower type $\Nature^{\follower}$ contains what is related to the follower player.
Exogenous Nature $\exo{\Nature}$ contains the rest, what is pure randomness.
Decomposing Nature helps to better represent the asymmetries of information between the leader and the follower.

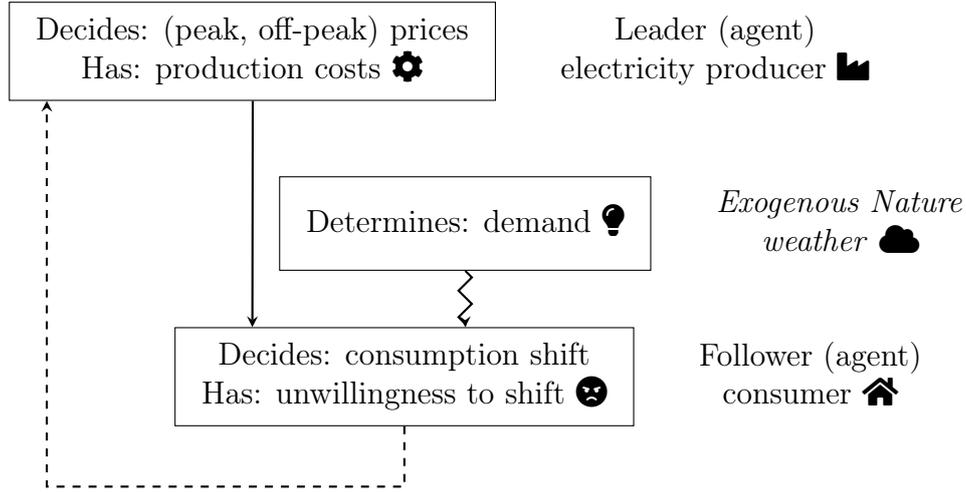
\begin{figure}[ht]
    \begin{center}
    \begin{tikzpicture}
    
    \tikzset{
        block/.style = {rectangle, draw, text centered, minimum height=3em, minimum width=10em},
        arrow/.style = {thick,->,>=stealth},
        darrow/.style = {thick, dashed,->,>=stealth},
        zigzag/.style = {decorate, decoration=zigzag, thick,->,>=stealth}
    }
    
    \node[block] (Electricity producer) {
        \begin{tabular}{c}
            Decides: (peak, off-peak) prices \\
            Has: production costs \faCog
        \end{tabular}
    };
    \node[block, below=3cm of Electricity producer, xshift = 2cm] (Consumer) {
        \begin{tabular}{c}
            Decides: consumption shift \\
            Has: unwillingness to shift \faAngry
        \end{tabular}
    };
    
    \node[block, below=1cm of Electricity producer, xshift=2.8cm] (Weather) {
        \begin{tabular}{c}
            Determines: demand \faLightbulb \\
        \end{tabular}
    };
       
    \draw[arrow] ($(Electricity producer.south east) + (-3.2,0)$) -- ($(Consumer.north) +(-2,0)$) ;
    
    \draw[darrow] (Consumer.south) |- ++(0,-0.8) -| ($(Electricity producer.south west) + (0.5,0)$)
    node[pos=0.25, below] {
    };

    \draw[zigzag] (Weather.south) -- ($(Consumer.north) + (0.8, 0)$);
    
    \node[right=0.5cm of Electricity producer] {
    \begin{tabular}{c}
            Leader (agent) \\ 
            electricity producer \faIndustry
        \end{tabular}
    };
    \node[right=0.5cm of Consumer] {
    \begin{tabular}{c}
            Follower (agent)\\ 
            consumer \faHome
        \end{tabular}
    };
    \node[right=0.5cm of Weather] {
    \begin{tabular}{c}
            \emph{Exogenous Nature} \\ 
            \emph{weather} \faCloud
        \end{tabular}
    };
    
    \end{tikzpicture}
    \end{center}
    \caption{Components of Nature}
    \label{fig:Nature_time-of-use_pricing_problem}
\end{figure}

\begin{illustration}
    A state of Nature can be written as
    \begin{equation}
        \nature = (\exo{\nature}, \nature^{\leader}, \nature^{\follower}) \in \Nature = \RR_+ \times \RR_+ \times \RR_+ = \RR_+^3 \eqsepv
    \label{eq:ex_Nature}
    \end{equation}
    where $\exo{\nature}$ is the demand (kWh), $\nature^{\leader}$ is the leader's unitary production cost (€/kWh) and $\nature^{\follower}$ is the consumer's unwillingness to shift consumption to off-peak hours (€/kWh).
    Fig.\ref{fig:Nature_time-of-use_pricing_problem} enhances the representation of our example by displaying these parameters and the exogenous Nature.
\end{illustration}

\subsubsection*{Configuration space}
In this case, the configuration space writes
\begin{subequations}
\begin{align}
    \HISTORY = \underbrace{\exo{\Nature} \times \Nature^{\leader} \times\Nature^{\follower}}_{\Nature} \times \; \CONTROL_{\leader} \times \CONTROL_{\follower} \eqsepv 
\end{align}
equipped with the $\sigma$-field
\begin{align}
    \tribu{\History} = \underbrace{\exo{\trib} \otimes \trib^{\leader} \otimes \trib^{\follower}}_{\trib} \otimes \; \tribu{\Control}_{\leader} \otimes \tribu{\Control}_{\follower} \eqfinp
\end{align}
\end{subequations}

\begin{illustration}
    From equations \eqref{eq:ex_action_set} and \eqref{eq:ex_Nature}, we find that the configuration space is given by
    \begin{subequations}
        \begin{align}
        \HISTORY = \RR^3_+ \times \left\{ (x, y) \in \RR^2 \mid x \geq y \right\} \times \left\{ (\alpha, \beta) \in \RR_+^2 \mid \alpha + \beta = 1 \right\} \eqsepv
        \label{ex_configuration_space}
        \end{align}
    equipped with the $\sigma$-field
        \begin{align}
        \tribu{\History} = \borel{\RR^3_+} \otimes \borel{\left\{ (x, y) \in \RR^2 \mid x \geq y \right\}} \otimes \borel{\left\{ (\alpha, \beta) \in \RR_+^2 \mid \alpha + \beta = 1 \right\}} \eqfinp
        \end{align}
     \end{subequations}
\end{illustration}

\subsubsection*{Information fields}
The rigorous definition is that the leader agent is the agent who does not observe more than Nature and the follower is the other one. 
In the general case, we can only say that the following inclusions always hold true:
\begin{subequations}
\begin{align}
     \tribu{\Information}_{\leader}
    \subset \overbrace{\exo{\trib} \otimes \trib^{\leader} \otimes \trib^{\follower}}^{\substack{\text{can see} \\ \text{parts of Nature}}}  &\otimes \overbrace{\{\emptyset, \CONTROL_{\leader} \}}^{\substack{\text{absence of} \\ \text{self-information}}} \otimes \overbrace{\{\emptyset, \CONTROL_{\follower} \} }^{\substack{\text{cannot see} \\ \text{the follower}}}  \eqsepv \\
    \tribu{\Information}_{\follower}
    \subset \underbrace{\exo{\trib} \otimes \trib^{\leader} \otimes \trib^{\follower}}_{\substack{\text{can see} \\ \text{parts of Nature}}}  &\otimes \underbrace{\tribu{\Control}_{\leader}}_{\substack{\text{can see} \\ \text{the leader}}} \otimes \underbrace{\{\emptyset, \CONTROL_{\follower} \} }_{\substack{\text{absence of} \\ \text{self-information}}} \eqfinp
\end{align}
\label{eq:information_structure_long}
\end{subequations}
To write the agent information fields in the most readable way, we will omit all trivial $\sigma$-fields. 
For instance, equation \eqref{eq:information_structure_long} writes more simply
\begin{subequations}
\begin{align}
    & \tribu{\Information}_{\leader}\subset \exo{\trib} \otimes \trib^{\leader} \otimes \trib^{\follower}  \eqsepv \\
   & \tribu{\Information}_{\follower} \subset \exo{\trib} \otimes \trib^{\leader} \otimes \trib^{\follower}  \otimes \tribu{\Control}_{\leader}   \label{eq:info_follower}\eqfinp
\end{align}
\label{eq:information_structure}
\end{subequations}
It is worth noticing that equation \eqref{eq:info_follower} is not sufficient.
Indeed, we cannot have $\tribu{\Information}_{\follower} = \trib^{\follower}$ or $\tribu{\Information}_{\follower} = \emptyset$, because the follower would be the leader.
In summary, the leader and the follower can see some parts of Nature and the follower sees some of what the leader is doing.

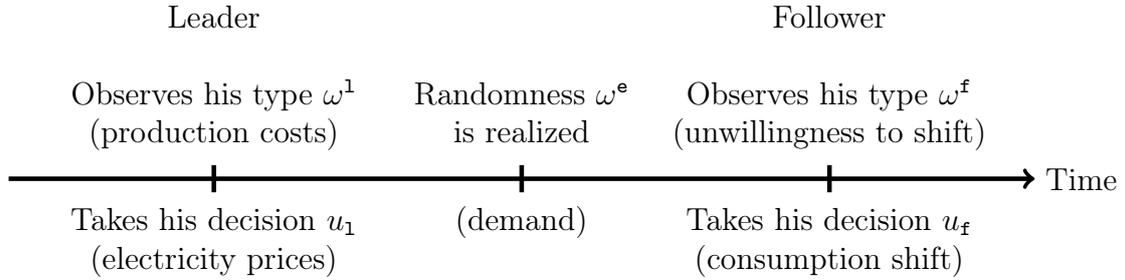
\begin{figure}[ht]
\begin{center}
    \begin{tikzpicture}[scale = 0.9]
    \draw[->, line width=0.6mm] (0,0) -- (15,0); 

    \draw[line width=0.6mm] (3,0.2) -- (3,-0.2); 
    \draw[line width=0.6mm] (7.5,0.2) -- (7.5,-0.2); 
    \draw[line width=0.6mm] (12,0.2) -- (12,-0.2); 
    
    \node[above] at (3, 0.2) {\begin{tabular}{c}
            Leader \\
            \\
            Observes his type $\nature^{\leader}$\\ 
            (production costs)
        \end{tabular}};
    \node[below] at (3,-0.2) {\begin{tabular}{c}
            Takes his decision $\control_{\leader}$ \\ 
            (electricity prices)
        \end{tabular}};
    \node[above] at (7.5, 0.2) {\begin{tabular}{c}
            Randomness  $\exo{\nature}$ \\ 
            is realized   \\
        \end{tabular}};
    \node[below] at (7.5, -0.2) {\begin{tabular}{c}
            (demand)
        \end{tabular}};
    \node[above] at (12, 0.2) {\begin{tabular}{c}
            Follower \\
            \\
            Observes his type $\nature^{\follower}$ \\ 
            (unwillingness to shift)
        \end{tabular}};
    \node[below] at (12,-0.2) {\begin{tabular}{c}
            Takes his decision $\control_{\follower}$ \\ 
            (consumption shift)
        \end{tabular}};
    
    \node[right] at (15,0) {Time};
    \end{tikzpicture}
    \end{center}
\caption{Visualization of the information structure}
\label{visualization_information_structure}
\end{figure}

\begin{illustration}
    The leader (the electricity producer) observes his type (production costs) but does not observe the rest (demand, the consumer's consumption distribution, and consumer's reluctance toshift consumption). 
    We write 
    \begin{subequations}
        \begin{align}
        \tribu{\Information}_{\leader} = \trib^{\leader}
        \end{align}
    And for the follower, he observes his type but also the leader's action and the exogenous parameters, that is  
        \begin{align}
        \tribu{\Information}_{\follower} = \exo{\trib} \otimes \trib^{\follower} \otimes \tribu{\Control}_{\leader}
        \end{align}
    \end{subequations}.
    Even though we never mentioned the notion of time, we can reconstruct the array of time from the information structure, see Fig~\ref{visualization_information_structure}.
\end{illustration}

\subsection{Strategies, playability and solution map}
\label{SLSF_strategies_playability_solution_map}

After being able to specify the W-model in the single-leader-single-follower static case, we focus on how the leader and the follower play.
While not much can be said in the general case of leader-follower problems, we have a special information structure here.

\subsubsection*{Strategies}
Recall that the leader's and the follower's strategies are functions
\begin{subequations}
\begin{align}
    \policy_{\leader} : \; 
    \overbrace{\exo{\Nature} \times \Nature^{\leader} \times\Nature^{\follower} \times \CONTROL_{\leader} \times \CONTROL_{\follower}}^{\HISTORY} &\to \CONTROL_{\leader} \eqsepv \\
    \policy_{\follower} : \; 
    \underbrace{\exo{\Nature} \times \Nature^{\leader} \times\Nature^{\follower} \times \CONTROL_{\leader} \times \CONTROL_{\follower}}_{\HISTORY} &\to \CONTROL_{\follower} \eqsepv
\end{align}
\end{subequations}
which have to be measurable with respect to the information fields, that is:
\begin{subequations}
\begin{align}
    (\policy_{\leader})^{-1}(\tribu{\Control}_{\leader}) &\subset \tribu{\Information}_{\leader} \eqsepv \\
     (\policy_{\follower})^{-1}(\tribu{\Control}_{\follower}) &\subset \tribu{\Information}_{\follower} \eqfinp
\end{align}
\end{subequations}

Given the specific information structure given in \eqref{eq:information_structure}, the strategies' domains can therefore be restricted.
By denoting with a tilde the restricted functions, we have:
\begin{subequations}
\begin{align}
    \policy_{\leader} (\exo{\nature}, \nature^{\leader}, \nature^{\follower}, \cancel{\control_{\leader}}, \cancel{\control_{\follower}}) &= \widetilde{\policy}_{\leader} (\exo{\nature}, \nature^{\leader}, \nature^{\follower})  \eqsepv \\
    \policy_{\follower} (\exo{\nature}, \nature^{\leader}, \nature^{\follower}, \control_{\leader}, \cancel{\control_{\follower}}) &= \widetilde{\policy}_{\follower} (\exo{\nature}, \nature^{\leader}, \nature^{\follower}, \control_{\leader})   \eqfinp
\end{align}
\label{eq:strategies_restrictions}
\end{subequations}
We draw the reader's attention to the symmetry between equations \eqref{eq:information_structure} and \eqref{eq:strategies_restrictions}.
The absence of self-information means a player's strategy cannot depend on their own decision.
Furthermore, the leader does not observe the follower's decision, so their strategy depends solely on Nature.
In the general case, no further variables can be eliminated, but once the information structure is known, each player's strategy may depend only on certain parts of Nature.

\begin{illustration}
    In our example, the information fields are given precisely.
    Thus, we know that the leader's decision depends only on his type, that is 
    \begin{subequations}
        \begin{align}
        \control_{\leader} = \widetilde{\policy}_{\leader}(\cancel{\exo{\nature}}, \nature^{\leader}, \cancel{\nature^{\follower}}, \cancel{\control_{\leader}}, \cancel{\control_{\follower}}) \eqfinp
        \end{align}
    And the follower observes his type, the leader's action and sees the exogenous part of Nature, which writes 
        \begin{align}
        \control_{\follower} = \widetilde{\policy}_{\follower}(\exo{\nature}, \cancel{\nature^{\leader}}, \nature^{\follower}, \control_{\leader}, \cancel{\control_{\follower}}) \eqfinp
        \end{align}
    \end{subequations}
\end{illustration}

\subsubsection*{Playability}
There is a natural order: the leader plays first, then the follower.
This makes the information structure sequential hence causal hence playable.
We can therefore define the solution map.

\subsubsection*{Solution map}
Fix a strategy profile $(\policy_{\follower}, \policy_{\leader})$.
Recall that the solution map in a mapping from Nature to the configuration space
\begin{equation}
\begin{aligned}
    \underbrace{\exo{\Nature} \times \Nature^{\leader} \times\Nature^{\follower}}_{\Nature} \xrightarrow{\SolutionMap_{\policy_{\follower}, \policy_{\leader}}} \underbrace{\exo{\Nature} \times \Nature^{\leader} \times\Nature^{\follower} \times \CONTROL_{\leader} \times \CONTROL_{\follower}}_{\HISTORY} 
    \eqsepv
\end{aligned}
\end{equation}
which satisfies, using the restricted versions of the strategies written in \eqref{eq:strategies_restrictions},
\begin{equation}
\begin{aligned}
    \SolutionMap_{\policy_{\follower}, \policy_{\leader}}
    (\exo{\nature}, \nature^{\leader}, \nature^{\follower}) 
    = (\exo{\nature}, \nature^{\leader}, \nature^{\follower}, \control_{\leader}, \control_{\follower})  \iff 
   \left\{
    \begin{array}{l}
    \control_{\leader} = \widetilde{\policy}_{\leader}(\exo{\nature}, \nature^{\leader}, \nature^{\follower}) \eqsepv \\
    \control_{\follower} = \widetilde{\policy}_{\follower}(\exo{\nature}, \nature^{\leader}, \nature^{\follower}, \control_{\leader}) \eqfinp
    \end{array}
    \right.
\end{aligned}
\label{def_solution_map}
\end{equation}

The particular information structure outlined in equation \eqref{def_solution_map} enables the decomposition of the solution map into two intermediate solution maps, as shown below:
\begin{equation}
\begin{aligned}
    \underbrace{\exo{\Nature} \times \Nature^{\leader} \times\Nature^{\follower}}_{\Nature} \xrightarrow{\widetilde{\SolutionMap}_{\policy_{\leader}}} \exo{\Nature} \times \Nature^{\leader} \times\Nature^{\follower} \times \CONTROL_{\leader}
    \xrightarrow{\widetilde{\SolutionMap}_{\policy_{\follower}}} \underbrace{\exo{\Nature} \times \Nature^{\leader} \times\Nature^{\follower} \times \CONTROL_{\leader} \times \CONTROL_{\follower}}_{\HISTORY} 
    \eqfinp
\end{aligned}
\end{equation}

The first solution map could be called the \emph{leader solution map} and gives the leader's decision for each state of Nature:
\begin{equation}
    \widetilde{\SolutionMap}_{\policy_{\leader}}(\exo{\nature}, \nature^{\leader}, \nature^{\follower})  = \left (\exo{\nature}, \nature^{\leader}, \nature^{\follower}, \widetilde{\policy}_{\leader}\left(\exo{\nature}, \nature^{\leader}, \nature^{\follower}\right)\right)
    \eqfinp
\end{equation}
The second solution map could be called the \emph{follower solution map} and gives the follower's decision for each state of Nature and according to the leader's decision:
\begin{equation}
    \widetilde{\SolutionMap}_{\policy_{\follower}}(\exo{\nature}, \nature^{\leader}, \nature^{\follower}, \control_{\leader})  = \left( \exo{\nature}, \nature^{\leader}, \nature^{\follower}, \control_{\leader}, \widetilde{\policy}_{\follower}\left (\exo{\nature}, \nature^{\leader}, \nature^{\follower}, \control_{\leader} \right) \right)
    \eqfinp
\end{equation}
It is easy to see that we have indeed
\begin{equation}
    \SolutionMap_{\policy_{\follower}, \policy_{\leader}} = \widetilde{\SolutionMap}_{\policy_{\follower}} \compo \widetilde{\SolutionMap}_{\policy_{\leader}}
    \eqfinp
\end{equation}

We are still dealing with the general case where we cannot say more.
However, when the information structures are given, that is when equation \eqref{eq:information_structure} is made of equality, we can be more precise in expressing the solution map. 

In what follows, we will use the same notation for the actual functions and their restrictions to the influential variables. 
There is no ambiguity in this, as it is sufficient to refer to the variables that appear as arguments of the functions to understand which function is being discussed. 

\begin{figure}[ht]
\begin{center}
    \begin{tikzpicture}[scale = 0.9]
    \draw[->, line width=0.6mm] (0,0) -- (15,0); 

    \draw[line width=0.6mm] (3,0.2) -- (3,-0.2); 
    \draw[line width=0.6mm] (7.5,0.2) -- (7.5,-0.2); 
    \draw[line width=0.6mm] (12,0.2) -- (12,-0.2); 
    
    \node[above] at (3, 0.2) {\begin{tabular}{c}
            Leader \\
            \\
            Observes his type $\nature^{\leader}$\\ 
        \end{tabular}};
    \node[below] at (3,-0.2) {\begin{tabular}{c}
            Takes his decision  \\ 
            $\policy_{\leader}(\nature^{\leader})$ \\ 
        \end{tabular}};
     \node[above] at (7.5, 0.2) {\begin{tabular}{c}
            Randomness  $\exo{\nature}$ 
        \end{tabular}};
    \node[above] at (12, 0.2) {\begin{tabular}{c}
            Follower \\
            \\
            Observes his type $\nature^{\follower}$\\ 
        \end{tabular}};
    \node[below] at (12,-0.2) {\begin{tabular}{c}
            Takes his decision \\
            $\policy_{\follower} \big(\exo{\nature}, \nature^{\follower}, \underbrace{\policy_{\leader} (\nature^{\leader})}_{\control_{\leader}} \big)$ \\ 
        \end{tabular}};
    
    \node[right] at (15,0) {Time};
    \end{tikzpicture}
    \end{center}
    \caption{Building the solution map}
    \label{building_solution_map}
\end{figure}
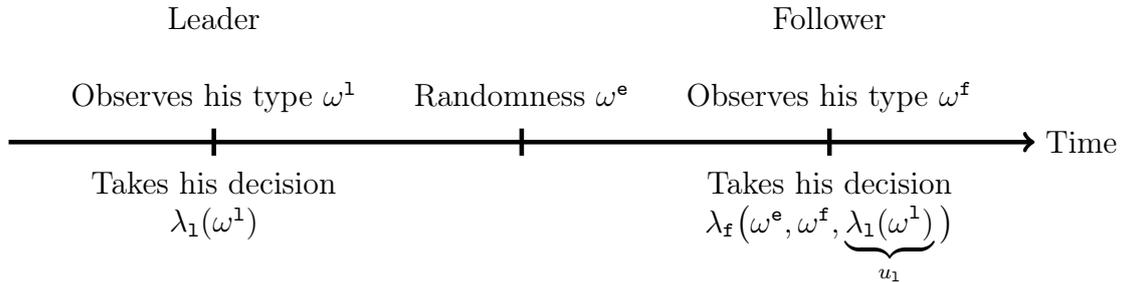

\begin{illustration}
    To easily obtain the expression for the solution map, we can take a look at Fig.~\ref{building_solution_map} and piece together all the elements of the configuration space, that is
    \begin{equation}
    \SolutionMap_{\policy_{\leader}, \policy_{\follower}}
        (\exo{\nature}, \nature^{\leader}, \nature^{\follower}) 
        = \Big(\exo{\nature}, \nature^{\leader}, \nature^{\follower},  \policy_{\leader} \big(\nature^{\leader}\big),  \policy_{\follower} \big(\exo{\nature}, \nature^{\follower}, \policy_{\leader} (\nature^{\leader})\big)\Big) \eqfinp    
    \end{equation}
\end{illustration}

\subsection{Players, objective functions and risk measures/beliefs}
\label{SLSF_players_objective_functions_beliefs}

We now examine the components of a leader-follower W-game.

\subsubsection*{Players}
Formally, the set of players is $\PLAYER = \{ \leader, \follower \}$, and the agents associated with each player are $\AgentSet^{\leader} = \{ \leader \}$ and $\AgentSet^{\follower} = \{ \follower \}$.

\begin{illustration}
    The leader is still the electricity producer and the follower is the consumer.
\end{illustration}

\subsubsection*{Objective functions}
Unless otherwise stated, the leader's objective function is always a payoff, meant to be maximized, and the follower's objective function is always a cost, meant to be minimized.

Since we will draw inspiration from optimization articles, we will need to translate the constraints that prevent the leader and the follower from making a decision within their entire decision set.
To do this, we will often introduce subsets of decisions $\Control_{\leader} \subset \CONTROL_{\leader}$ and $ \Control_{\follower} \subset \CONTROL_{\follower}$ modeling the constraints on the decision variables. To impose the decision to belong to these subsets, we use the indicator function.
Recall that for a given set $\Control$, the indicator function $\Indicator{\Control}$ is defined as:
\begin{equation}
    \Indicator{\Control}(u) =
    \begin{cases}
    0 & \text{if } u \in \Control \eqsepv \\
    +\infty & \text{if } u \notin \Control \eqfinp
    \end{cases}
\end{equation}

\begin{figure}[ht]
    \begin{center}
    \begin{tikzpicture}
    
    \tikzset{
        block/.style = {rectangle, draw, text centered, minimum height=3em, minimum width=10em},
        arrow/.style = {thick,->,>=stealth},
        darrow/.style = {thick, dashed,->,>=stealth},
        zigzag/.style = {decorate, decoration=zigzag, thick,->,>=stealth}
    }
    
    \node[block] (Electricity producer) {
        \begin{tabular}{c}
            Maximizes: (sales - production costs) \\
            Decides: (peak, off-peak) prices \\
            Has: production costs \faCog
        \end{tabular}
    };
    \node[block, below=3cm of Electricity producer, xshift = 1.4cm] (Consumer) {
        \begin{tabular}{c}
            Minimizes: (bills + unwillingness cost) \\
            Decides: consumption profile \\
            Has: unwillingness to shift \faAngry
        \end{tabular}
    };
    
    \node[block, below=1cm of Electricity producer, xshift=2.4cm] (Weather) {
        \begin{tabular}{c}
            Determines: demand \faLightbulb\\
        \end{tabular}
    };
       
    \draw[arrow] ($(Electricity producer.south east) + (-4.35,0)$) -- ($(Consumer.north) +(-2,0)$) ;
    
    \draw[darrow] (Consumer.south) |- ++(0,-0.8) -| ($(Electricity producer.south west) + (0.5,0)$)
    node[pos=0.25, below] {
    };

    \draw[zigzag] (Weather.south) -- ($(Consumer.north) + (1, 0)$);
    
    \node[right=0.5cm of Electricity producer] {
    \begin{tabular}{c}
            Leader (player) \\ 
            Leader (agent) \\
            electricity producer \faIndustry
        \end{tabular}
    };
    \node[right=0.5cm of Consumer] {
    \begin{tabular}{c}
            Follower (player) \\
            Follower (agent) \\ 
            consumer \faHome
        \end{tabular}
    };
    \node[right=0.5cm of Weather] {
    \begin{tabular}{c}
            \emph{Exogenous Nature} \\ 
            \emph{weather} \faCloud
        \end{tabular}
    };
    
    \end{tikzpicture}
    \end{center}
    \caption{Illustration of the objective functions}
    \label{illustration_objective_functions}
\end{figure}
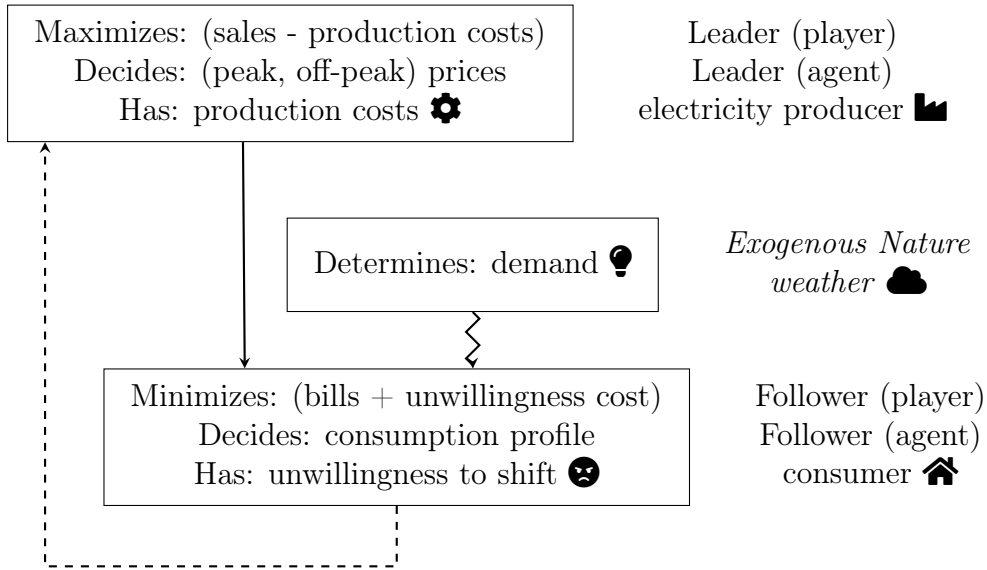

\begin{illustration}
    The leader's payoff is made of the profits made by selling electricity minus her production cost:
    \begin{equation}
    \criterion^{\leader}(\exo{\nature}, \nature^{\leader}, \cancel{\nature^{\follower}} \control_{\leader}, \control_{\follower}) = 
    \underbrace{\overbrace{\exo{\nature} \overline{\control}_{\follower} }^{\scriptsize{\substack{\text{peak}\\ \text{demand}}}}  
    \overbrace{\overline{\control}_{\leader}}^{\scriptsize{\substack{\text{peak} \\ \text{price}}}}
    + \overbrace{\exo{\nature}\underline{\control}_{\follower}}^{\scriptsize{\substack{\text{off-peak} \\ \text{demand}}}} 
    \overbrace{\underline{\control}_{\leader}}^{\scriptsize{\substack{\text{off-peak} \\ \text{price}}}}}_{\text{sales}}
    - \underbrace{\overbrace{\exo{\nature}}^{\scriptsize{\substack{\text{total} \\ \text{demand}}}}  
    \overbrace{\nature^{\leader}}^{\scriptsize{\substack{\text{unitary} \\ \text{cost}}}}}_{\text{production cost}} \eqfinp
    \end{equation}
    And the follower's cost is made of the cost of electricity minus her unwillingness to change consumption:
    \begin{equation}
    \criterion^{\follower}(\exo{\nature}, \cancel{\nature^{\leader}}, \nature^{\follower} \control_{\leader}, \control_{\follower}) = 
    \underbrace{\overbrace{ \overline{\control}_{\follower} \exo{\nature}}^{\scriptsize{\substack{\text{peak}\\ \text{demand}}}}  
    \overbrace{\overline{\control}_{\leader}}^{\scriptsize{\substack{\text{peak} \\ \text{price}}}}
    + \overbrace{\exo{\nature}\underline{\control}_{\follower}}^{\scriptsize{\substack{\text{off-peak} \\ \text{demand}}}}  
    \overbrace{\underline{\control}_{\leader}}^{\scriptsize{\substack{\text{off-peak} \\ \text{price}}}}}_{\text{bills}}
    + \underbrace{\overbrace{\exo{\nature}\underline{\control}_{\follower}}^{\scriptsize{\substack{\text{off-peak} \\ \text{demand}}}}  \overbrace{\nature^{\follower}}^{\scriptsize{\substack{\text{unwillingness} \\ \text{to shift}}}}}_{\text{inconvenience cost}} \eqfinp
    \end{equation}
     Including all of this, we obtain Fig.\ref{illustration_objective_functions}.
\end{illustration}

\subsubsection*{Risk measures/beliefs}

\subsubsubsection{With risk measures}
The leader faces the random variable $\criterion^{\leader} 
    \compo \SolutionMap_{\policy_{\leader}} 
    \compo \SolutionMap_{\policy_{\follower}} : \exo{\Nature} \times \Nature^{\leader} \times \Nature^{\follower} \to \HISTORY$ then is equipped with a risk measure $\RiskMeasure^{\leader}$ over an implicit space $\RiskSpace^{\leader}$.
    The same goes for the follower endowed with a risk measure $\RiskMeasure^{\follower}$ over a space $\RiskSpace^{\follower}$.

\subsubsubsection{With beliefs}
When working with beliefs, we will assume that a player's beliefs about each part of Nature are independent of one another.
This allows us to express the beliefs as a product of probability measures on the different parts of Nature like so:
\begin{subequations}
\begin{align}
    \Nature &= \exo{\Nature} \times
    \Nature^{\leader} \times \Nature^{\follower} \eqsepv \\
    \Belief^{\leader} &= \exog{\Belief}^{\leader} \otimes
    \Belief^{\leader}_{\leader} \otimes
    \Belief^{\leader}_{\follower} \eqsepv \\
   \Belief^{\follower} &= \exog{\Belief}^{\follower} \otimes
    \Belief^{\follower}_{\leader} \otimes
    \Belief^{\follower}_{\follower} \eqfinp
\end{align}
\end{subequations}
It is crucial to specify which player each belief belongs to, since there is no reason to assume that the players share the same beliefs on the same parts of Nature.
When a player knows a part of Nature, this means that they are equipped with a Dirac measure on an element of that part, denoted as $\delta_{\overline{\nature}}$.

\begin{illustration}
    We consider a simple case where the leader and the follower base their decisions on their expected gain.
    In other words, we work with beliefs and not risk measures.
    Since the leader knows his type,  he is endowed with a belief about the exogenous part of Nature and about the type of the follower.
    His belief writes 
    \begin{subequations}
        \begin{align}
        \Belief^{\leader} &= \exog{\Belief}^{\leader} \otimes \delta_{\overline{\nature}^{\leader}} \otimes \Belief^{\leader}_{\follower} \eqfinp
        \end{align}
    For the follower, he knows exogenous Nature and he knows his type.
    Nevertheless, he has a belief about the type of the leader, thus we write
        \begin{align}
        \Belief^{\follower} &= \delta_{\overline{\exo{\nature}}} \otimes \Belief^{\follower}_{\leader} \otimes \delta_{\overline{\nature}^{\follower}} \eqfinp
        \end{align} 
    \end{subequations}
\end{illustration}

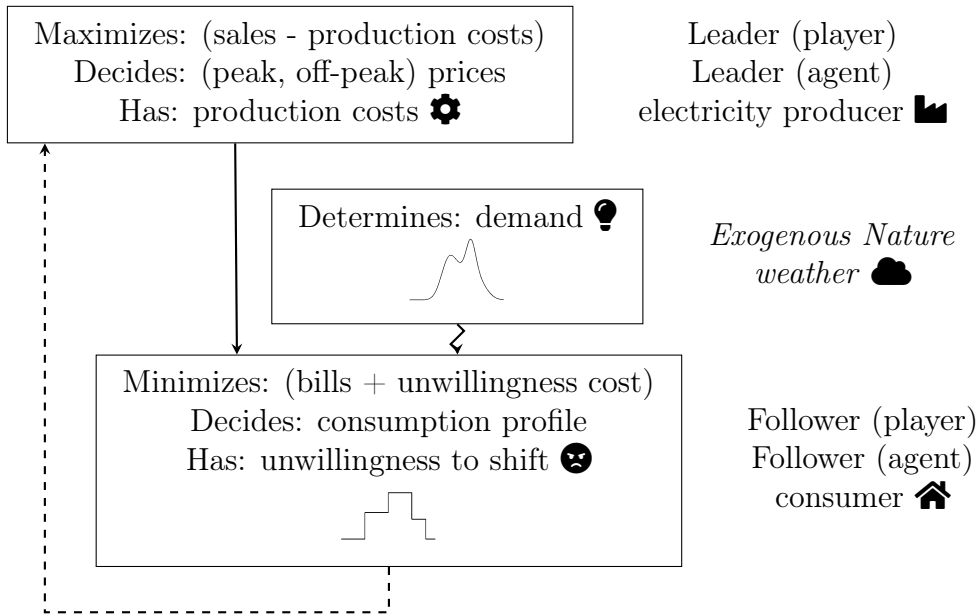
\begin{figure}[ht]
    \centering
    \begin{tikzpicture}
        
        \tikzset{
            block/.style = {rectangle, draw, text centered, minimum height=3em, minimum width=10em},
            arrow/.style = {thick,->,>=stealth},
            darrow/.style = {thick, dashed,->,>=stealth},
            zigzag/.style = {decorate, decoration=zigzag, thick,->,>=stealth}
        }
        
        \node[block] (Electricity producer) {
            \begin{tabular}{c}
                Maximizes: (sales - production costs) \\
                Decides: (peak, off-peak) prices \\
                Has: production costs \faCog
            \end{tabular}
        };
        \node[block, below=2.8cm of Electricity producer, xshift = 1.3cm] (Consumer) {
            \begin{tabular}{c}
                Minimizes: (bills + unwillingness cost) \\
                Decides: consumption profile \\
                Has: unwillingness to shift \faAngry \\
                \includegraphics[width=1.3cm]{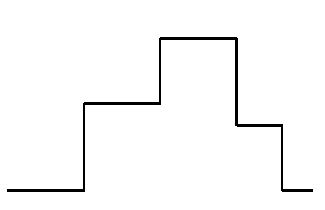}
            \end{tabular}
        };
        
        \node[block, below=0.6cm of Electricity producer, xshift=2.2cm] (Weather) {
            \begin{tabular}{c}
                Determines: demand \faLightbulb \\
                \includegraphics[width=1.3cm]{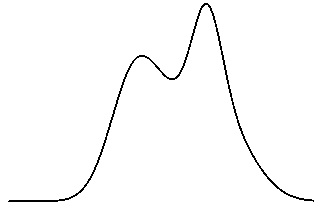}
            \end{tabular}
        };
           
        \draw[arrow] ($(Electricity producer.south east) + (-4.45,0)$) -- ($(Consumer.north) +(-2,0)$) ;
        
        \draw[darrow] (Consumer.south) |- ++(0,-0.6) -| ($(Electricity producer.south west) + (0.5,0)$)
        node[pos=0.25, below] {
        };
    
        \draw[zigzag] (Weather.south) -- ($(Consumer.north) + (0.9, 0)$);
        
        \node[right=0.5cm of Electricity producer] {
        \begin{tabular}{c}
                Leader (player) \\
                Leader (agent) \\ 
                electricity producer \faIndustry
            \end{tabular}
        };
        \node[right=0.5cm of Consumer] {
        \begin{tabular}{c}
                Follower (player) \\
                Follower (agent) \\ 
                consumer \faHome
            \end{tabular}
        };
        \node[right=0.5cm of Weather] {
        \begin{tabular}{c}
                \emph{Exogenous Nature} \\
                \emph{weather} \faCloud
            \end{tabular}
        };
        
        \end{tikzpicture}
    \caption{Illustration of the leader's belief}
    \label{fig:illustration_leader_belief}
\end{figure}

\begin{figure}[ht]
    \centering
    \begin{tikzpicture}
    
    \tikzset{
        block/.style = {rectangle, draw, text centered, minimum height=3em, minimum width=10em},
        arrow/.style = {thick,->,>=stealth},
        darrow/.style = {thick, dashed,->,>=stealth},
        zigzag/.style = {decorate, decoration=zigzag, thick,->,>=stealth}
    }
    
    \node[block] (Electricity producer) {
        \begin{tabular}{c}
            Maximizes: (sales - production costs) \\
            Decides: (peak, off-peak) prices \\
            Has: production costs \faCog \\
            \includegraphics[width=1.3cm]{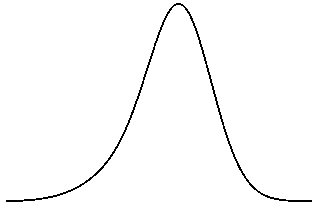}
        \end{tabular}
    };
    \node[block, below=2.2cm of Electricity producer, xshift = 1.3cm] (Consumer) {
        \begin{tabular}{c}
            Minimizes: (bills + unwillingness cost) \\
            Decides: consumption profile \\
            Has: unwillingness to shift \faAngry 
        \end{tabular}
    };
    
    \node[block, below=0.6cm of Electricity producer, xshift=2.2cm] (Weather) {
        \begin{tabular}{c}
            Determines: demand \faLightbulb \\
        \end{tabular}
    };
       
    \draw[arrow] ($(Electricity producer.south east) + (-4.45,0)$) -- ($(Consumer.north) +(-2,0)$) ;
    
    \draw[darrow] (Consumer.south) |- ++(0,-0.6) -| ($(Electricity producer.south west) + (0.5,0)$)
    node[pos=0.25, below] {
    };

    \draw[zigzag] (Weather.south) -- ($(Consumer.north) + (1, 0)$);
    
    \node[right=0.5cm of Electricity producer] {
    \begin{tabular}{c}
            Leader (player) \\
            Leader (agent) \\
            electricity producer \faIndustry
        \end{tabular}
    };
    \node[right=0.5cm of Consumer] {
    \begin{tabular}{c}
            Follower (player) \\ 
            Follower (agent) \\
            consumer \faHome
        \end{tabular}
    };
    \node[right=0.5cm of Weather] {
    \begin{tabular}{c}
            \emph{Exogenous Nature} \\ 
            \emph{weather} \faCloud
        \end{tabular}
    };
    
    \end{tikzpicture}
    \caption{Illustration of the follower's belief}
    \label{fig:illustration_follower_belief}
\end{figure}
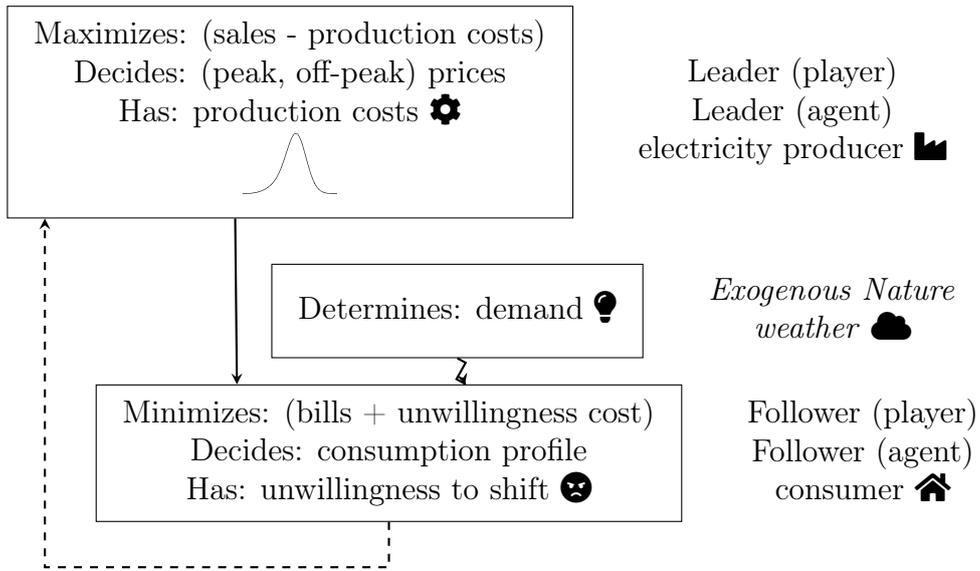


\section{Dynamic and multi-leader-multi-follower setup}
\label{MLMFS_setup}

In this section, we explain how we will address more complex problems.
In \S.~\ref{dynamic_single-leader-single-follower_problems}, we will consider cases where each player has multiple agents, and in \S.~\ref{MLMF_setup}, we will consider cases involving multiple leaders and multiple followers.


\subsection{Dynamic single-leader-single-follower problems}
\label{dynamic_single-leader-single-follower_problems}

In a dynamic setup, the player makes multiple decisions, each with a different information field.
This means that each player controls multiple agents, and exogenous Nature is revealed over time.

\subsubsection*{Agents and players}
From now on, the leader and the follower perform a series of actions over a time period $\TIME$.
We will therefore express their agents in the form of a pair (player, time), that is,
\begin{subequations}
    \begin{align}
    \AgentSet^{\leader} &= \{ \leader \} \times \TIME^{\leader} \eqsepv \\
    \AgentSet^{\follower} &= \{ \follower \} \times \TIME^{\follower} \eqfinp
    \end{align}
\end{subequations}

As before, we adopt a notation that highlights which player each agent belongs to.

\begin{illustration}
    We extend the previous example. 
    The electricity producer (leader) sets the electricity prices for one year, so we still have $\AgentSet^{\leader} = \{ \leader \}$.
    However, the consumer (follower) can decide each day how he will shift his consumption, which we write $\AgentSet^{\follower} = \{ (\follower, \mytime) \eqsepv \mytime \in \{ 1,\ldots 365 \} \}$.
    In other words, we set $\TIME^\leader = \emptyset$ and $\TIME^\follower = \{ 1,\ldots 365 \}$ : only the follower operates within the multi-agent framework.
\end{illustration}

\subsubsection*{Nature}
To represent the evolution of knowledge regarding exogenous uncertainties, exogenous Nature can be equipped with a filtration\footnote{A filtration is an increasing (in the sense of inclusion) sequence of $\sigma$-fields.} 
\begin{equation}
    (\exo{\trib}_{\mytime})_{\mytime \in \TIME}
\end{equation}
where the time indexes $\mytime$ belong to a time span $\exo{\TIME}$.
However, we do not modify the rest of Nature so that we always have
\begin{equation}
    \trib = \exo{\trib} \otimes \trib^\leader \otimes \trib^\follower \eqfinp
\end{equation}

In practice, the two most common structures encountered for exogenous Nature are the Cartesian product and the tree. 
In the first case, this amounts to writing
\begin{equation}
    \exo{\Nature} = \prod_{\mytime \in \exo{\TIME}} \exo{\Nature}_{\mytime} \eqsepv
\end{equation}
where each set $\exo{\Nature}_{\mytime}$ is equipped with a $\sigma$-field $\exo{\trib}_{\mytime}$.

When working with a tree, we consider the natural filtration on a tree.
In other words, when we are at a given node in the tree, the natural filtration contains all the information accumulated along the path leading to that node from the root.

\begin{illustration}
    We write exogenous Nature as a Cartesian product so that a state of Nature writes 
    \begin{equation}
        \nature = \bigl((\exo{\nature}_{\mytime})_{\mytime \in \exo{\TIME}}, \nature^{\leader}, \nature^{\follower}\bigr) \in 
        \Nature = \RR_+^{\exo{\TIME}} \times \RR_+ \times \RR_+ \eqsepv
    \end{equation} 
    where $\nature^{\leader}$ is still the producer's unitary production cost (€/kWh), $\nature^{\follower}$ is still the consumer's unwillingness to shift consumption but now $\exo{\nature}_{\mytime}$ is the demand at day $\mytime$ (kWh).
\end{illustration}

\subsubsection*{Actions}
Each leader agent $(\leader, \mytime)$ comes with an action set $(\CONTROL_{(\leader, \mytime)}, \tribu{\Control}_{(\leader, \mytime)})$.
The same goes for the follower player whose agents $(\follower, \mytime)$ make their decision in the action set $(\CONTROL_{(\follower, \mytime)}, \tribu{\Control}_{(\follower, \mytime)})$.


\begin{illustration}
    For the leader, nothing changes, we have 
    \begin{subequations}
        \begin{align}
            \control_{\leader} = (\overline{\control}_{\leader}, \underline{\control}_{\leader}) \in \CONTROL_{\leader} = \left\{ (x, y) \in \RR^2 \mid x \geq y \right\} \eqsepv
        \end{align}
        but for the follower, we simply consider the sequence of his actions, which writes
        \begin{align}
        \control_{(\follower, \mytime)} = (\overline{\control}_{(\follower, \mytime)}, \underline{\control}_{(\follower, \mytime)}) \in \CONTROL_{(\follower, \mytime)} = \left\{ (\alpha_\mytime, \beta_\mytime)_\mytime \in (\RR_+^2)^\TIME \mid \forall \mytime \in \TIME \eqsepv \alpha_\mytime + \beta_\mytime = 1 \right\}  \eqsepv
        \end{align}
    \end{subequations}
    and represents the fraction of consumption between peak and off-peak hours for each day $\mytime$.
\end{illustration}

\subsubsection*{Configuration space}
The configuration space is now bigger because made of product spaces but keeps the same structure
\begin{subequations}
\begin{align}
    \HISTORY = \exo{\Nature} \times \; \Nature^{\leader} \times\Nature^{\follower} \times \; \prod_{\mytime \in \TIME^{\leader}} \CONTROL_{(\leader, \mytime)} \times \prod_{\mytime \in \TIME^{\follower}} \CONTROL_{(\follower, \mytime)} \eqsepv 
\end{align}
equipped with the $\sigma$-field
\begin{align}
    \tribu{\History} = \exo{\trib} \otimes \; \trib^{\leader} \otimes \trib^{\follower} \otimes \; \bigotimes_{\mytime \in \TIME^{\leader}} \tribu{\Control}_{(\leader, \mytime)} \otimes \bigotimes_{\mytime \in \TIME^{\follower}} \tribu{\Control}_{(\follower, \mytime)}\eqfinp 
\end{align}
\end{subequations}

\begin{illustration}
    We simply update equation $\eqref{ex_configuration_space}$
    \begin{equation}
    \begin{split}
    \HISTORY &= \RR^{\TIME}_+ \times \RR_+ \times \RR_+ \times \left\{ (x, y) \in \RR^2 \mid x \geq y \right\} \times \\ &\left\{ (\alpha_\mytime, \beta_\mytime)_\mytime \in (\RR_+^2)^\TIME \mid \forall \mytime \in \TIME \eqsepv \alpha_\mytime + \beta_\mytime = 1 \right\} \eqsepv 
    \end{split}
    \end{equation}
    and we leave it to the reader to rewrite the corresponding $\sigma$-field.
\end{illustration}

\subsubsection*{Information fields}
For the information structures, we must first revisit the definition of the leader and the follower.
Formally, we first introduce the notion of a \emph{static team}, which refers to the set of agents who do not know more than the state of Nature.
Then, we say that the leader player is the player whose group of agents has a non-empty intersection with the static team. 
Finally, we say that the follower player is the player whose group of agents has an empty intersection with the static team.




\begin{illustration}
\begin{subequations}
    The leader's information field remains the same, that is 
    \begin{align}
       \tribu{\Information}_{\leader} = \trib^{\leader} \eqfinp
    \end{align}
    For the follower, as before, each of her agent knows the leader's action and the follower's type.
    Furthermore, he is  we can assume that each agent knows only the demand for the upcoming day, 
    \begin{align}
        \tribu{\Information}_{(\follower, \mytime)} = \exo{\trib}_{\mytime} \otimes \trib^{\follower} \otimes \tribu{\Control}_{\leader} \eqsepv
    \end{align} 
    or remembers all past demands, 
    \begin{align}
        \tribu{\Information}_{(\follower, \mytime)} = \bigotimes_{\mytime' \leq \mytime} \exo{\trib}_{\mytime'} \otimes \trib^{\follower} \otimes \tribu{\Control}_{\leader} \eqfinp
    \end{align}
\end{subequations}
\end{illustration}

\subsubsection*{Strategies}
The strategy set for any leader agent (resp. follower) writes $\Policy_{(\leader, \mytime)}$ (resp. $\Policy_{(\leader, \mytime)}$).
When dealing with games, we rather consider the strategy sets for the leader player and the follower player that correspond to the following Cartesian products:
\begin{subequations}
    \begin{align}
    \Policy^{\leader} &= \prod_{\mytime \in \TIME^{\leader}} \Policy_{(\leader, \mytime)} \eqsepv \\ 
    \Policy^{\follower} &= \prod_{\mytime \in \TIME^{\follower}} \Policy_{(\follower, \mytime)} \eqfinp
    \end{align}
\end{subequations}

\begin{illustration}
    In the example, we have $\control_{\leader} = \policy_{\leader}(\nature^{\leader})$ and $\control_{(\follower, \mytime)} = \policy_{(\follower, \mytime)}(\exo{\nature}, \nature^{\follower}, \control_{\leader})$.
\end{illustration}


\subsection{Multi-leader-multi-follower problems}
\label{MLMF_setup}

When dealing with problems where the leader is actually a set of players, and the follower is as well, we need new notations and adjustments to the configuration space.

\subsubsection*{Agents and players}
From now on, $\Leader$ will no longer denote a leader, but rather the set of players $\leader$ of leader type.
The same applies to $\Follower$.
Thus, we can write
\begin{equation}
    \PLAYER = \Leader \sqcup \Follower \eqsepv
\end{equation}
and we specify that each player can have multiple agents, meaning that we have
\begin{subequations}
\begin{align}
    \forall \leader \in \Leader \eqsepv \AgentSet^{\leader} &= \{  \leader \} \times \TIME^{\leader} \eqsepv \\
    \forall \follower \in \Follower \eqsepv \AgentSet^{\follower} &= \{  \follower \} \times \TIME^{\follower}  \eqfinp
\end{align}
\end{subequations}

\begin{illustration}
    We still consider a single electricity producer, hence a single leader, but we will account for the heterogeneity in consumption behaviors by introducing a set of followers $\Follower$, where each element $\follower$ is a consumer.
\end{illustration}

\subsubsection*{Nature}
We divided Nature to highlight the type of each player.
Here, since we have several leaders and several followers, we will therefore write
\begin{subequations}
\begin{equation}
    \Nature = \exo{\Nature} \times \prod_{\leader \in \Leader}\Nature^{\leader} \times \prod_{\follower \in \Follower} \Nature^{\follower} \eqsepv
\end{equation}
equipped with the $\sigma$-field
\begin{equation}
    \trib = \exo{\trib} \otimes \bigotimes_{\leader \in \Leader} \trib^{\leader} \bigotimes_{\follower \in \Follower} \trib^{\follower} \eqfinp
\end{equation}
\end{subequations}

\begin{illustration}
    In total, a state of Nature writes 
    \begin{equation}
        \nature = \big((\exo{\nature}_{\mytime})_{\mytime \in \exo{\TIME}}, \nature^{\leader}, (\nature^{\follower})_{\follower \in \Follower}\big) \in \Nature = \RR_+^{\exo{\TIME}} \times \RR_+ \times \RR_+^\Follower  \eqsepv
    \end{equation}
    where $\nature^{\leader}$ is still the producer's unitary production cost (€/kWh), $\nature^{\follower}$ is the unwillingness to shift consumption of consumer $\follower$ and $\exo{\nature}_{\mytime}$ is still the demand at day $\mytime$ (kWh).
\end{illustration}

\subsubsection*{Actions}
Each player has a bunch of agents, each coming with its own action set. 

\begin{illustration}
    The leader plays
    \begin{subequations}
        \begin{align}
            \control_{\leader} = (\overline{\control}_{\leader}, \underline{\control}_{\leader}) \in \CONTROL_{\leader} = \left\{ (x, y) \in \RR^2 \mid x \geq y \right\} \eqsepv
        \end{align}
        but then each agnet of each follower decides 
        \begin{align}
        \control_{(\follower, \mytime)} = (\overline{\control}_{(\follower, \mytime)}, \underline{\control}_{(\follower, \mytime)}) \in \CONTROL_{(\follower, \mytime)} = \left\{ (\alpha_\mytime, \beta_\mytime)_\mytime \in (\RR_+^2)^\TIME \mid \forall \mytime \in \TIME \eqsepv \alpha_\mytime + \beta_\mytime = 1 \right\} \eqfinp
        \end{align}
    \end{subequations}
    and represents the fraction of consumption between peak and off-peak hours for each day $\mytime$.
\end{illustration}

\subsubsection*{Configuration space}
The modularity of the W-model enables us to write the configuration space in a simple way, that is
\begin{subequations}
\begin{align}
    \HISTORY = \exo{\Nature} \times \; \prod_{\leader \in \Leader} \Nature^{\leader} \times \prod_{\follower \in \Follower} \Nature^{\follower} \times \; \prod_{\leader \in \Leader} \prod_{\mytime \in \TIME} \CONTROL_{(\leader, \mytime)} \times \prod_{\follower \in \Follower} \prod_{\mytime \in \TIME} \CONTROL_{(\follower, \mytime)} \eqsepv 
\end{align}
equipped with the $\sigma$-field
\begin{align}
    \tribu{\History} = \exo{\trib} \otimes \; \bigotimes_{\leader \in \Leader} \trib^{\leader} \otimes \bigotimes_{\follower \in \Follower} \trib^{\follower} \otimes \; \bigotimes_{\leader \in \Leader} \bigotimes_{\mytime \in \TIME} \tribu{\Control}_{(\leader, \mytime)} \otimes \bigotimes_{\follower \in \Follower} \bigotimes_{\mytime \in \TIME} \tribu{\Control}_{(\follower, \mytime)} \eqfinp 
\end{align}
\end{subequations}

We draw the reader's attention to the fact that $\exo{\Nature}$ can be itself a product space over the time period considered.


\section{Nash-Stackelberg equilibria in W-games}
\label{Nash-Stackelberg_equilibria_W-games}

In the specific case of leader-follower problems, a new notion of equilibrium is introduced, namely the Nash-Stackelberg equilibria, which is based on the idea that the Leader seeks to anticipate the Follower's actions by playing a Stackelberg strategy.
We first introduce this concept in the case of two players and then generalize these definitions to the multi-leader-multi-follower case. 


\subsection{Single-leader-single-follower}

Let us now turn to the behaviors of the leader and the follower. We will detail their normal form payoffs and then address the concepts of best response, Nash equilibrium, and introduce the notion of Stackelberg equilibrium.

\subsubsection*{Normal form payoffs}
We provide the expression of the normal form payoffs based on whether we are working with risk measures or in the specific case of beliefs.
Since we are dealing with only two players, we can represent the game in matrix form in Table.~\ref{tab:normal_form_representation_W-game}.
\begin{table}[ht]
\begin{center}
    \begin{tabular}{|c|c|c|c|}
    \hline
     $\leader \; , \; \follower$ & $\ldots$ & $\policy^{\follower}$ & $\ldots$ \\ \hline
     $\ldots$ & &  &  \\ \hline
     $\policy^{\leader}$ &  & $\Criterion^{\leader}(\policy^{\leader}, \policy^{\follower} ; \data^{\leader}) \; , \; \Criterion^{\follower}(\policy^{\leader}, \policy^{\follower} ; \data^{\follower})$ &  \\ \hline
     $\ldots$ &  &  &  \\ \hline
    \end{tabular}
    \caption{Normal form representation of a W-game}
    \label{tab:normal_form_representation_W-game}
\end{center}
\end{table}

\subsubsubsection{With risk measures}
We can write the leader's normal form payoff like so
\begin{subequations}
    \begin{align}
    \Criterion^{\leader}(\policy^{\leader}, \policy^{\follower} ; \data^{\leader}) =
    \RiskMeasure^{\leader}[\criterion^{\leader} 
    \compo \SolutionMap_{\policy^{\leader}} 
    \compo \SolutionMap_{\policy^{\follower}}] \eqsepv
    \end{align}
and for the follower we can write 
     \begin{align}
    \Criterion^{\follower}(\policy^{\leader}, \policy^{\follower} ; \data^{\follower}) =
    \RiskMeasure^{\follower}[\criterion^{\follower} 
    \compo \SolutionMap_{\policy^{\leader}} 
    \compo \SolutionMap_{\policy^{\follower}}] \eqfinp
    \end{align}
\end{subequations}

\subsubsubsection{With beliefs}
In the particular case where players seek to maximize the expected value of what they obtain using their beliefs, we can write the leader's normal form payoff as
\begin{subequations}
    \begin{align}
    \Criterion^{\leader}(\policy^{\leader}, \policy^{\follower} ; \data^{\leader}) =
    \expect{\Belief^{\leader}}[\criterion^{\leader} 
    \compo \SolutionMap_{\policy^{\leader}} 
    \compo \SolutionMap_{\policy^{\follower}}] \eqsepv
    \end{align}
and for the follower's normal form cost as
     \begin{align}
    \Criterion^{\follower}(\policy^{\leader}, \policy^{\follower} ; \data^{\follower}) =
    \expect{\Belief^{\follower}}[\criterion^{\follower} 
    \compo \SolutionMap_{\policy^{\leader}} 
    \compo \SolutionMap_{\policy^{\follower}}] \eqfinp
    \end{align}
\end{subequations}

\subsubsection*{Follower's best response}
Since this is going to be useful, we explicitly write the set of best responses of the follower to a leader's strategy $\underline{\policy}^{\leader}$:
\begin{equation}
    \Nash{\Policy}^{\follower}(\underline{\policy}^{\leader}; \data^{\follower}) = \argmin_{\policy^{\follower} \in \Policy^{\follower}} \Criterion^{\follower}(\underline{\policy}^{\leader}, \policy^{\follower}; \data^{\follower})
    \subset \Policy^{\follower}
    \eqfinp
    \label{eq:follower_best_response}
\end{equation}

\subsubsection*{Nash equilibrium}
\begin{figure}[ht]
    \centering
    \begin{tikzpicture}
    
    \tikzset{
        block/.style = {rectangle, draw, text centered, minimum height=3em, minimum width=10em},
        arrow/.style = {thick,->,>=stealth},
        darrow/.style = {thick, dashed,->,>=stealth}
    }
    
    \node[block] (Electricity producer) {  
        \begin{tabular}{c}
            Leader
        \end{tabular}
    };
    \node[block, right=3.8cm of Electricity producer] (Consumer) {
        \begin{tabular}{c} 
            Follower 
        \end{tabular}
    };
    
    \draw[arrow] ($(Electricity producer.east) + (0,0.2cm)$) -- ($(Consumer.west) + (0,0.2cm)$) 
    node[midway, above] {%
        \begin{tabular}{c}
            Best response 
        \end{tabular}};
    \draw[arrow] ($(Consumer.west) + (0,-0.2cm)$) -- ($(Electricity producer.east) + (0,-0.2cm)$) 
    node[midway, below] {%
        \begin{tabular}{c}
            Best response 
        \end{tabular}};

    \end{tikzpicture}
    \caption{Graphical representation of a Nash equilibrium in a leader-follower game}
    \label{fig:Nash_equilibrium}
\end{figure}
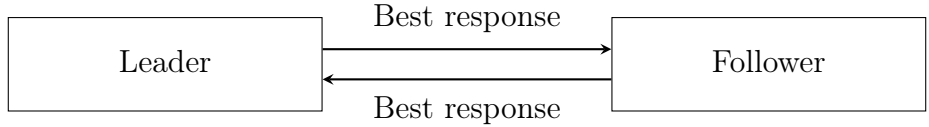

The reader can easily adapt equation \eqref{eq:follower_best_response} to define a Nash equilibrium, but this may not necessarily be the most relevant option, as it does not accurately reflect the information structure. 
Fig.~\ref{fig:Nash_equilibrium} allows for visualizing the concept of Nash equilibrium between the two players. 

\subsubsection*{Leader's Stackelberg strategy}

A natural approach for the leader is to anticipate that the follower will respond by choosing a best response. 
However, it is possible that several best responses are available to the follower.
The leader must therefore decide which strategy she believes the follower will choose from her set of best responses. 
The most common approach in bilevel optimization is to adopt an optimistic point of view. 
However, we also present the pessimistic approach, as well as other, more realistic formulations that fall between the optimistic and pessimistic approaches.

\subsubsubsection{Optimistic}

We say the leader plays an \emph{optimistic Stackelberg strategy} if she thinks the follower will choose the best option for her.
That is, the follower plays the best response maximizing the leader's normal form payoff. 
The set of optimistic Stackelberg strategies for the leader is denoted  
\begin{subequations}
    \begin{align}
    \Stackelberg{\Policy}^{\leader}(\data^{\leader}, \data^{\follower}) = \argmax_{\policy^{\leader} \in \Policy^{\leader}} \sup_{\policy^{\follower} \in \Nash{\Policy}^{\follower}(\policy^{\leader}; \data^{\follower})}\Criterion^{\leader}(\policy^{\leader}, \policy^{\follower}; \data^{\leader})
    \subset \Policy^{\leader}
    \eqfinp
    \end{align}

It is important to note that this set depends not only on the leader's data but also on the follower's data, as they are necessary to determine the follower's set of best responses.

\subsubsubsection{Pessimistic}

We now present the opposite situation when the leader thinks the follower will choose the worst option for her. 
Thus, we define the set of \emph{pessimistic Stackelberg strategies} as:
    \begin{align}
    \Stackelberg{\Policy}^{\leader}(\data^{\leader}, \data^{\follower}) = \argmax_{\policy^{\leader} \in \Policy^{\leader}} \inf_{\policy^{\follower} \in \Nash{\Policy}^{\follower}(\policy^{\leader}; \data^{\follower})}\Criterion^{\leader}(\policy^{\leader}, \policy^{\follower}; \data^{\leader})
    \subset \Policy^{\leader}
    \eqfinp
    \end{align}

\subsubsubsection{Intermediate formulations}
As mentioned as in \citet{Beck-Ljubic-Schmidt:2023}, the pessimistic and optimistic formulations are extreme situations.
We could want to model the level of cooperation between players by adding a probability of cooperation $\theta \in \ClosedIntervalClosed{0}{1}$ and consider a convex combination between the pessimistic and optimistic formulations.
It would give us the following:
\begin{align}
\begin{split}
    \Stackelberg{\Policy}^{\leader}(\data^{\leader}, \data^{\follower}) = \argmax_{\policy^{\leader} \in \Policy^{\leader}}
    &\biggl(
    \theta \sup_{\policy^{\follower} \in \Nash{\Policy}^{\follower}(\policy^{\leader}; \data^{\follower})}
    \Criterion^{\leader}(\policy^{\leader}, \policy^{\follower}; \data^{\leader})
    \\ &+ (1 - \theta) \inf_{\policy^{\follower} \in \Nash{\Policy}^{\follower}(\policy^{\leader}; \data^{\follower})}
    \Criterion^{\leader}(\policy^{\leader}, \policy^{\follower}; \data^{\leader})
    \biggr)
    \eqfinp
\end{split}
\end{align}

\subsubsubsection{Further formulation}
We could even imagine that the leader has a measure $\riskmeasure^{\leader}$ over a subset of random variables of the space $\barRR^{\Policy^{\follower}}$ and consider
\begin{align}
    \Stackelberg{\Policy}^{\leader}(\data^{\leader}, \data^{\follower}) = 
    \argmax_{\policy^{\leader} \in \Policy^{\leader}}
    \riskmeasure^{\leader}[\Criterion^{\leader}(\policy^{\leader}, \cdot \; ; \data^{\leader})] \eqfinp
\end{align}
\label{eq:Stackelberg_strategy}
\end{subequations}

\begin{illustration}
    The majority of bi-level optimization problems adopt an optimistic approach. The electricity supplier will thus consider that the consumer chooses the best response that is most favorable to the electricity supplier.
\end{illustration}

\subsubsection*{Nash-Stackelberg equilibrium}
Now, we can define a new type of equilibrium.
Indeed, we say that a strategy profile $(\underline{\policy}^{\leader}, \underline{\policy}^{\follower})$ is a \emph{Nash-Stackelberg equilibrium} if the follower plays a best response and the leader plays a Stackelberg strategy, that is:
\begin{equation}
    (\underline{\policy}^{\leader}, \underline{\policy}^{\follower}) \in \Stackelberg{\Policy^{\leader}}(\data^{\leader}, \data^{\follower}) \times \Nash{\Policy}^{\follower}(\underline{\policy}^{\leader}; \data^{\follower})
    \eqfinp
\end{equation}

\begin{figure}[ht]
    \centering
    \begin{tikzpicture}
    
    \tikzset{
        block/.style = {rectangle, draw, text centered, minimum height=3em, minimum width=10em},
        arrow/.style = {thick,->,>=stealth},
        darrow/.style = {thick, dashed,->,>=stealth}
    }
    
    \node[block] (Electricity producer) {
        \begin{tabular}{c}
            Leader
        \end{tabular}
    };
    \node[block, below=0.8cm of Electricity producer, xshift = 2cm] (Consumer) {
        \begin{tabular}{c} 
            Follower
        \end{tabular}
    };
    
    \draw[arrow] ($(Electricity producer.south west) + (3.15,0)$) -- ($(Consumer.north) + (-0.9,0)$)
    node[midway, right] {Best response};
    
    \draw[darrow] (Consumer.south) |- ++(0,-0.5) -| ($(Electricity producer.south west) + (1,0)$)
    node[pos=0.25, below] {Stackelberg strategy};
        
    \end{tikzpicture}
    \caption{Graphical representation of a Nash-Stackelberg equilibrium}
    \label{fig:Nash-Stackelberg_equilibrium}
\end{figure}
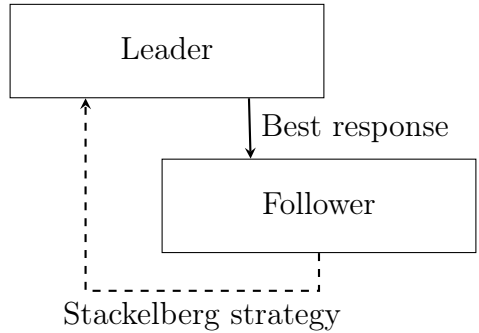

This definition holds regardless of the expression of the equation \eqref{eq:Stackelberg_strategy} chosen for the leader's Stackelberg strategy.
Fig.\ref{fig:Nash-Stackelberg_equilibrium} provides a graphical representation of the concept of Stackelberg equilibrium.

\subsubsection*{Link with bilevel optimization}

A Nash-Stackelberg equilibrium is the game theory concept that reflects the structure of bi-level problems.
Indeed, the leader's problem writes
\begin{subequations}
    \begin{align}
    & \underset{\policy^{\leader} \in \Policy^{\leader}}{\text{max}} \quad 
    \underset{\policy^{\follower} \in \Nash{\Policy}^{\follower}(\policy^{\leader} ; \data^{\follower})}{\text{sup}} \quad
    \Criterion^{\leader}(\policy^{\leader}, \policy^{\follower} ; \data^{\leader}) \tag{UL} \label{eq:UL} \\
    & \text{where} \quad
    \Nash{\Policy}^{\follower}( \policy^{\leader} ; \data^{\follower}) = \argmin_{\policy^{\follower} \in \Policy^{\follower}} \Criterion^{\follower}(\policy^{\leader}, \policy^{\follower} ; \data^{\follower})  \tag{LL} \label{eq:LL}
    \end{align}
\end{subequations}

The upper-level problem (UL) corresponds to the leader's problem (maximization) and the lower-level problem (LL) corresponds the follower's problem (minimization).
Thanks to the objects we introduced, it shows the ambiguous knowledge of the follower's data $\data^{\follower}$ which are necessary for the computation of the leader's strategy.


\subsection{Multi-leader-multi-follower}
Writing a Nash-Stackelberg equilibrium does not pose any conceptual difficulties but it requires new notations to define the notion of the best response for a group of players.

\subsubsection*{Specific notation}
In a multi-leader-multi-follower context, the notation $-\leader$ (resp. $-\follower$) does not denote the set of all players except leader $\leader$ (resp. follower $\follower$) but rather the set of leaders $\Leader \setminus \leader$ (resp. the set of followers $\Follower \setminus \follower$).

\subsubsection*{Followers' best response}
The set of best response for the group of followers to a strategy of the leaders $\underline{\policy}^{\Leader} \in \Policy^{\Leader}$ is defined by

\begin{equation}
    \Nash{\Policy}^{\Follower}( \underline{\policy}^{\Leader}; \data^{\Follower}) = 
    \Bset{\underline{\policy}^{\Follower} \in \Policy^{\Follower}}
    {\forall \follower \in \Follower \eqsepv
    \underline{\policy}^{\follower} \in \argmin_{\policy^{\follower} \in \Policy^{\follower}}
    \Criterion^{\follower}(\policy^{\follower}, \underline{\policy}^{-\follower}, \underline{\policy}^{\Leader}; \data^{\follower})}
    \eqfinp
\end{equation}

\subsubsection*{Leaders' Stackelberg strategy}

\subsubsubsection{Optimistic}

Now that we have extended the concept of best response to the group of followers, we can define the optimistic response of the leaders.
Thus, we define the set of optimistic best response of the leaders by

\begin{equation}
    \Stackelberg{\Policy^{\Leader}}(\data^{\Leader}, \data^{\Follower}) =
    \Bset{\underline{\policy}^{\Leader} \in \Policy^{\Leader}}
    {\forall \leader \in \Leader \eqsepv
    \underline{\policy}^{\leader} \in \argmax_{\policy^{\leader} \in \Policy^{\leader}}
    \sup_{\policy^{\Follower} \in \Nash{\Policy}^{\Follower}(\policy^{\leader}, \underline{\policy}^{-\leader}; \data^{\Follower})}
    \Criterion^{\leader}(\policy^{\leader}, \underline{\policy}^{-\leader}, \policy^{\Follower}; \data^{\leader})}
    \eqfinp
\end{equation}
\label{eq:MLMF_Stackelberg_strategy}

\subsubsubsection{Other formulations}
We also could have written equation \eqref{eq:MLMF_Stackelberg_strategy} for the other formulations by using equation \eqref{eq:Stackelberg_strategy} as a reference.


\section*{Conclusion}

After all these theoretical considerations first on the W-model and W-games, then on leader-follower problems within this framework, we will conduct a literature review of energy management issues that we will express in the form of W-games.


\part[Reformulation of energy management problems as W-games]{Reformulation \\ of energy management problems \\ as W-games}
\label{reformulation_energy_management_problems_W-games}


\chapter{Single-leader-single-follower problems}
\label{Single-leader-single-follower_problems}

In Chap.~\ref{Single-leader-single-follower_problems}, we examine the literature on single-leader-single-follower problems.
We begin by reformulating in-depth two different articles as W-games.
Then we propose a short overview of the literature on this topic 
Finally, we present the various conclusions that can be drawn from these analyses.

For each paper, we provide a general overview, then we detail Nature and the actions of the players before addressing the information structures to establish the W-model of this problem. 
Then, successively, we will associate an objective function and a belief to each player in order to have a W-game.


\section{Supplier-consumer electricity pricing \citep{Alekseeva-Brotcorne-Lepaul-Montmeat:2019}}

An electricity supplier operates various technologies to produce electricity and has contracts with a group of consumers.
They offer two time-based rates: peak hours and off-peak hours, with different prices for each period. 
The goal is to optimize hourly electricity prices while considering consumer behavior and minimizing energy generation costs.
To achieve this, the supplier proposes a new contract that encourages consumers to shift their electricity usage from peak to off-peak hours. 
This contract includes a monetary bonus for each kilowatt-hour (kWh) shifted and sets a consumption limit to prevent overconsumption.
By adopting this new pricing strategy, consumers can reduce their electricity bills, and overall energy generation costs may also decrease.
The overall model assumes that customers will shift their consumption without increasing or decreasing their total usage. 


\subsection{W-model}

\subsubsection*{Agents, actions and Nature}
To begin with, the leader is the energy supplier because he fixes the new contract's prices and the follower is the consumers because he reacts to this decision.
We take a macroscopic view, meaning we do not focus on individual decisions but on the aggregate behaviour of all consumers.
In Table \ref{tab:nature_actions_Alekseeva}, the reader can find the concepts related to the W-model on the left side and their corresponding terms from the article on the right side.
To attempt to have a compact model, all the parameters that are exogenous and known are grouped under the term \emph{common knowledge}, considering them, in a way, as the data of the problem.
We do not introduce any exogenous nature since there are no exogenous parameters that will asymmetrically influence an agent's decision.

\begin{table}[h!]
\centering
\begin{tabular}{|l|l||c|c|}
\hline
\multicolumn{2}{|c||}{\textbf{W-model}} & \multicolumn{2}{c|}{\textbf{Supplier-consumer electricity pricing}} \\ \hline \hline
\multirow{6}{*}{Common knowledge} & \multirow{6}{*}{} & Time span & $\mathbb{H}$ \\ \cline{3-4}
                                  &                   & Consumers' segments & $\mathbb{S}$ \\ \cline{3-4}
                                  &                   & Old pricing policy & $P_h$ \\ \cline{3-4}
                                  &                   & Bonus to shift to the new contract & $B$ \\ \cline{3-4}
                                  &                   & Consumption limit & $\overline{C}_{sh}$  \\ \cline{3-4}
                                  &                   & Demand & $D_{sh}$ \\ \hline \hline
\multirow{3}{*}{Leader's type} & \multirow{3}{*}{$\nature^{\leader}$} & Technologies & $\mathbb{T}$ \\ \cline{3-4} 
                             &                                      & Technologies' production cost & $F_t$\\ \cline{3-4}
                             &                                      & Technologies' power capacity & $C_t$\\ \cline{3-4} \hline
\multirow{2}{*}{Leader's decision} & \multirow{2}{*}{$\control_{\leader}$} & New pricing policy & $p_h$ \\ \cline{3-4}
                                 &                                       & Technologies' management & $a_{th}, z_{th}$ \\ \hline \hline
\multirow{1}{*}{Follower's type} & \multirow{1}{*}{$\nature^{\follower}$}   & Unwillingness to shift consumption & $W$ \\
                                 \hline
\multirow{2}{*}{Follower's decision} & \multirow{2}{*}{$\control_{\follower}$} & Contract shift & $r_s$ \\ \cline{3-4}
                                   &                                         & Consumption shift & $y_{sh}, q$ \\ \hline
\end{tabular}
\caption{Detailed actions, Nature and common knowledge \citep{Alekseeva-Brotcorne-Lepaul-Montmeat:2019}}
\label{tab:nature_actions_Alekseeva}
\end{table}

\subsubsection*{Configuration space, information fields and solution map}
Consumers adapt their consumption regarding the supplier's prices, hence a leader-follower situation.
Then, regarding knowledge of Nature, each player is assumed to know their type.
Therefore, the information fields write:
\begin{subequations}
\begin{align}
    \tribu{\Information}_{\leader} &= \trib^{\leader}  \eqsepv \\
    \tribu{\Information}_{\follower} &= \trib^{\follower} \otimes \tribu{\Control}_{\leader} \eqfinp
\end{align}
\label{info_structure_Alekseeva}
\end{subequations}

Finally, the whole W-model is summed up in Table \ref{tab:WIM_Alekseeva}.

\begin{table}[h!]
    \centering
    \footnotesize
    \begin{tabular}{|l|l||l|l|}
    \hline
    \multicolumn{2}{|c||}{\textbf{W-model}} & \multicolumn{2}{c|}{\textbf{Supplier-consumer electricity pricing }} \\ \hline \hline
    Leader & $\leader$ & Energy supplier &  \\ \hline
    - type & $\Nature^{\leader}$ & Technologies & $\RR^{\mathbb{T}} \times \RR^{\mathbb{T}} = \RR^{2\mathbb{T}}$ \\ \hline
    - decision set & $\CONTROL_{\leader}$ & Pricing and producing & $\RR^{\mathbb{H}} \times \RR^{\mathbb{T} \times \mathbb{H}} \times \RR^{\mathbb{T} \times \mathbb{H}} = \RR^{\mathbb{H}(1 + 2\mathbb{T})}$ \\ \hline
    - information field & $\tribu{\Information}_{\leader}$ & Own type & $\borel{\RR^{2\mathbb{T}}}$ \\ \hline \hline
    Follower & $\follower$ & Consumers &  \\ \hline
    - type & $\Nature^{\follower}$ & Unwillingness to change & $\RR$ \\ \hline
    - decision set & $\CONTROL_{\follower}$ & Consumption shifts & $\RR^{\mathbb{S}} \times \RR^{\mathbb{S} \times \mathbb{H}} \times \RR = \RR^{\mathbb{S}(1 + \mathbb{H}) + 1}$\\ \hline
    - information field & $\tribu{\Information}_{\follower}$ & Own type, leader's decision & $\borel{\RR} \otimes \borel{\RR^{\mathbb{H}(1 + 2\mathbb{T})}}$ \\ \hline \hline
    Configuration space & $\HISTORY$ &  & $\RR^{2\mathbb{T}} \times \RR \times \RR^{\mathbb{H}(1 + 2\mathbb{T})} \times \RR^{\mathbb{S}(1 + \mathbb{H}) + 1}$\\  \hline
    \end{tabular}
    \caption{Detailed W-model \citep{Alekseeva-Brotcorne-Lepaul-Montmeat:2019}}
    \label{tab:WIM_Alekseeva}
\end{table}

The solution map derives from equation \eqref{info_structure_Alekseeva}:
\begin{subequations}
\begin{align}
    \SolutionMap_{\policy_{\leader}, \policy_{\follower}} : 
     & \;\Nature^{\leader} \times \Nature^{\follower} \to \Nature^{\leader} \times \Nature^{\follower} \times \CONTROL_{\leader} \times \CONTROL_{\follower}  \eqsepv \\
    & \; (\nature^{\leader}, \nature^{\follower}) \mapsto \left(\nature^{\leader}, \nature^{\follower}, \policy_{\leader}(\nature^{\leader}), \policy_{\follower}(\nature^{\follower}, \policy_{\leader}(\nature^{\leader}) ) \right) \eqfinp
\end{align}
\end{subequations}

After classifying the variables from the article into a W-model and displaying the information structures, the objective functions and beliefs are defined to create a W-game.


\subsection{W-game}

We remind that we are only discussing two players here.
Indeed, the article does not consider that each segment of consumers will minimize its objective function.
Instead, it focuses on minimizing the sum of the costs of each consumer.
In the modeling, if there is any doubt about the number of players to consider, simply look at the number of optimization problems presented in the article.

\subsubsection{Objective functions}

For both the leader and the follower, the objective function considers consumption, whether at the old or new price, as well as the bonus given to consumers who have switched contracts.
For the leader, electricity production, its costs, and the technological constraints on production must also be taken into account.
To represent the constraints on production, we introduce without specifying it a set $\Control_{\leader}(\nature^{\leader}) \subset \CONTROL_{\leader}$.
Hence, the leader's payoff writes:
\begin{subequations}
    \begin{align}
        \criterion^{\leader}(\nature^{\leader},
        \cancel{\nature}^{\follower}, \control_{\leader}, \control_{\follower}) &= 
        \text{bilinear}(\control_{\leader}, \control_{\follower}) : \text{consumption payoff and bonus cost} \eqsepv \\ 
        &- \text{bilinear}(\nature^{\leader}, \control_{\leader}) : \text{production cost} \eqsepv \\
        &- \Indicator{\Control_{\leader}(\nature^{\leader})}(\control_{\leader}) : \text{constraints on use of technologies} \eqfinp
    \end{align}
\end{subequations}

For the follower, the reluctance to shift to the new contract is modeled through the unwillingness cost.
And we also introduce a set of constraints  $\Control_{\follower} \subset \CONTROL_{\follower}$ to reflect the constraints on consumption (fulfilling demand and not exceeding consumption limit).
Consequently, the follower's cost can be expressed as:
\begin{subequations}
\begin{align}
    \criterion^{\follower}(\cancel{\nature}^{\leader}, \nature^{\follower}, \control_{\leader}, \control_{\follower}) &= 
    \text{bilinear}(\control_{\leader}, \control_{\follower}) : \text{consumption cost and bonus payoff} \eqsepv \\
    &+ \text{bilinear}(\nature^{\follower}, \control_{\follower}) : \text{unwillingness cost} \eqsepv \\
    &+ \Indicator{\Control_{\follower}}(\control_{\follower}) : \text{constraints on consumption}
    \eqfinp
\end{align}
\end{subequations}

\subsubsection*{Beliefs}
The supplier's belief on Nature looks like this:
\begin{subequations}
\begin{align}
    \Nature &= \Nature^{\leader} \times \Nature^{\follower} \eqsepv\\
    \Belief^{\leader} &=  \delta_{\overline{\nature}^{\leader}} \otimes \Belief^{\leader}_{\follower} \eqfinp
\end{align}
\end{subequations}

However, the article makes an implicit assumption that the electricity supplier is aware of the consumers' unwillingness.
This can be expressed by writing
\begin{equation}
    \Belief^{\leader}_{\follower} = \delta_{\overline{\nature}^{\follower}} \eqfinp
\end{equation}

As for the follower, her belief can be expressed as:
\begin{subequations}
\begin{align}
    \Nature &= \Nature^{\leader} \times \Nature^{\follower} \eqsepv\\
    \Belief^{\follower} &=  \Belief^{\follower}_{\leader} \otimes \delta_{\overline{\nature}^{\follower}} \eqfinp
\end{align}
\end{subequations}

We then notice that the question of the follower's knowledge of the leader is not specified anywhere. 
We thus have, in a sense:
\begin{equation}
    \Belief^{\follower}_{\leader} = \; ? \eqfinp
\end{equation}

\subsubsection*{Normal form payoffs}

The normal form payoff for the follower can therefore be expressed as:
\begin{equation}
    \Criterion^{\follower}(\policy^{\follower}, \policy^{\leader}; \data^{\follower}) =
    \int_{\Nature^{\leader}} \criterion^{\follower} \left (\overline{\nature}^{\follower}, \policy^{\leader}(\nature^{\leader}), \policy^{\follower}(\overline{\nature}^{\follower}, \policy^{\leader}(\nature^{\leader}) \right)
    \dd \Belief^{\follower}_{\leader}(\nature^{\leader})
    \eqfinp
\end{equation}

And the normal form payoff of the leader writes
\begin{equation}
    \Criterion^{\leader}(\policy^{\leader}, \policy^{\follower}; \data^{\leader}) =
    \criterion^{\leader} \left (\overline{\nature}^{\follower}, \policy^{\leader}(\overline{\nature}^{\leader}), \policy^{\follower}(\overline{\nature}^{\follower}, \policy^{\leader}(\overline{\nature}^{\leader}) \right)
    \eqfinp
\end{equation}

We write it for the first article we reformulate but it has the same structure in the other problems. 
Moreover, artcles always consider players play a Nash-Stackelberg equilibrium.



\section[Stochastic SGO-clients energy pricing \citep{Brotcorne-Lepaul-VonNiederhausern:2021}]{Grid Operator-clients energy pricing with scenario tree \citep{Brotcorne-Lepaul-VonNiederhausern:2021}}

An energy supplier buys energy on the spot market and sells it to a smart grid operator (SGO).
The SGO manages its power grid to meet the energy demands of a set of clients, each with specific devices that require a certain amount of energy within a given time window.
The SGO has to decide how to source energy for each device, choosing between purchasing from the supplier, a competitor or using distributed generation (like solar panels).
The SGO also manages a battery, deciding how much energy to store from different sources.
The energy supplier and competitor can provide unlimited energy, but distributed generation has a maximum production limit. 
To model this, we introduce weather scenarios and thus it becomes a dynamic problem. 
Additionally, there is a cost associated with delaying the use of a device, which the SGO must consider when making decisions.


\subsection{W-model}
We discussed situations where the information available to the players does not change over time, what was referred to as "static" framework.
We will therefore address situations where the available information evolves over a time period $\exo{\TIME} = \TIME^{\follower} = \{ 1, 2, \ldots T \} $.

\subsubsection{Agents, actions and Nature}
We still consider two players: a leader $\leader$ and a follower $\follower$.
Nevertheless, in order to represent the fact that they are taking several decisions with different levels of information, we need to introduce several agents for each player.

\begin{table}[h]
\centering
\small
\begin{tabular}{|l|l||c|c|}
\hline
\multicolumn{2}{|c||}{\textbf{WIM}} & \multicolumn{2}{c|}{\textbf{Stochastic time-of-use pricing}} \\ \hline \hline
\multirow{2}{*}{Common knowledge} & \multirow{2}{*}{} & Demand's components & $A_n, E_{(n,a)}, T_{(n,a)}, \beta^{\text{max}}_{(n,a)}$ \\ \cline{3-4}
                                  &                   & Competitor's price & $\overline{p}^h$ \\ \hline \hline
\multirow{1}{*}{Exogenous Nature} & \multirow{1}{*}{$\exo{\om}_{\mytime}$} & Weather scenario & $\sigma$ \\ \hline \hline
\multirow{1}{*}{Leader's type} & \multirow{1}{*}{$\nature^{\leader}$} & Spot market's price & $K^h$\\ \cline{3-4} \hline
\multirow{1}{*}{Leader's decision} & \multirow{1}{*}{$\control_{\leader}$} & Selling price & $p^h$ \\ \hline \hline
\multirow{3}{*}{Follower's type} & \multirow{3}{*}{$\nature^{\follower}$} & Battery's characteristics & $S^{\text{start}}, S^{\text{min}}, S^{\text{max}}, \rho^d, \rho^c $ \\ \cline{3-4}
                                  &                   & Inconvenience coefficient & $C^h_{(n,a)}$ \\ \cline{3-4} 
                                  &                   & DG capacity & $\lambda_{\text{max}}$ \\ \hline

\multirow{2}{*}{Follower's decision} & \multirow{2}{*}{$\control_{(\follower, \mytime)}$} & Energy consumption & $x_{(n,a)}, \overline{x}_{(n,a)}, \lambda_{(n,a)}, s_{(n,a)} $ \\ \cline{3-4}
                                   &                                         & Energy storage & $x_s, \overline{x}_s, \lambda_s, S$ \\ \hline
\end{tabular}
\caption{Detailed actions and Nature \citep{Brotcorne-Lepaul-VonNiederhausern:2021}}
\label{tab:nature_actions_Brotcorne}
\end{table}

\subsubsection{Information fields and solution map}
The information structure is such that the leader fixes the prices at the beginning of the time horizon and then the follower acts according to the weather scenario he sees.
It can be expressed as:
\begin{subequations}
    \begin{align}
    \tribu{\Information}_{\leader} &= \trib^{\leader} \eqsepv \\
    \forall \mytime \in \TIME \eqsepv \tribu{\Information}_{(\follower, \mytime)} &= \exo{\trib}_{\mytime} \otimes \trib^{\follower} \otimes \tribu{\Control}_{\leader} \eqfinp
    \end{align}
\end{subequations}

The solution map writes
\begin{equation}
    \SolutionMap_{\policy^{\leader}, \policy^{\follower}}(\exo{\nature}, \nature^{\leader}, \nature^{\follower}) = \Big(\exo{\nature}, \nature^{\leader}, \nature^{\follower}, \policy^{\leader}\big(\nature^{\leader}\big), \policy^{\follower}\big(\exo{\nature}, \nature^{\follower}, \policy^{\leader}(\nature^{\leader}) \big) \Big) \eqfinp
\end{equation}


\subsection{W-game}

\subsubsection*{Objective functions}
The leader's payoff writes 
\begin{subequations}
    \begin{align}
    &\criterion^{\leader}(\nature^{\leader}, \control_{\leader}, (\control_{(\follower, \mytime)})_{\mytime \in \TIME}) \\ &= 
    \text{bilinear}(\control_{\leader}, (\control_{(\follower, \mytime)})_{\mytime \in \TIME}): \text{sales} \eqsepv \\
    &- \text{bilinear}(\nature^{\leader}, (\control_{(\follower, \mytime)})_{\mytime \in \TIME}): \text{production cost}\eqsepv 
    \end{align}
and the follower's cost is
    \begin{align}
    &\criterion^{\follower}(\nature^{\follower}, \control_{\leader}, (\control_{(\follower, \mytime)})_{\mytime \in \TIME}) \\ &= 
    \text{bilinear}(\control_{\leader}, (\control_{(\follower, \mytime)})_{\mytime \in \TIME}): \text{purchases
    (supplier and competitor)
    } \eqsepv \\ 
    &+ \text{bilinear}(\nature^{\follower}, (\control_{(\follower, \mytime)})_{\mytime \in \TIME}): \text{inconvenience cost} \eqsepv \\
    &+ \Indicator{\Control_{\follower}(\exo{\nature}, \nature^{\follower})}((\control_{(\follower, \mytime)})_{\mytime \in \TIME}): \text{battery and DG
    (weather dependent)
    constraints}\eqfinp
    \end{align}
\end{subequations}

\subsubsection*{Beliefs}
The beliefs write as usual:
\begin{subequations}
\begin{align}
    \Belief^{\leader} &= \exog{\Belief}^{\leader} \otimes \delta_{\overline{\nature}^{\leader}} \otimes \delta_{\overline{\nature}^{\follower}} \eqsepv \\
    \Belief^{\follower} &= \exog{\Belief}^{\follower} \otimes \Belief^{\follower}_{\leader} \otimes \delta_{\overline{\nature}^{\follower}}  \eqfinp
\end{align}
\end{subequations}

We emphasize the following implicit assumption: both players are assumed to share a probability distribution over the scenarios.
According to the notations of the article, it writes
\begin{equation}
    \exog{\Belief}^{\leader} = \exog{\Belief}^{\follower} = P \eqfinp
\end{equation}



\section{Other problems}

\subsubsection*{
\citep{Beraldi:2024}}

In the model by \citet{Beraldi:2024}, multiple leaders, represented by energy suppliers, strategically set prices, while multiple followers, composed of regional distributors, adjust their purchasing decisions to optimize supply costs for end consumers.

\subsubsection*{\citep{Beraldi-Khodaparasti:2022}}

This article has the particularity to deal with the notion of aggregator.
An aggregator controlling tariffs and a battery, with prosumers managing their resources and scheduling flexible loads.
A distinction is made between shiftable and interruptible loads.

\subsubsection*{\citep{Grimm-Orlinskaya-Schewe-Schmidt-Zottl:2021}}

The article models an electricity retailer offering time-of-use pricing contracts to prosumers, who own PV systems, batteries, heat pumps, and heat storage units. 
The model assumes perfect foresight, with both players knowing market prices, PV resale prices, and future demand.
It is the same framework than \cite{Alekseeva-Brotcorne-Lepaul-Montmeat:2019} but it takes into account a greater number of variables.

\subsubsection*{
\citep{Besancon-Anjos-Brotcorne-Gomez-Herrera:2020}}
This article examines a supplier’s challenge in determining pricing components while anticipating consumer booking capacities. 
Consumption is unknown to both players, who minimize their objectives based on expectations. 
The model also accounts for the follower’s reluctance to achieve optimality by including constraints in the objective function (though the explanation is somewhat unclear).

\subsubsection*{
\citep{Zugno-Morales-Pinson-Madsen:2013}}
This article models the interaction between an energy retailer and consumers, with randomness introduced in prices, weather-related variables, and must-serve loads.
These are divided into two stages.
Market competition is not explicitly modeled.
The leader purchases energy on the spot market and sets prices for end-consumers, who then make purchasing decisions. 
No risk measure is applied.


\section*{Conclusion}
After examining several single-leader-single-follower problems, we highlighted different hidden assumptions on beliefs, information fields and game data.
We can now move to more complex problems where there are multiple leader and multiple followers.


\chapter{Multi-leader-multi-follower problems}
\label{Multi-leader-multi-follower_problems}

This chapter mirrors the structure of Chap.~\ref{Single-leader-single-follower_problems} by first analyzing articles with multiple leaders and multiple followers, then offering a brief overview of the multi-leader-multi-follower literature before drawing conclusions on the modeling that can be established with W-games. 


\section{Retailer-prosumer groups pricing \citep{Kovacs:2019}}

In the model presented by \citet{Kovacs:2019}, a single leader, the electricity retailer, interacts with multiple followers, who are prosumers (producers-consumers) adjusting their electricity consumption in response to feed-in and purchased tariffs in the smart grid determined by the retailer.
This article does not introduce stochastic considerations or information revelation over time.


\subsection{W-model}

\subsubsection{Agents, actions and Nature}
The leader is an electricity retailer that faces a group of followers. 
Each follower represents a group of prosumers, which can be either a household, an office.
There is no exogenous Nature.
Common knowledge is not detailed.

\begin{table}[ht]
    \centering
    \begin{tabular}{|c|l|}
    \hline
    \multirow{1}{*}{$\nature^{\leader}$} & Wholesale market prices  \\ 
    \hline 
    \multirow{2}{*}{$\control_{\leader}$} & Electricity purchased/fed-in on the wholesale market  \\ \cline{2-2}
    & Purchase/feed-in prices for prosumers \\ \cline{2-2} 
    \hline \hline
    \multirow{3}{*}{$\nature^{\follower}$} & Battery's characteristics \\ \cline{2-2}
    &  Uncontrollable production/consumption \\ \cline{2-2}
    &  Controllable load requirements \\ \cline{2-2}
    &  Utility of scheduling a controllable load \\ 
    \hline 
    \multirow{3}{*}{$\control_{\follower}$} & Electricity purchased/fed-in with the prosumers \\ \cline{2-2}
    & Controllable load \\ \cline{2-2}
    & Battery management \\ 
    \hline 
    \end{tabular}
    \caption{Details of the decisions' sets and of the Nature}
    \label{tab:details_retailer_futures_market_trading}
\end{table}

\subsubsection{Information fields and solution map}
If each leader knows her type and each follower knows her type, there is no reason for each follower to know the type of the other followers.
Therefore we write
\begin{subequations}
\begin{align}
    \tribu{\Information}_{\leader} &= \trib^{\leader} \eqsepv \\
    \forall \follower \in \Follower \eqsepv \tribu{\Information}_{\follower} &= \trib^{\follower} \otimes \tribu{\Control}_{\leader} \eqfinp
\end{align}
\end{subequations}

If we write the solution map concisely, we get
\begin{equation}
    \SolutionMap_{\policy^{\leader}, \policy^{\Follower}}(\nature^{\Leader}, \nature^{\Follower}) = \Big(\nature^{\leader}, (\nature^{\follower})_{\follower \in \Follower}, \policy^{\leader}\big( \nature^{\leader} \big), \big(\policy^{\follower}(\nature^{\follower}, \policy^{\leader}(\nature^{\leader})) \big)_{\follower \in \Follower} \Big) \eqfinp
\end{equation}


\subsection{W-game}

We emphasize that the followers are indeed players minimizing their own cost function and not agents of the same player who would minimize some form of social welfare.

\subsubsection{Objective functions}
The leader's payoff writes 
\begin{subequations}
    \begin{align}
    &\criterion^{\leader}(\nature^{\leader}, \control_{\leader}, (\control_{\follower})_{\follower \in \Follower}) \\&= 
    \text{bilinear}(\control_{\leader}, (\control_{\follower})_{\follower \in \Follower})): \text{exchanges with the prosumers} \eqsepv \\
    &+ \text{bilinear}(\nature^{\leader}, \control_{\leader}): \text{exchanges with the wholesale market} \eqsepv \\
    &- \Indicator{\Control_{\leader}(\nature^{\leader}, (\control_{\follower})_{\follower \in \Follower}))}(\control_{\leader}): \text{price regulation and supply constraints} \eqsepv 
    \end{align}
and the follower's cost is
    \begin{align}
    \forall \follower \in \Follower \eqsepv &\criterion^{\follower}(\nature^{\follower}, \control_{\leader}, (\control_{\follower})_{\follower \in \Follower})) \\&= 
    \text{bilinear}(\control_{\leader}, (\control_{\follower})_{\follower \in \Follower}): \text{exchanges with the retailer} \eqsepv \\ 
    &- \text{bilinear}(\nature^{\follower}, (\control_{\follower})_{\follower \in \Follower}): \text{utility payoff} \eqsepv \\
    &+ \Indicator{\Control^{\follower}(\nature^{\follower})}((\control_{\follower})_{\follower \in \Follower}): \text{battery's and uncontrollable requirements} \eqfinp
    \end{align}
\end{subequations}

\subsubsection{Beliefs}
In this article, it is explicitly written that "the retailer is aware of the decision model and the parameters of PGs", leading to the following equations:
\begin{subequations}
\begin{align}
    \Nature &= \Nature^{\leader} \times \prod_{\follower \in \Follower} \Nature^{\follower} \eqsepv \\
    \Belief^{\leader} &= \delta_{\overline{\nature}^{\leader}} \otimes \delta_{\overline{\nature}^{\Follower}} \eqsepv \\
    \forall \follower \in \Follower \eqsepv \Belief^{\follower} &= \Belief^{\follower}_{\leader} \otimes \left( \delta_{\overline{\nature}^{\follower}} \otimes \Belief^{\follower}_{-\follower} \right) \eqfinp
\end{align}
\end{subequations}

The belief $\Belief^{\follower}_{-\follower}$ is the probability distribution of the follower about the types of the other followers.


\section[Retailer futures market trading with scenarios \citep{Carrion-Arroyo-Conejo:2009}]{Retailer futures market trading with scenarios \\ \citep{Carrion-Arroyo-Conejo:2009}}

A power retailer faces a medium-term decision-making problem.
Firstly, he determines if he signs forward contracts, which are ways of buying electricity on the futures market, and proposes simultaneously selling prices to his potential clients.
The model accounts for competitors and model their participation as random prices which realize after the principal has chosen his price.
Then, the consumers, without knowing their demand decide the proportion of energy to buy to the retailer and to his competitors. 
Only now, the demand of the consumers realizes and the retailer completes his need in energy by buying in real time on the pool market.


\subsection{W-model}

\subsubsection*{Agents, actions and Nature}
The leader is the retailer. We will denote time 0 as the moment when the retailer makes its first decision, which we denote as $\control^{\Leader}_0$.
This decision involves signing the futures contracts and determining the sale prices for consumers.
Then, at time 1, the retailer makes the decision $\control^{\Leader}_1$ and decides how much to purchase from the pool market and for consumers.
Formally, if we want to revisit the notations used in the section, we have $\exo{\TIME} = \TIME^{\follower} = \{ 0, 1 \}$.

\begin{table}[h]
    \centering
    \begin{tabular}{|c|l|}
    \hline
    \multirow{1}{*}{$\nature^e_0$} & Competitors' prices \\ 
    \hline 
    \multirow{2}{*}{$\nature^e_1$} & Pool market prices \\ \cline{2-2}
    &  Energy demand \\ 
    \hline \hline
    \multirow{2}{*}{$\nature^{\leader}$} & Characteristics of forward contracts  \\ \cline{2-2}
    & Prior forward contracts \\ 
    \hline 
    \multirow{2}{*}{$\control_{(\leader, 0)}$} & Quantity of energy bought with forward contracts \\ \cline{2-2}
    & Selling price to the consumers \\ \hline
    \multirow{2}{*}{$\control_{(\leader, 1)}$} & Quantity of energy sold to the consumers \\ \cline{2-2}
    &  Quantity of energy bought on the pool market \\ 
    \hline \hline
    \multirow{2}{*}{$\nature^{\follower}$} & Initial proportion of energy supplied by 
    each retailer \\ \cline{2-2}
    &  Unwillingness to switch retailers \\ 
    \hline 
    \multirow{1}{*}{$\control_{\follower}$} & Proportion of energy supplied by each retailer \\ 
    \hline 
    \end{tabular}
    \caption{Details of the decisions' sets and of the nature}
    \label{tab:details_retailer_futures_market_trading}
\end{table}

\subsubsection*{Information fields, strategies and solution map}

When taking its first decision, the retailer does not know any part of uncertainty
Reading Table \ref{tab:details_retailer_futures_market_trading} from top to bottom gives us the information structure.
The game is causal so we can decompose the solution map:
\begin{subequations}
\begin{align}
     \tribu{\Information}_{(\leader, 0)} &= \trib^{\Leader} \eqsepv \\
     \forall \follower \in \Follower \eqsepv \tribu{\Information}_{\follower} &= \exo{\trib}_0 \otimes \trib^{\follower} \otimes \exo{\trib}_1  \eqfinp
\end{align}
\end{subequations}


\subsection{W-game}

\subsubsection*{Objective functions}
The follower's payoff can be written under the following form. 
\begin{equation}
\criterion^{\follower}(\exo{\nature},\nature^{\follower}, \control_{(\leader, 0)}, \control_{\follower}) = - \underbrace{\alpha(
\overbrace{\control_{(\leader, 0)},\exo{\nature}_0}^{\text{prices}},
\overbrace{\nature^{\follower}}^{\substack{\text{unwillingness} \\ \text{to switch}}},
\overbrace{\control_{\follower}}^{\substack{\text{supply} \\ \text{distribution}}},
\overbrace{\exo{\nature}_1}^{\text{demand}}}_{\text{energy costs}}
)
- \underbrace{\Indicator{\Control_{\follower}(\nature^{\follower})}(\control^{\follower}}_{\text{constraints}})
\end{equation}

We decompose the leader's payoff according to his agents:
\begin{subequations}
\begin{align}
    \criterion^{\leader}_0(\exo{\nature}_1,\nature^{\leader}, \control_{(\leader, 0)}, \control_{\follower})  &= 
    \underbrace{\phi_0(\control_{(\leader, 0)}, \exo{\nature}_1, \control_\follower)}_{\text{sells' profits}} -
    \underbrace{\psi_0(\nature^{\leader}, \control_{(\leader, 0)}}_{\text{forward contracts costs}} -
    \underbrace{\Indicator{\Control_{(\leader, 0)}(\nature^{\leader})}(\control_{(\leader, 0)}}_{\text{constraints}})
    \\
    \criterion^{\leader}_1(\exo{\nature}_1,\nature^{\leader}, \control^{\Leader}_1) &= -
    \underbrace{\phi_1(\exo{\nature}_1, \control_{(\leader, 1)}}_{\text{pool's costs}}) -
    \underbrace{\Indicator{\Control_{(\leader, 1)}(\nature^{\leader})}(\control^{(\leader, 1)}}_{\text{constraints}}) 
    \\
    \criterion^{\leader}(\exo{\nature},\nature^{\leader}, \control_{\leader}, \control_{\follower}) &= 
    \criterion^{\leader}_0(\exo{\nature}_1,\nature^{\leader}, \control_{(\leader, 0)}, \control_{\follower}) +
     \criterion^{\leader}_1(\exo{\nature}_1,\nature^{\leader}, \control_{(\leader, 1)})
\end{align}
\end{subequations}

\subsubsection*{Beliefs}
Now, let's take a look at the follower's belief.
He knows the first stage of the exogenous nature.
And he is endowed with a probability distribution over the second stage of the nature.
It is implicit that he has a belief over the leader nature, even though it will not take part in his criterion.
And he knows his part of the nature.
\begin{subequations}
\begin{align}
    \Nature &= \exo{\Nature}_0 \times \exo{\Nature}_1  \times \Nature^{\leader} \times \Nature^{\follower} \eqsepv \\
    \Belief^{\follower} &= \delta_{\overline{\exo{\nature}_0}}  \otimes 
    \underbrace{{\exog{\Belief}^{\follower}}_{1}}_{\substack{\text{uncertainty in} \\ \text{demand}}} \otimes 
    \underbrace{\Belief^{\follower}_{\leader}}_{\substack{\text{not} \\ \text{precised}}}  \otimes \;
    \delta_{\overline{\nature}^{\follower}}
\end{align}
\end{subequations}

The leader's belief can also be decomposed.
He is endowed with a belief on the two stages of the exogenous nature. 
And he shares the same belief on the second stage of the exogenous nature with the follower.
Moreover, it is implicit but he has to know the follower nature to predict his reaction.
\begin{subequations}
\begin{align}
    \Nature &= \exo{\Nature}_0 \times \exo{\Nature}_1 \times \Nature^{\leader} \times \Nature^{\follower} \eqsepv \\
    \Belief^{\leader} &= \underbrace{\Belief^\leader_{e, 0}}_{\substack{\text{uncertainty in} \\ \text{competitors}}}  \otimes 
    \underbrace{\Belief^\leader_{e, 1}}_{\substack{\text{uncertainty in demand} \\ \text{and pool prices}}} \otimes \; 
    \delta_{\overline{\nature}^{\leader}} \otimes 
    \underbrace{\delta_{\overline{\nature}^{\follower}}}_{\substack{\text{implicitly} \\ \text{supposed known}}} \eqfinp
\end{align}
\end{subequations}






\section{Other problems}

\subsubsection*{\citep{Aussel-Brotcorne-Lepaul-VonNiderhausern:2020}}
This article is interesting because it considers interactions between three actors of the electricity market.
Trilevel problems can be modelled by writing two consecutive bilevel problems.

\subsubsection*{\citep{Fochesato-Cenedese-Lygeros:2022}}

The leader is the DSO and the followers are a set of prosumers. 
Each prosumer optimizes its own behavior in response to the DSO’s incentives, as the DSO aims to stabilize the grid through demand-side incentive management.

\subsubsection*{\citep{Oggioni-Schwartz-Wiertz-Zottl:2024}}

The multi-leader-single-follower framework is explored where multiple electricity retailers (multi-leaders) compete to set prices in a way that attracts local agents, the single follower. Each retailer adjusts prices to maximize profit while influencing the demand response of the local agents, who act as a collective entity aiming to meet their energy needs at the lowest cost.

\subsubsection*{\citep{Wogrin-Pineda-Tejada-Arango:2020}}

This article classifies various problem types encountered in the energy sector and proposes an investment problem.

\subsubsection*{\citep{Kallabis-Gabriel-Weber:2020}}

The focus of this article is on investment, splitting the problem into short-term/long-term variables aligned with the UL/LL problem, and incorporating scenario trees related to demand, with a long-term perspective.



\subsubsection*{\citep{Aussel-Lepaul-VonNiederhausern:2022}}
This article presents a multi-leader-multi-follower model in energy demand-side management.
In this setup, multiple leaders (energy suppliers) simultaneously interact with multiple followers (local agents or consumers). 
Each leader aims to set energy supply terms to maximize profit, while each follower independently optimizes their energy consumption and potential production to minimize costs or maximize utility, all while responding to the multiple leaders' strategies.


\section*{Conclusion}

In this chapter, we have explored how the W-model demonstrates flexibility and adaptability when applied to problems involving multiple leaders and multiple followers. 
Similar underlying principles were observed as in the single-leader single-follower scenario. 
In the following part, we aim to establish a W-game not by relying on an existing model from the literature but by constructing one ourselves, based on the real-world dynamics of demand response in Thailand.


\part[Design of W-games in energy management]{Design of W-games \\ in energy management}
\label{Design_W-games_energy_management}


\include{Thai.tex}



\clearpage           
\chapter*{Acknowledgments} 
\thispagestyle{plain}  
This research benefited from the support of the FMJH Program Gaspard Monge for optimiza-
tion and operations research and their interactions with data science.

\newpage

\thispagestyle{plain}




\bibliographystyle{plainnat}  
\bibliography{references_demand_response_modelling}

\end{document}

%% file: Thai.tex

\newcommand{\exogen}{\mathtt{e}}
\newcommand{\Sol}{S}
\newcommand{\price}{p}
\newcommand{\consumption}{x}
\newcommand{\reward}{r}
\newcommand{\target}{\tau}
\newcommand{\baseline}{B}
\newcommand{\leaderparam}{\nature^\leader}
\newcommand{\followerparam}{\nature^\follower}
\newcommand{\mylocaltime}{s}

\chapter{Demand Response Program in Thailand}

The authors thank Parin Chaipunya\footnote{Department of Mathematics, Faculty of Science, King Mongkut’s University of Technology Thonburi, Bangkok, parin.cha@kmutt.ac.th} for writing the first draft of this chapter.

We use all the framework developed in the previous parts to
formulate games in product form, of different levels of complexity, for the Demand Response Program of Thailand.
The chapter starts by describing, in Sect.~\ref{sec:Thai-DR}, the background and
state-of-the-art
of the Demand Response Program that has been implemented in Thailand.
Then, in Sect.~\ref{sec:W-model-Thai-DR}, we develop a series of games in product form with
increasing complexity.

\section{The Thai system of demand response}\label{sec:Thai-DR}

Thailand has been operating, on and off, demand response programs since mid 1990s. Starting from 2022, it has initialized a pilot Demand Response program (DR program in the subsequel) to request load reduction of industries and buildings during certain hours of the day. 
The main actors include:
\begin{itemize}[label=$\circ$, leftmargin=*]
\item
  The retailer, which is the Energy Generation Authority of Thailand
  (EGAT). It happens that EGAT is both the main electricity producer and
  retailer of Thailand. Except in some very restricted cases, EGAT is the only
  buyer, meaning that all the distributed productions are sold to EGAT, before being resaled.
\item
  The load aggregators (LA). There are two LAs, namely the Metropolitan
  Electricity Authority (MEA) which manages the area around Bangkok, and
  Provincial Electricity Authority (PEA) which manages the rest of the country. 
\item
  The consumers that participate in the DR program. It is currently impossible for household users to join the program. Hence the participants are companies or buildings.
\end{itemize}

The operation procedures of the DR program in Thailand can be summarized as follows.
\begin{itemize}[label=$\circ$, leftmargin=*]
\item
  At the beginning of the year, EGAT launches a program to the destination of
  companies and buildings.
  The program fixes the proposed incentive, which is a monetary reward per unit reduced. 
\item
  Participants register to the program and provide information regarding their reduction capacity.
\item
  Once the contracts have been made, every day EGAT makes a forecast of
  the next day demand. Depending whether the forecast exceeds a certain
  threshold or not,
  EGAT decides the amount of energy to be saved globally and communicates it to the Load Aggregators.
\item
  The LAs decides to call some participants and assign them a target reduction, depending on their capacity.
\item
  Whether a called participant complies or not with the assigned reduction, he gets a reward proportional to the actual reduction.
\end{itemize}

\section{Games in product form for Thai demand response program}\label{sec:W-model-Thai-DR}


In this section, we have built several games in product form and W-games of the Thai demand response program.
To keep the situation simple, we neglect the LAs in the model since they only
act as a medium of communication between the producer and the consumers.
In \S\ref{subsec:SLSF-ST}, we consider the simplest case where there is one
leader (EGAT) and only one follower, and a single call (no time dimension). 
From there, we increase the realism of the model by adding timesteps and
multiple followers.
In \S\ref{subsec:SLSL-MT}, multiple timesteps enter into the model. 
In \S\ref{subsec:SLMF-MT}, we keep the multiple timesteps and introduce more consumers into the game in product form.

\subsection{Single leader, single follower with a single timestep}
\label{subsec:SLSF-ST}

\subsubsection{Agents, actions and Nature}
Here we consider a very simple model that fits into the scope of DR as described in
Sect.~\ref{sec:Thai-DR},
with one leader agent (energy provider) and one follower agent (consumer).
The unitary reward~$\reward \in \RR_{+}$ and the baseline~$\baseline \in
\RR_{+}$ for the follower are part of the contract with the leader.

First, we describe the action sets.
On the one hand, the leader agent chooses a target reduction~$\control_\leader \in \RR_{+}$.
Thus, we write a typical action and the action set of the leader agent as
\begin{subequations}
\begin{equation}
  \control_{\leader} \in \CONTROL_{\leader} = \RR_{+}
  \eqfinv
\end{equation}
equipped with the corresponding Borel field
 \(\tribu{\Control}_{\leader}=\borel{\RR_{+}}\).
On the other hand, the follower agent plans his consumption level.
This is represented by the follower agent's action set
\begin{equation}
  \control_\follower \in \CONTROL_{\follower} = \RR_{+}\eqfinv
\end{equation}  
\end{subequations}
equipped with the corresponding Borel field \(\tribu{\Control}_{\follower}=\borel{\RR_{+}}\).

After describing the two action sets, we turn to Nature.
At this stage, we neglect the exogenous Nature (like weather conditions
affecting demand), and we focus on the types of the two players.
The sets~$\Nature^{\leader} = \RR^{m^{\leader}}$ and~$\Nature^{\follower}
= \RR^{m^{\follower}}$ describe the types of the players, and are equipped with 
the corresponding Borel fields \( \trib^{\leader} = \borel{\RR^{m^{\leader}}} \)
and \( \trib^{\follower} = \borel{\RR^{m^{\follower}}} \).
Then, we describe Nature as 
\begin{subequations}
  \begin{align}
    \Nature &= \Nature^{\leader} \times \Nature^{\follower} \\
    \trib &= \trib^{\leader} \otimes \trib^{\follower},\quad  \trib^{\leader} = \borel{\RR^{m^{\leader}}}, \quad \trib^{\follower} = \borel{\RR^{m^{\follower}}}.
  \end{align}
\end{subequations}
The types appear in the (net) production cost function
$\phi^{\leader} \colon \HISTORY \to \RR$
of the leader player,
and in the (net) utility function
$\phi^{\follower} \colon \HISTORY \to \RR$
of the follower player.
In generality, we define these two functions on the whole configuration space~\( \HISTORY\) but later we only write the influential variables as arguments.
Elements of the game in product form can be found in Table~\ref{tab:first_W-model_DR_Thailand}.

\begin{table}[H]
  \centering
  \small 
  \begin{tabular}{|l||l|c|}
    \hline
    Common knowledge
                      & Baseline consumption  &~$\baseline$ \\\hline
                      & Electricity price  &~$\price$ \\\hline
                      & Unitary reward  &~$\reward$ \\\hline\hline
                      Leader player's type 
                      & (Net) production cost function's parameters &~$\nature^{\leader}$  \\ \hline
    Leader agent's decision  
                                              & Target reduction  &~$\control_\leader$ \\ \hline\hline
                      Follower player's type 
                      & (Net) utility function's parameters &~$\nature^{\follower}$  \\ \hline
    Follower agent's decision 
                                                & Consumption   &~$\control_\follower$ \\\hline
  \end{tabular}
  \caption{Variables of the Thai DR program and their first expressions in a game in product form
  \label{tab:first_W-model_DR_Thailand}}
\end{table}

\subsubsection{Information fields}

The configuration space is given by
\begin{subequations}
  \begin{align}
  \label{equation:history-single-agent}
    \HISTORY
    &=
      \Nature^{\leader} \times \Nature^{\follower} \times
  \CONTROL_{\leader} \times \CONTROL_{\follower}
  \eqfinv
\intertext{equipped with the~$\sigma$-field}
      \tribu{\History}
    &=
      \trib^{\leader} \otimes \trib^{\follower} \otimes
  \tribu{\Control}_{\leader} \otimes \tribu{\Control}_{\follower}
  \eqfinp
  \end{align}
\end{subequations}
To represent the information structure, 
we stick to the leader-follower (Nash-Stackelberg) scheme.
The leader agent observes his own type (that is, more rigorously, the type of the leader player)
\begin{subequations}
  \begin{align}
    \tribu{\Information}_{\leader}
    &=
      \trib^{\leader} \otimes \{\emptyset,\Nature^{\follower}\} \otimes
      \{\emptyset, \CONTROL_{\leader} \} \otimes \{\emptyset,
      \CONTROL_{\follower}\}
      \eqfinv
      \intertext{whereas the follower agent observes his own type
      (that is, more rigorously, the type of the follower player) and the leader's
      decision}
      \tribu{\Information}_{\follower}
    &=
      \{\emptyset, \Nature^{\leader} \} \otimes \trib^{\follower} \otimes
      \tribu{\Control}_{\leader} \otimes \{\emptyset, \CONTROL_{\follower}\}
      \eqfinp 
  \end{align}
   \label{eq:first_W-model_DR_Thailand_Information}
\end{subequations}

\subsubsection{W-strategies}

As a consequence of~\eqref{eq:first_W-model_DR_Thailand_Information},
the leader's and follower's agents W-strategies
\( \policy_{\leader} \colon \HISTORY \to \CONTROL_{\leader} \),
\( \policy_{\follower} \colon \HISTORY \to \CONTROL_{\follower} \) 
are given, respectively, by
\begin{subequations}
  \begin{align}
      \policy_{\leader}\np{\nature^\leader,\cancel{\nature^\follower},\cancel{\control_\leader},\cancel{\control_\follower}}
    &=
    \tilde{\policy}_{\leader}\np{\nature^\leader} \eqsepv \forall\nature^\leader \in \Nature^{\leader}
      \eqfinv
    \\
   \policy_{\follower}\np{\cancel{\nature^\leader},\nature^\follower,\control_\leader,\cancel{\control_\follower}}
    &=
    \tilde{\policy}_{\follower}\np{\nature^\follower,\control_\leader} \eqsepv
      \forall\nature^\follower \in \Nature^{\follower}
      \eqsepv \forall\control_\leader \in \CONTROL_{\leader} 
      \eqfinp
  \end{align}
\end{subequations}
The solution map \( \Sol_{\policy_{\leader},\policy_{\follower}} \colon \Nature
\to \HISTORY \) is given by
\begin{equation}\label{eq:solutionmap_slsfst}
  \Sol_{\policy_{\leader},\policy_{\follower}}(\underbrace{\nature^\leader,\nature^\follower}_{\nature})
  =
  (\underbrace{\nature^\leader,\nature^\follower}_{\nature},\underbrace{\control_\leader}_{\tilde{\policy}_{\leader}(
    \nature^\leader )},
  \tilde{\policy}_{\follower}\np{\nature^\follower,\control_\leader})
  \eqfinp 
\end{equation}

\subsubsection{Objective functions}

The leader's and follower's players objective functions are given, respectively, by
\begin{subequations}\label{equation:obj-single-agent-case}
  \begin{align}
        \criterion^{\leader}(\nature^\leader,\cancel{\nature^\follower},\control_\leader,\control_\follower)
    &=
      \price \overbrace{\control_\follower}^{\textrm{consumption}}
      - \reward \overbrace{\min\{ \control_\leader, \baseline -
    \control_\follower \}}^{\textrm{effective reduction}} -
      \overbrace{\phi^{\leader}(\nature^\leader,\control_\follower)}^{\textrm{production costs}}
      \eqfinv
    \\
   \criterion^{\follower}(\cancel{\nature^\leader},\nature^\follower,\control_\leader,\control_\follower)    
    &=
      \reward \min\{ \control_\leader, \baseline - \control_\follower \} +
      \phi^{\follower}(\nature^\follower,\control_\follower) - \price \control_\follower
      \eqfinp
  \end{align}
\end{subequations}
In the above definitions, the leader player aims at minimizing her
cost~$\criterion^{\leader}$,
which is composed of the net sales subtracted by the paid reward and the (net)
production cost,
whereas the follower player aims at maximizing her payoff~$\criterion^{\follower}$,
which is composed of the reward for reduction plus utility minus energy bought.

\subsubsection{Beliefs}

We suppose that both the leader and the follower players know
their own type.
Moreover, they each have a subjective assessment of the other's type as follows
\begin{subequations}\label{eq:beliefs_slsfst}
  \begin{align}
    \Belief^{\leader}
    &=
      \delta_{\bar{\nature}^\leader}
      \otimes
      \overbrace{\Belief^{\leader}_{\follower}
}^{\substack{\textrm{subjective
      assessment}\\ \textrm{of the follower's type}}}
      \eqfinv
      \\
    \Belief^{\follower}
    &=
                           \underbrace{\Belief^{\follower}_{\leader}
                           }_{\substack{\textrm{subjective
      assessment}\\ \textrm{of the leader's type}}}
    \otimes \,
    \delta_{\bar{\nature}^\follower}
                          \eqfinp
  \end{align}
\end{subequations}

\subsubsection{Normal-form objective functions}

Finally, the normal form objective functions are given ---
based on the beliefs~\eqref{eq:beliefs_slsfst} and on the solution
map~\eqref{eq:solutionmap_slsfst}
--- as follows
\begin{subequations}
  \begin{align}
    \Criterion^{\leader}(\policy^{\leader},\policy^{\follower}) &=
                                                                  \int_{\Omega}\
                                                                  \criterion^{\leader}
                                                                  \circ
                                                                  \Sol_{\policy^{\leader},\policy^{\follower}}
                                                                  (\nature^\leader,\nature^\follower)\
                                                                  \dd\Belief^{\leader}(\nature^\leader,\nature^\follower)
                                                                  \eqfinv
    \\
    \Criterion^{\follower}(\policy^{\follower},\policy^{\leader})
                                                                &=
                                                                  \int_{\Omega}\
                                                                  \criterion^{\follower}
                                                                  \circ
                                                                  \Sol_{\policy^{\follower},\policy^{\leader}}
                                                                  (\nature^\leader,\nature^\follower)\
                                                                  \dd\Belief^{\follower}(\nature^\leader,\nature^\follower)
                                                                  \eqfinp 
  \end{align}
\end{subequations}

\subsection{Single leader, single follower with multiple timesteps}\label{subsec:SLSL-MT}

Till now, we have considered the DR problem without the time dimension.
Here, we consider  the case where the target reduction is time-dependent with timesteps~$\mytime \in \TIME = \{1,2,\dots,\Time\}$.
Typically, in the Thai system $\Time = 365$ represents the number of days in a year.

\subsubsection{Agents, actions and Nature}

Similarly to the previous case in~\S\ref{subsec:SLSF-ST},
we suppose that the unitary reward~$\reward \in \RR_{+}$ and the baseline~$\baseline \in \RR_{+}$ for the follower are provided.
The leader agent chooses, at each timestep~$\mytime \in \TIME$, a target reduction~$\control_{(\leader, \mytime)} \in \RR_{+}$.
We write a typical action and the action set of the leader agent~$(\leader,\mytime)$ at~$\mytime \in \TIME$ as
\begin{subequations}
\begin{equation}\label{equation:CONTROL-leader-mult}
  \control_{(\leader,\mytime)} \in
  \CONTROL_{(\leader,\mytime)} = \RR_{+}
  \eqfinp
\end{equation}
On the other hand, the follower agent at time~$\mytime$ decides his consumption level.
This is represented by the follower agent's action set
\begin{equation}\label{equation:CONTROL-follower-mult}
  \control_{(\follower,\mytime)} \in \CONTROL_{(\follower, \mytime)} = \RR_{+}. 
\end{equation}
These action sets are equipped with the Borel $\sigma$-fields~$\tribu{\Control}_{(\leader,\mytime)}$ and~$\tribu{\Control}_{(\follower,\mytime)}$ respectively.

Unlike in the previous subsection~\S\ref{subsec:SLSF-ST}, we introduce exogenous
nature as a product
\begin{equation}\label{equation:exo-nature}
  \exo{\Nature} = \prod_{\mytime \in \TIME} \exo{\Nature}_{\mytime} 
\end{equation}
\end{subequations}
to depict the fact that some elements may change depending on the time~$\mytime$.
Each~$\exo{\Nature}_{\mytime}$ is equipped with a~$\sigma$-field~$\exo{\trib}_{\mytime}$ and~$\exo{\Nature}$ with~$\exo{\trib} = \bigotimes_{\mytime \in \TIME} \exo{\trib}_{\mytime}$.
For instance, the (net) production cost and the (net) utility may be affected by the season.

\begin{table}[H]
  \centering
  \small
  \begin{tabular}{|l||l|c|}
    \hline
    Common knowledge  & Baseline consumption  &~$\baseline_{\mytime}$ \\\hline
                    & Electricity price  &~$\price_{\mytime}$ \\\hline
                      & Unitary reward  &~$\reward$ \\\hline\hline
    Exogenous Nature            & Weather, production units' availability, demand &~$\exo{\omega}_{\mytime}$ \\ \hline\hline
    Leader player's type 
                      & (Net) production cost function's parameters &~$\nature^{\leader}$  \\ \hline
    Leader agent's decision 
                      & Target reduction  &~$\control_{(\leader, \mytime)}$ \\ \hline\hline
                      Follower player's type 
                      & (Net) utility function's parameters &~$\nature^{\follower}$  \\ \hline
    Follower agent's decision 
                      & Consumption   &~$\control_{(\follower, \mytime)}$ \\\hline
  \end{tabular}
  \caption{Variables of the Thai DR program and their first expressions in a
    game in product form with multiple timesteps}
\end{table}

The configuration space is defined in a similar manner to that
of~\eqref{equation:history-single-agent}, but with~$\exo{\Nature}$, $\CONTROL_{\leader}$
and~$\CONTROL_{\follower}$ replaced with products
\begin{subequations}
\begin{equation}\label{equation:history-multiple-time}
  \HISTORY = \prod_{\mytime \in \TIME} \exo{\Nature}_{\mytime} 
  \times \Nature^{\leader} \times \Nature^{\follower}
  \times \prod_{\mytime \in \TIME} \CONTROL_{(\leader,\mytime)} \times
  \prod_{\mytime \in \TIME} \CONTROL_{(\follower,\mytime)}
  \eqfinp
\end{equation}
Its~$\sigma$-field is also expressed with a product as follows
\begin{equation}
  \tribu{\History} = \bigotimes_{\mytime \in \TIME} \exo{\trib}_{\mytime}
  \otimes \trib^{\leader} \otimes
  \trib^{\follower} \otimes \bigotimes_{\mytime \in \TIME}
  \tribu{\Control}_{(\leader,\mytime)} \otimes \bigotimes_{\mytime \in \TIME}
  \tribu{\Control}_{(\follower,\mytime)}
  \eqfinp
\end{equation}  
\end{subequations}

\subsubsection{Information fields}

In the framework of multiple timesteps, an agent may observe at most what happened prior to the current time~$\mytime \in \TIME$.
More precisely, the leader agent at time~$\mytime \in \TIME$ observes at most
the historical record of exogenous nature as well as the decision of each agent
up to the time~$\mytime - 1$, that is,
\begin{subequations}
\begin{equation}
  \label{equation:information-multi-timesteps-leader-general}
  \tribu{\Information}_{(\leader,\mytime)} \subset
  \bigotimes_{s=1}^{\mytime-1} \exo{\trib}_{s} \otimes \trib^{\leader}
  \otimes \bigotimes_{s=1}^{\mytime-1} \tribu{\Control}_{(\leader,s)} \otimes
  \bigotimes_{s=1}^{\mytime-1} \tribu{\Control}_{(\follower,s)}
    \eqfinp
\end{equation}
The same is true for the follower agent at time~$\mytime$ except that he could
also observe the action~$\control_{(\leader,\mytime)}$ of the leader agent, giving
\begin{equation}\label{equation:information-multi-timesteps-follower-general}
  \tribu{\Information}_{(\follower,\mytime)} \subset
  \bigotimes_{s=1}^{\mytime-1} \exo{\trib}_{s}  \otimes \trib^{\follower}
  \otimes \bigotimes_{s=1}^{\mytime} \tribu{\Control}_{(\leader,s)}
  \otimes \bigotimes_{s=1}^{\mytime-1} \tribu{\Control}_{(\follower,s)} 
\end{equation}
\end{subequations}

One of the trivial examples of the above information fields
\eqref{equation:information-multi-timesteps-leader-general} and
\eqref{equation:information-multi-timesteps-follower-general} are the one found
in the open-loop case where 
both~$\tribu{\Information}_{(\leader,\mytime)}$ and~$\tribu{\Information}_{(\follower,\mytime)}$ are trivial fields.

\subsubsection{W-strategies}

Now that we have multiple timesteps, with the information fields given by
\eqref{equation:information-multi-timesteps-leader-general} and
\eqref{equation:information-multi-timesteps-follower-general}, the leader's and
follower's agents W-strategies
\( \policy_{( \leader,\mytime)} \colon \HISTORY \to \CONTROL_{( \leader,\mytime)} \),
\( \policy_{(\follower,\mytime)} \colon \HISTORY \to \CONTROL_{(\follower,\mytime)} \) 
are given by
\begin{subequations}
  \begin{align}
    \policy_{( \leader,\mytime)}\np{\exo{\nature},\nature^\leader,\cancel{\nature^\follower},\cancel{\{\control_{(\leader, \mytime)} \}_{\mytime \in \TIME}},\cancel{\{ \control_{(\follower, \mytime)} \}_{\mytime \in \TIME}}}
    &=
    \tilde{\policy}_{(\leader, \mytime)}\np{\exo{\nature},\nature^\leader}
      \eqfinv
    \\
    \policy_{(\follower,\mytime)}\np{\exo{\nature},\cancel{\nature^\leader},\nature^\follower,\{ \control_{(\leader, \mytime)} \}_{\mytime \in \TIME},\cancel{\{ \control_{(\follower, \mytime)} \}_{\mytime \in \TIME}}}
    &=
    \tilde{\policy}_{(\follower,\mytime)}\np{\exo{\nature},\nature^\follower,\{
      \control_{(\leader, \mylocaltime)} \}_{\mylocaltime < \mytime}}
      \eqfinp
  \end{align}
\end{subequations}
The solution map is given by 
and 
\begin{subequations}
  \begin{align}
    \policy^{\leader} &= \{\policy_{(\leader,\mytime)}\}_{\mytime \in \TIME}
                        \eqfinv
    \\
    \policy^{\follower} &= \{\policy_{(\follower,\mytime)}\}_{\mytime \in \TIME}
                          \eqfinv
    \\
    \Sol_{\policy^{\leader},\policy^{\follower}}(\underbrace{\exo{\nature},\nature^\leader,\nature^\follower}_{\nature})
                      &=
                        (\underbrace{\exo{\nature},\nature^\leader,\nature^\follower}_{\nature},\underbrace{\control_\leader}_{\tilde{\policy}_{\leader}(\exo{\nature},
                        \nature^\leader
                        )},\tilde{\policy}_{\follower}\np{\exo{\nature},\nature^\follower,\control_\leader})
                              \eqfinp
  \end{align}
\end{subequations}

\subsubsection{Objective functions}

The objective functions are now additive in time:
\begin{subequations}
  \begin{equation}
  \begin{split}
  &\criterion^{\leader}(\exo{\nature},\nature^{\leader},\cancel{\nature^{\follower}},\{\control_{(\leader, \mytime)}\}_{\mytime
    \in \TIME},\{\control_{(\follower, \mytime)}\}_{\mytime
    \in \TIME})
    \\&=
      \sum_{\mytime \in \TIME} \price_{\mytime}x_{\mytime} - \reward \sum_{\mytime
      \in \TIME} \min\{ \control_{(\leader, \mytime)}, \baseline_{\mytime} -
      \control_{(\follower, \mytime)} \} - \sum_{\mytime \in
      \TIME}\phi^{\leader}(\exo{\nature}_{\mytime},\control_{(\follower, \mytime)})
      \eqfinv
      \end{split}
      \end{equation}
\begin{equation}
\begin{split}
&\criterion^{\follower}(\exo{\nature},\cancel{\nature^{\leader}},\nature^{\follower},\{\control_{(\leader, \mytime)}\}_{\mytime
    \in \TIME},\{\control_{(\follower, \mytime)}\}_{\mytime
    \in \TIME})
    \\ &=
      \reward \sum_{\mytime \in \TIME}\min\{ \control_{(\leader, \mytime)},
      \baseline_{\mytime} - \control_{(\follower, \mytime)} \} + \sum_{\mytime \in
      \TIME}\phi^{\follower}(\exo{\nature}_{\mytime},\control_{(\follower, \mytime)}) -
      \sum_{\mytime \in \TIME} \price_{\mytime}\control_{(\follower, \mytime)}
      \eqfinp
  \end{split}
\end{equation}
\end{subequations}
Notice how exogenous Nature plays a role in these functions.

\subsubsection{Beliefs and normal-form objective functions}

Similar assumption is made on the leader and follower players, that is,
they know their own types and they have a subjective assessment of the other's type.
Together with the multiple timesteps, their beliefs are presented as
\begin{subequations}
  \begin{align}
    \Belief^{\leader}
    &=
                                                                      \Belief^{\leader}_{\exogen}
                                                                      \otimes
                                                                      \delta_{\bar{\nature}^\leader}
                                                                      \otimes
                                                                      \Belief^{\leader}_{\follower}
                                                                      \eqfinv
    \\ 
    \Belief^{\follower}
    &= \Belief^{\follower}_{\exogen}
    \otimes \Belief^{\follower}_{\follower}
    \otimes \delta_{\bar{\nature}^\follower}
                                                                        \eqfinp
  \end{align}
\end{subequations}

The normal form objective function are 
\begin{subequations}
  \begin{align}
    \Criterion^{\leader}(\policy^{\leader},\policy^{\follower})
    &=
      \int_{\Nature} \criterion^{\leader} \circ
      \Sol_{\policy^{\leader},\policy^{\follower}}(\exo{\nature},\nature^\leader,\nature^\follower)
      \dd\Belief^{\leader}(\exo{\nature},\nature^\leader,\nature^\follower)
      \eqfinv
    \\
    \Criterion^{\follower}(\policy^{\follower},\policy^{\leader})
    &=
      \int_{\Nature} \criterion^{\follower} \circ
      \Sol_{\policy^{\follower},\policy^{\leader}}(\exo{\nature},\nature^\leader,\nature^\follower)
      \dd\Belief^{\leader}(\exo{\nature},\nature^\leader,\nature^\follower)
      \eqfinp
  \end{align}
\end{subequations}

\subsection{Single leader, multiple followers with multiple timesteps}\label{subsec:SLMF-MT}

Here, we consider the case where there are multiple followers
(consumers)
and the leader's reduction target is to be met by the total reduction of all the followers.

\subsubsection{Agents, actions and nature}

The single leader remains the same with the earlier subsections, but we increase the number of followers (as players).
Suppose that~$\Follower$ denotes the set of followers and each follower~$\follower \in \Follower$ possesses agents at different timesteps~$\mytime \in \TIME$, denoted by~$(\follower,\mytime)$.
Exploiting the modularity of games in product form, at each~$\mytime \in \TIME$, 
the action sets of~$(\leader,\mytime)$ and~$(\follower,\mytime)$ are given as in
\eqref{equation:CONTROL-leader-mult} and \eqref{equation:CONTROL-follower-mult}
and the exogenous nature are given by~\eqref{equation:exo-nature}.

The configuration space is defined in a similar manner to that
of~\eqref{equation:history-multiple-time}, but with $\Nature^{\follower}$ and
$\CONTROL_{(\follower,\mytime)}$
replaced with the products
\( \prod_{\follower \in \Follower}\Nature^{\follower} \)
and \( \prod_{\follower \in \Follower}\CONTROL_{(\follower,\mytime)} \), giving 
\begin{subequations}
\begin{equation}
  \HISTORY = \prod_{\mytime \in \TIME} \exo{\Nature}_{\mytime} 
  \times \Nature^{\leader} \times
  \prod_{\follower \in \Follower}\Nature^{\follower}
  \times \prod_{\mytime \in \TIME} \CONTROL_{(\leader,\mytime)} \times
  \prod_{\follower \in \Follower} \prod_{\mytime \in \TIME} 
\CONTROL_{(\follower,\mytime)} 
  \eqfinv
\end{equation}
with $\sigma$-field also expressed with a product as follows
\begin{equation}
  \tribu{\History} = \bigotimes_{\mytime \in \TIME} \exo{\trib}_{\mytime} 
  \otimes \trib^{\leader} \otimes
  \bigotimes_{\follower \in \Follower} \trib^{\follower} \otimes \bigotimes_{\mytime \in \TIME}
  \tribu{\Control}_{(\leader,\mytime)} \otimes \bigotimes_{\follower \in \Follower} \bigotimes_{\mytime \in \TIME}
  \tribu{\Control}_{(\follower,\mytime)}
  \eqfinp
\end{equation}  
\end{subequations}

\subsubsection{Information fields and objective functions}

Using again the modularity of games in product form,
the information fields~$\tribu{\Information}_{(\leader,\mytime)}$
and~$\tribu{\Information}_{(\follower,\mytime)}$ are defined by
(compare with~\eqref{equation:information-multi-timesteps-leader-general}
and \eqref{equation:information-multi-timesteps-follower-general})
\begin{subequations}
  \begin{align}
    \tribu{\Information}_{(\leader,\mytime)} \subset &
    \bigotimes_{s=1}^{\mytime-1} \exo{\trib}_{s} \otimes \trib^{\leader}
    \otimes \bigotimes_{s=1}^{\mytime-1} \tribu{\Control}_{(\leader,s)} \otimes \bigotimes_{\follower \in \Follower}\bigotimes_{s=1}^{\mytime-1} \tribu{\Control}_{(\follower,s)} \\
    \tribu{\Information}_{(\follower,\mytime)} \subset &
    \bigotimes_{s=1}^{\mytime-1} \exo{\trib}_{s}  \otimes \trib^{\follower}
    \otimes \bigotimes_{s=1}^{\mytime} \tribu{\Control}_{(\leader,s)}
    \otimes \bigotimes_{\follower \in \Follower} \bigotimes_{s=1}^{\mytime-1} \tribu{\Control}_{(\follower,s)}.
  \end{align}
\end{subequations}
Notice that the increasing number of followers turns 
the information fields from \eqref{equation:information-multi-timesteps-leader-general} and \eqref{equation:information-multi-timesteps-follower-general} into the above products.

On the other hand, the introduction of multiple followers induces the additivity of cost terms over both time~$\mytime \in \TIME$ and followers~$\follower \in \Follower$:
\begin{subequations}
  \begin{equation}
  \begin{split}
    &\criterion^{\leader}(\exo{\nature},\nature^{\leader},\cancel{\nature^{\follower}},\{\control_{(\leader, \mytime)}\}_{\mytime \in \TIME},\{\control_{(\follower, \mytime)}\}_{\substack{\follower \in \Follower \\ \mytime \in \TIME }})
    \\ &=
      \sum_{\substack{\mytime
      \in \TIME \\ \follower \in
    \Follower}} \price_{\mytime}\control_{(\follower, \mytime)} - \reward \sum_{\mytime
      \in \TIME} \min\{ \control_{(\leader, \mytime)}, \baseline_{\mytime} -
      \control_{(\follower, \mytime)} \} - \sum_{\mytime \in \TIME}\phi^{\leader}(\exo{\nature}_{\mytime},\nature^\leader,\control_{(\follower, \mytime)})
    \eqfinv
\end{split}
\end{equation}
\begin{equation}
\begin{split}
    &\criterion^{\follower}(\exo{\nature},\cancel{\nature^{\leader}},\nature^{\follower},\{\control_{(\leader, \mytime)}\}_{\mytime \in \TIME}, \{\control_{(\follower, \mytime)}\}_{\substack{\follower \in \Follower \\ \mytime \in \TIME }})
    \\ &= \reward \sum_{\mytime \in \TIME}\min\{ \control_{(\leader, \mytime)},
      \baseline_{\mytime} - \control_{(\follower, \mytime)} \} + \sum_{\substack{\mytime
      \in \TIME \\ \follower \in
    \Follower}}\phi^{\follower}(\exo{\nature}_{\mytime},\nature^\follower,\control_{(\follower, \mytime)})
    - \sum_{\mytime \in \TIME} \price_{\mytime}\control_{(\follower, \mytime)}
        \eqfinp
  \end{split}
  \end{equation}
  
\end{subequations}

\subsubsection{W-strategies}

Now that we have multiple timesteps, with the information fields given by
\eqref{equation:information-multi-timesteps-leader-general} and
\eqref{equation:information-multi-timesteps-follower-general}, the leader's and
follower's agents W-strategies
\( \policy_{( \leader,\mytime)} \colon \HISTORY \to \CONTROL_{( \leader,\mytime)} \),
\( \policy_{(\follower,\mytime)} \colon \HISTORY \to \CONTROL_{(\follower,\mytime)} \) 
are given by
\begin{subequations}
  \begin{equation}
  \begin{split}
  &\policy_{( \leader,\mytime)}\np{\{\exo{\nature}_{\mytime}\}_{\mytime \in \TIME},\nature^\leader,\cancel{\{\nature^\follower\}_{\follower \in \Follower}},\cancel{\{\control_{(\leader, \mytime)} \}_{\mytime \in \TIME}},\cancel{\{\control_{(\follower, \mytime)}\}_{\substack{\follower \in \Follower \\ \mytime \in \TIME }}}}
    \\ &=
    \tilde{\policy}_{(\leader,\mytime)}\np{\{\exo{\nature}_{\mytime}\}_{\mytime \in \TIME},\nature^\leader}
      \eqfinv
    \end{split}
    \end{equation}
    \begin{equation}
    \begin{split}
    &\policy_{(\follower,\mytime)}\np{\{\exo{\nature}_{\mytime}\}_{\mytime \in \TIME},\cancel{\nature^\leader},\nature^\follower,\cancel{\{\nature^{\mathtt{g}}\}_{\mathtt{g} \in \Follower\setminus\{\follower\}}},\{ \control_{(\leader, \mytime)} \}_{\mytime \in \TIME},\cancel{\{\control_{(\follower, \mytime)}\}_{\substack{\follower \in \Follower \\ \mytime \in \TIME }}}}
    \\ &=
    \tilde{\policy}_{(\follower,\mytime)}\np{\{\exo{\nature}_{\mytime}\}_{\mytime
      \in \TIME},\nature^\follower,\{ \control_{(\leader, \mytime)} \}_{\mytime \in
      \TIME}}
      \eqsepv \forall\follower \in \Follower
      \eqfinp
  \end{split}
  \end{equation}
\end{subequations}
The corresponding solution map is given by
\begin{subequations}
  \begin{align}
    \policy^{\leader} &= \{\policy_{(\leader,\mytime)}\}_{\mytime \in \TIME}
                        \eqfinv
    \\
    \policy^{\follower} &= \{\policy_{(\follower,\mytime)}\}_{\mytime \in \TIME}
                          \eqsepv \forall\follower \in \Follower \eqfinv
                          \\
    \policy &= \{\policy^{p} \}_{p \in \{\leader\} \cup \Follower}
              \eqfinv
    \\
    \Sol_{\policy}(\underbrace{\{\exo{\nature}_{\mytime}\}_{\mytime \in
    \TIME},\nature^\leader,\{\nature^\follower\}_{\follower \in
    \Follower}}_{\nature}) &= (\underbrace{\{\exo{\nature}_{\mytime}\}_{\mytime
                             \in
                             \TIME},\nature^\leader,\{\nature^\follower\}_{\follower
                             \in
                             \Follower}}_{\nature},\underbrace{\control_\leader}_{\tilde{\policy}_{\leader}(\exo{\nature},
                             \nature^\leader
                             )},\{\tilde{\policy}_{\follower}\np{\exo{\nature},\nature^\follower,\control_\leader}\}_{\follower
                             \in \Follower})
                             \eqfinp
  \end{align}
\end{subequations}

\subsubsection{Beliefs}

One can simply mimic the belief from the previous~\S\ref{subsec:SLSL-MT},
with a modification on the time-dependent exogenous nature~$\{\exo{\nature}_{\mytime}\}_{\mytime \in \TIME}$:
\begin{subequations}
  \begin{align}
    \Belief^{\leader}
    &=
                                                                               \bigotimes_{\mytime
                                                                               \in
                                                                               \TIME}
                                                                               \Belief^{\leader}_{(\exogen,\mytime)}
                                                                               \otimes
                                                                               \delta_{\tilde{\nature^\leader}}
                                                                               \otimes
                                                                               \bigotimes_{\follower
                                                                               \in
                                                                               \Follower}\Belief^{\leader}_{\follower}
                                                                               \eqfinv
    \\ 
    \Belief^{\follower}
    &=
                                                                               \bigotimes_{\mytime
                                                                               \in
                                                                               \TIME}
                                                                               \Belief^{\follower}_{(\exogen,\mytime)}
                                                                               \otimes
                                                                               \Belief^{\follower}_{\follower}
                                                                               \otimes
                                                                               \delta_{\bar{\nature}^\follower}
                                                                               \bigotimes_{\mathtt{g}
                                                                               \in
                                                                               \Follower}\Belief^{\leader}_{\mathtt{g}}
                                                                               \eqfinp
  \end{align}
\end{subequations}
The normal form objective functions are still given by the same formulae as before but with the new beliefs described above.

\section*{Conclusion}

Till now, there has been no mathematical formulation of the Thai demand response
system in the literature. To design it, we have used the (product) modularity of
games in product form starting with the simple case with a single leader, and
then including time, and finally multiple followers.

From the process of building a game in product form throughout this section,
we suggest a guideline in the building.
One should start with identifying players and agents,
together with their decision variables, and then their objective functions.
From here, one should get an idea for the required elements that could be
structured as exogenous nature,
common knowledge, and players' types.
Once we have decided which (and possibly when) the information is revealed to an
agent, his information field could then be set up.
The belief of a player is then finally formed, based on what she knows.